# Path distributions for describing eigenstates of orbital angular momentum

Randall M. Feenstra
Dept. Physics, Carnegie Mellon University, Pittsburgh, PA 15213

## Abstract

The manner in which probability amplitudes of paths sum up to form wave functions of orbital angular momentum eigenstates is described. Using a generalization of stationary-phase analysis, distributions are derived that provide a measure of how paths contribute towards any given eigenstate. In the limit of long travel-time, these distributions turn out to be real-valued, non-negative functions of a momentum variable that describes classical travel between the endpoints of a path (with the paths explicitly including nonclassical ones, described in terms of *elastica*). The distributions are functions of both this characteristic momentum as well as a polar angle that provides a tilt, relative to the $z$-axis of the chosen coordinate system, of the geodesic that connects the endpoints. The resulting description provides a replacement for the well-known "vector model" for describing orbital angular momentum, and importantly, it includes treatment of the case when the quantum number $\ell$ is zero (*i.e.*, $s$-states).

## I. Introduction

Understanding eigenstates of orbital angular momentum serves as a foundational problem in any undergraduate quantum mechanics course. The associated eigenstates apply to motion in 3 dimensions ($\mathbb{R}^3$) of a rigid rotator in which all masses concentrated along a single axis (so that the moment of inertia associated with that axis is zero and the two other moments of inertia, denoted $I$, are identical). Equivalently, the same problem can be viewed in terms of the motion of a particle on the surface of a sphere, a 2-dimensional curved space known as a "two-sphere" (denoted by $S^2$), so long as we take the mass of the particle $M$ and the radius of the sphere $R$ to be related to the moment of inertia $I$ of the rotator by

$$MR^2 = I \, . \tag{1}$$

Operator-based methods for studying this problem, *e.g.* using raising and lowering operators, provide a compact means of analyzing the problem. Alternatively, we can seek to describe it in terms of quantum paths [1,2,3,4,5,6], *i.e.* following the call from a number authors for greater incorporation of path integrals in undergraduate quantum mechanics education [7,8,9,10,11].

A closed-form, path-based expression for the time-dependent propagator describing motion in $S^2$ is known [12] (by closed-form, we mean a propagator that is written in terms of a relatively small set of paths – a finite or countably infinite number, as opposed to an uncountably infinite number in a time-sliced expression). Utilizing this form, we derive here *distributions* of paths that provide an exact description of the eigenstates of orbital angular momentum. The expressions that we obtain are valid for all values of travel time, but they are most easily interpreted in the limit of long travel time. In that case, they turn out to be real-valued and non-negative, thereby providing an exact and rigorous *measure* [5] of the contribution that paths make towards forming a given wave function. These distributions provide a replacement for the *ad hoc* "vector model" that is sometimes employed for describing eigenstates of angular momentum [13,14,15,16]. Importantly, we obtain a distribution for the case of $s$-states (having value of quantum number $\ell = 0$), something that is *not* described in any way by the vector model.

The methodology that we utilize for obtaining these distributions of paths is a generalization of the method of stationary phase, as developed in our recent work and applied there to various 1-dimensional problems [17]. For free-particle motion in $\mathbb{R}^1$, the path distribution trivially turns out to be a delta-function, $\delta(p_{\mathrm{c}} - \hbar k)$, where $\hbar k$ is the energy eigenvalue and $p_{\mathrm{c}}$ is a classical momentum that is obtained from the endpoints of the path, $x_0$ and $x_{\mathrm{f}}$, according to

$$p_{\mathrm{c}} = \frac{M(x_{\mathrm{f}} - x_0)}{T} \tag{2}$$

with $T$ being the travel time. The path distributions for motion around a circle ($S^1$) similarly takes the form of a single delta-function, whereas for reflection off of a hard wall and for motion in an infinite square-well the distributions consist of two delta-functions [17]. For a harmonic oscillator, the $p_{\mathrm{c}}$ variable that the distributions are expressed in terms of refers to the *maximum* classical momentum along a trajectory that connects the endpoints of a path, and the path distribution in this case was found to consists of a somewhat broad, continuous function of $p_{\mathrm{c}}$, peaked at a value that corresponds to a classical energy which equals the energy eigenvalue for the state being investigated [17].

Unlike those simple 1-dimensional problems, for which path-based propagators are well known and extensively discussed in the literature, it turns out for motion in $S^2$ that the propagator we utilize in our analysis, although known [12], has *not* been closely examined in the literature. Hence, we discuss the properties of this propagator in some detail, both in Section II and in a series of Appendices. We compare the exact propagator to the corresponding semiclassical form (which is only approximate) [18], and we demonstrate that nonclassical paths contained within the exact form can be described as *elastica* [19], elastic curves that minimize the bending energy integrated along the path. We also examine two types of prefactors that occur in the propagator, one that varies with winding number $n$, and another that varies with travel time $T$. Both of these prefactors play an absolutely essential role in determining the time-dependence of the eigenstates.

In Section III we proceed with the derivation of path distributions that describe eigenstates of orbital angular momentum. These distributions must be expressed in terms of two variables, one of which we denote by $L_{\mathrm{c}}$, a classical angular momentum determined by the endpoints of the path (analogous to Eq. (2)). The other variable we take to be the polar angle (tilt) of the plane containing the geodesic that connects the endpoints of the travel. Importantly, even though the paths are described as having *endpoints* whose separation varies classically with $T$, the entire paths themselves are *not* required to be classical trajectories. Indeed, as just mentioned, the exact propagator for motion in $S^2$ explicitly contains nonclassical paths (*elastica*), and these are included in our analysis. In addition, various prefactors within the propagator implicitly incorporate the effects of all possible paths that differ (fluctuate) from the *elastica*. In Section IV we discuss some aspects of the distributions that we obtain, and in Section V we summarize our work.

Obtaining the paths distributions that describe eigenstates of orbital angular momentum turns out to be rather involved. We therefore provide a preview of our final result here, as motivation for the lengthy analysis. We obtain distributions that are somewhat analogous to those in the "vector model", in that they portray eigenstates in terms of distributions of angular momenta vectors [13,14,15]. Even though we find that there are two significant errors in the vector model, we nonetheless consider the underlying representation used therein – picturing the motion within a given state as somehow composed of a distribution of motions along great-circle geodesics – to be an appealing and useful one.

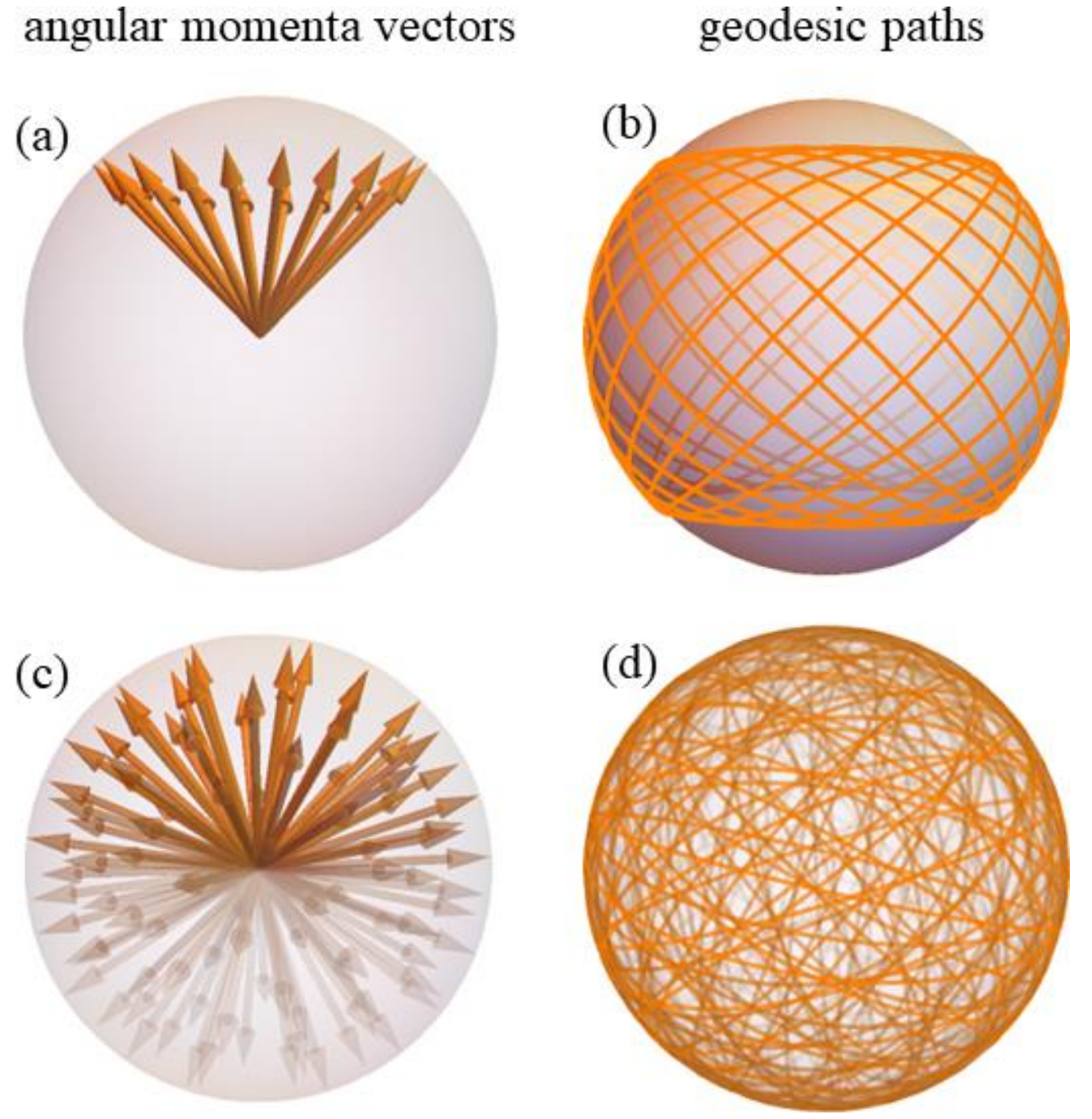

FIG 1. Comparison of distributions of geodesic (great circle) orbits for describing an eigenstates with $\ell = 1$ and $m = 1$: (a) conventional result from the vector model, showing a distribution of vectors $\mathbf{L_c}$ all with polar angle of $\theta = \pi/4$; (b) position-space orbits corresponding to the set of $\mathbf{L_c}$ vectors shown in (a); (c) set of $\mathbf{L_c}$ vectors as determined by the path distribution derived in the present work; (d) position-space orbits corresponding to the set of $\mathbf{L_c}$ vectors shown in (c).

In Fig. 1(a) we picture the usual sort of diagram obtained from the vector model, for a specific eigenstate with $\ell = 1$ and $m = 1$. Within this model, the motion is envisioned as being composed of constant-speed motion along great-circle (geodesic) paths. The associated angular momentum vectors form a cone, as shown, with angle of $\cos^{-1}\left(m/\sqrt{\ell(\ell+1)}\right)$ relative to a vertical, $L_z$-axis. Each great-circle path pictured by a $\mathbf{L_c}$ vector in Fig. 1(a) would apply to only a certain set of possible initial and final locations, *i.e.* along the great-circle path, but integrating such paths over all initial and final locations (as done for the type of distributions that we construct) might then yield at least a first step towards forming a meaningful distribution. It turns out that we must supplement these great-circle paths with the aforementioned *elastica* curves associated with the $S^2$ propagator, but nonetheless it suffices to specify a distribution just in terms of the *endpoints* of the paths, with these endpoints then being located on great circles. In this sense, the $\mathbf{L_c}$ vectors displayed in Fig. 1(a) (and in Fig. 1(c)) can be referred to as *characteristic* vectors associated with the *endpoints* of the travel. However, it is clear that limiting ourselves to the $\mathbf{L_c}$ vectors of Fig. 1(a) leads to a very substantial limitation in the types of motion that we can describe. In particular, in Fig. 1(b) we draw great-circle paths associated with these $\mathbf{L_c}$ vectors (yielding a picture nearly identical to that shown by Thaller [15]); clearly these paths do not span all of $S^2$.

To better describe the motion, the distribution of $\mathbf{L_c}$ vectors must be expanded beyond the set shown in Fig. 1(a). Labeling those vectors in terms of their angle relative to the *z*-axis, which we will refer to as $\theta_{\mathbf{L_c}}$, we thus recognize that a distribution containing just a single value (delta-function distribution) in $\theta_{\mathbf{L_c}}$ is inadequate; rather, the distribution as a function of $\theta_{\mathbf{L_c}}$ must be allowed to broaden. In Fig. 1(c) we picture the distribution of paths that we obtain in Section III(B) for describing the $\ell = 1$ and $m = 1$ eigenstate. In this picture, we use a dense set of $\mathbf{L_c}$ vectors with randomly chosen directions having endpoints that uniformly extend over the sphere, with the *intensity* of the respective vectors then representing the *magnitude* of the distribution at the particular value of $\theta_{\mathbf{L_c}}$ (in Section III we will show this distribution more quantitatively by utilizing a polar plot). This distribution does indeed have substantial breadth in $\theta_{\mathbf{L}_c}$, as required to fully describe the eigenstate. The position-space picture corresponding to Fig. 1(c) is shown in Fig. 1(d); clearly these paths extend over the entire $S^2$ space.

## II. The propagator for motion in $S^2$

### A. Definition of the propagator

To describe free-particle motion in $S^2$, we utilize spherical coordinates $(\theta, \phi)$. Since we are discussing free-particle motion, the $S^2$ propagator will depend only on the angular *difference* between initial and final locations. Denoting the angle between initial and final locations by $\gamma$, it can be obtained from unit vectors extending to the initial and final locations, $\mathbf{u}_0$ and $\mathbf{u}_\mathrm{f}$ respectively, according to

$$\gamma = \cos^{-1}(\mathbf{u}_0 \cdot \mathbf{u}_\mathrm{f}) \tag{3a}$$

$$= \cos^{-1}(\cos\theta_0 \cos\theta_\mathrm{f} + \sin\theta_0 \sin\theta_\mathrm{f} \cos(\phi_0 - \phi_\mathrm{f})) \tag{3b}$$

where the principal branch of $\cos^{-1}$, varying between 0 and $\pi$, is utilized. We denote the resulting propagator as $K^{(S^2)}(\gamma, T)$.

The exact form of the propagator for motion in $S^2$ was obtained by Camporesi in 1990 [12]; it can be derived from the propagator for motion in the "group space" SO(3) as obtained by Schulman in 1968 [20], using the topological method of Dowker [21] to relate that propagator to the one for $S^2$ by means of the homotopy $S^2 \cong$ SO(3)/SO(2). There is little mention of this form of the $S^2$ propagator in the literature (it does not appear in the otherwise very comprehensive *Handbook of Feynman Path Integrals* [4], although it *is* discussed in Ref. [22] in connection with coherent states). This propagator was derived using the language of "heat kernels", as applicable to quantum field theory, and as such a factor of 2 revision in the time variable (as mentioned *e.g.* in [23]), is needed such that this propagator applies to the single-particle, nonrelativistic situation that we are considering here. With that, and including factors of $I$ and $\hbar$ as appropriate, the exact propagator appears as

$$K^{(S^2)}(\gamma, T) = \sqrt{2}\, e^{i\hbar T/8I} \left(\frac{I}{2\pi\hbar i T}\right)^{3/2} \sum_{n=-\infty}^{\infty} (-1)^n \int_{\gamma}^{\pi} \frac{d\alpha\ \alpha_n\ e^{iI\alpha_n^2/2\hbar T}}{(\cos\gamma - \cos\alpha)^{1/2}} \tag{4a}$$

with

$$\alpha_n \equiv \alpha + 2\pi n\ . \tag{4b}$$

The integer $n$ in these expressions is a *winding number*. Some explanation of this winding number is necessary, since it differs significantly from what occurs for motion around a circle ($\mathbb{S}^1$). To explain this difference, it is convenient to first introduce the semiclassical form of the propagator for motion in $S^2$, which we now turn to.

### B. Comparison with semiclassical propagator

By definition, the probability amplitude terms of a semiclassical propagator are written only using paths that are classical trajectories. Prefactors of these terms are then given by a product of a fluctuation factor – the square-root of a Van Vleck-Pauli-Morette (VPM) determinant [2,3,24,25,26] – which describes the influence of all paths other than classical trajectories, along with a homotopy factor if needed [27,28]. The semiclassical propagator for $S^2$ was derived by Marinov and Terentyev [18] for the usual case of short travel-times (for which the semiclassical form is invariably exact [2]). Using Eq. (3) for the angular difference $\gamma$, and including appropriate factors of $I$ and $\hbar$, the short-time expression appears as

$$K_{\rm sc}^{(S^2)}(\gamma,T) = \frac{I}{2\pi\hbar iT}\sum_{n=-\infty}^{\infty}\left[\frac{\gamma_n}{\sin\gamma_n}\right]^{1/2} e^{iI\gamma_n^2/2\hbar T} \tag{5a}$$

with

$$\gamma_n \equiv \gamma + 2\pi n \tag{5b}$$

where the integer $n$ is, again, a winding number for the path.

In principle a semiclassical propagator can always be applied consecutively to a series of small travel times in order to obtain a result which is valid for long times [2]. The resulting expression for $S^2$ as provided by Marinov and Terentyev [18] is completely intractable. Alternatively, we can just extend Eq. (5a) by generalizing the $(\gamma_n/\sin\gamma_n)^{1/2}$ factor there such that it applies to travel times that are not necessarily small ($\gamma$ values that are not necessarily small), thereby necessitating a more complete specification of the branch of the square root. When $\gamma_n/\sin\gamma_n < 0$ we must utilize $(\gamma_n/\sin\gamma_n)^{1/2} = \pm i|\gamma_n/\sin\gamma_n|^{1/2}$, and even when $\gamma_n/\sin\gamma_n > 0$ we should, in general, use $\pm|\gamma_n/\sin\gamma_n|^{1/2}$ rather than just $+|\gamma_n/\sin\gamma_n|^{1/2}$ for $(\gamma_n/\sin\gamma_n)^{1/2}$. As a means of specifying the phase, we tentatively re-express this factor in the conventional manner [2,3,29,30]

$$\left(\frac{\gamma_n}{\sin\gamma_n}\right)^{1/2} \rightarrow \left|\frac{\gamma_n}{\sin\gamma_n}\right|^{1/2} e^{-i\pi\nu(\gamma_n)/2} \tag{6}$$

where the integer $\nu$, known as the Maslov index, is a function of $\gamma_n$. Generally speaking, this integer increases by a step of some size each time the prefactor, as a function of $\gamma_n$, passes through a divergence. Consistent with the possible phases for $(\gamma_n/\sin\gamma_n)^{1/2}$ just stated, the value of this step size is simply 1 for the case of motion in $S^2$.

In this way, the value of $\nu(\gamma_n)$ is given by the number of times the path followed by a particle, starting from an initial location and following a great-circle trajectory extending over an angle $|\gamma_n|$, extends either through the antipodal point (on the opposite side of the sphere from the initial point) or through the initial point itself. Thus, we can write

$$\nu(\gamma_n) = \left\lfloor\frac{|\gamma_n|}{\pi}\right\rfloor \tag{7}$$

where $\lfloor u\rfloor$ is the floor function, defined as the largest integer less than or equal to $u$. With this specification, a revised form of the semiclassical propagator, appropriate for long times, is

$$K_{\rm sc}^{(S^2)}(\gamma,T) = \frac{I}{2\pi\hbar iT}\sum_{n=-\infty}^{\infty}\left|\frac{\gamma_n}{\sin\gamma_n}\right|^{1/2} e^{-i\pi\lfloor|\gamma_n|/\pi\rfloor/2}\, e^{iI\gamma_n^2/2\hbar T} \quad . \tag{8}$$

To confirm the correctness of Eqs. (6) and (7), we perform a term by term comparison of the semiclassical and the exact propagators. For this purpose, we write the former as

$$K_{\mathrm{sc}}^{(S^2)}(\gamma,T) = \frac{I}{2\pi\hbar i T}\sum_{n=-\infty}^{\infty} f_{n,\mathrm{sc}}(\gamma,T) \tag{9a}$$

with

$$f_{n,\mathrm{sc}}(\gamma,T) = \left|\frac{\gamma_n}{\sin\gamma_n}\right|^{1/2} e^{-i\pi\lfloor|\gamma_n|/\pi\rfloor/2}\; e^{iI\gamma_n^2/2\hbar T} \tag{9b}$$

and the latter as

$$K^{(S^2)}(\gamma,T) = \sqrt{2}\,\left(\frac{I}{2\pi\hbar i T}\right)^{3/2}\sum_{n=-\infty}^{\infty} f_{n,\mathrm{ex}}(\gamma,T) \tag{10a}$$

with

$$f_{n,\mathrm{ex}}(\gamma,T) = (-1)^n\, e^{i\hbar T/8I}\int_{\gamma}^{\pi}\frac{d\alpha\;(\alpha+2\pi n)e^{iI(\alpha+2\pi n)^2/2\hbar T}}{(\cos\gamma-\cos\alpha)^{1/2}}\quad . \tag{10b}$$

In comparing the terms in Eqs. (9b) and (10b), we plot them as a function of $|\gamma_n| \equiv |\gamma + 2\pi n|$, using an ordering of the $n$ values of $n = 0, -1, 1, -2, 2, \ldots$. This ordering leads to a monatomic increase in $|\gamma_n|$ values, *i.e.*, with $0 \le \gamma \le \pi$ then values of $n = 0, -1, 1$, *etc*. lead, respectively, to ranges of $|\gamma_n|$ ranging between 0 to $\pi$, $\pi$ to $2\pi$, $2\pi$ to $3\pi$, *etc*. Figure 2 shows the magnitude and phase of both Eqs. (9) and (10), as a function of $|\gamma + 2\pi n|$. Comparing the results for these propagators, we see that roughly comparable values for both the magnitude and the phase of the terms is obtained: The semiclassical propagator displays $-\pi/2$ discontinuous changes in phase at each integer value of $|\gamma + 2\pi n|/\pi$, whereas the exact form displays discontinuities of $\pm\pi$ at even-integer values of $|\gamma + 2\pi n|/\pi$ (*i.e.* arising from the $(-1)^n$ term combined with the fact that $(\alpha + 2\pi n)$ reverses sign between positive and negative $n$). Choosing for convenience of discussion the lower sign of the $\pm\pi$, we see that both propagators display the same overall amount of phase change, $-\pi$, per $2\pi$ change in $|\gamma + 2\pi n|$. Based on this agreement, we conclude that our specification of phase (Eqs. (6) and (7)) for the semiclassical propagator appears to be appropriate, and furthermore, that the ordering of winding numbers on the abscissa of Fig. 2 is also correct. Regarding the detailed difference between the phase of the semiclassical and the exact propagators in Fig. 2, we find that this arises from the fact that the former is valid only for large values of angular momentum, as discussed in Appendix I.

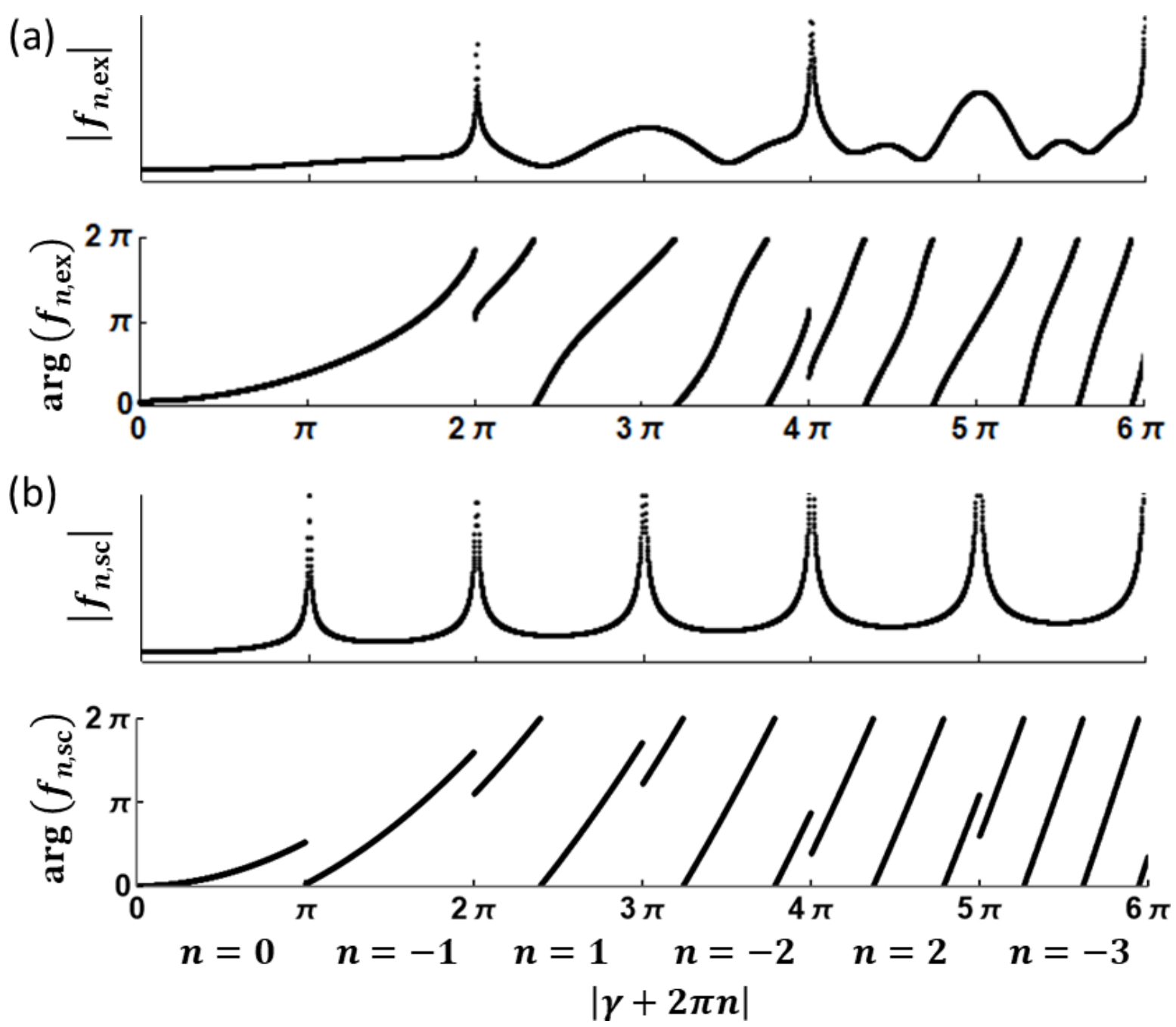

FIG 2. Comparison of (a) the exact propagator, and (b) the semiclassical propagator, for free-particle motion in $S^2$. Magnitude and phase (modulo $2\pi$) of each propagator is shown, plotted as a function of $|\gamma + 2\pi n|$ where $\gamma$ is the angle subtended between initial and final locations ($0 \le \gamma \le \pi$) and $n$ is the winding number. The travel time is taken to be $T = 3$, using units of $\hbar = I = 1$.

### C. Winding numbers for motion on a sphere

As illustrated in Fig. 2, the sequence of winding number for motion on a sphere is given by $n = 0, -1, 1, -2, 2, \ldots$ as a function of the total angle traveled. Let us refer to this total angle traveled as $\tilde{\gamma}$, and furthermore, let us consider this to be a *signed* variable (as needed for our analysis in Section III). Hence, for a given azimuthal direction leaving some point on the sphere (*i.e.* a given value of $\phi$ if leaving from $\theta = 0$), a small positive increment in $\tilde{\gamma}$ corresponds to motion in the direction of that chosen azimuth, whereas a small negative increment in $\tilde{\gamma}$ corresponds to motion in the opposite direction. Thus, for the results of Fig. 2 which are plotted as a function of $|\gamma_n| = |\gamma + 2\pi n|$, we clearly have $|\tilde{\gamma}| = |\gamma_n|$. More generally, varying $\tilde{\gamma}$ from some negative value (say, $-5\pi$), through 0, and then to some positive value (say, $5\pi$), we expect winding numbers that vary according to $n = 2, -2, 1, -1, 0, 0, -1, 1, -2, 2$ as the value of $\tilde{\gamma}$ changes by $\pi$. This result is of such paramount importance in all of our analysis that we further examine it here. We compare it to what occurs for motion on a circle ($S^1$), and in so doing, we arrive at the same result just stated for the sphere, but by an argument that is slightly different than the one presented in connection with Fig. 2.

For motion in $S^1$, winding numbers simply increase by integer values for each $2\pi$ increase in total angular distance traveled between initial and final locations along a path, since measurements of an angular distance in either a clockwise or counterclockwise manner are topologically distinct [2,18,20,31,32]. However, the situation for $S^2$ is different. Consider measuring the minimum angular separation between two points in $S^2$ utilizing a polar angle $\theta$ (*i.e.* with one point located at $\theta = 0$, so that the $\theta$ value for the other point is equal to $\gamma$ from Eq. (3)). We are free to change the azimuthal angle $\phi$ when measuring this minimum angular separation. Consider a point moving continuously around a sphere, on a great-circle trajectory. With $\tilde{\gamma}$ denoting the *total* angular distance traveled, then the angle $\gamma$ as given by Eq. (3) will vary as a triangle wave, as shown in Fig. 3(a). Using that variation, we can obtain values of the winding number as a function of $\tilde{\gamma}$ simply from the relationship $|\tilde{\gamma}| = |\gamma + 2\pi n|$. The resultant $n$ values are shown in Fig. 3(b), varying in just the same manner as in Fig. 2. Expressions for $\gamma$ and $n$ as a function of $\tilde{\gamma}$ that match the behavior shown in Fig. 3 are found by inspection to be

$$\gamma = |[(\tilde{\gamma} - \pi) \text{ modulo } 2\pi] - \pi| \tag{11a}$$

$$n = \left\lfloor \frac{\tilde{\gamma}}{2\pi} \right\rfloor - \left(1 + 2\left\lfloor \frac{\tilde{\gamma}}{2\pi} \right\rfloor\right)\left(\left\lfloor \frac{\tilde{\gamma}}{\pi} \right\rfloor \text{ modulo } 2\right) \tag{11b}$$

where modulo of a negative number is defined as being the *positive* remainder upon division (*e.g.*, $-2.5 \text{ modulo } 2 = +1.5$).

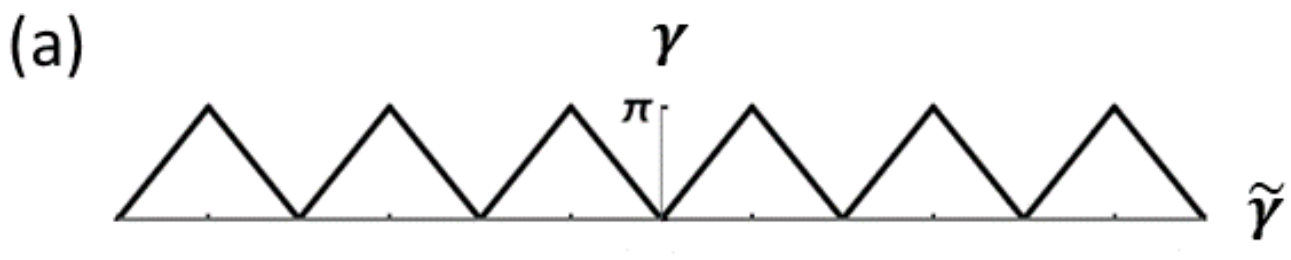

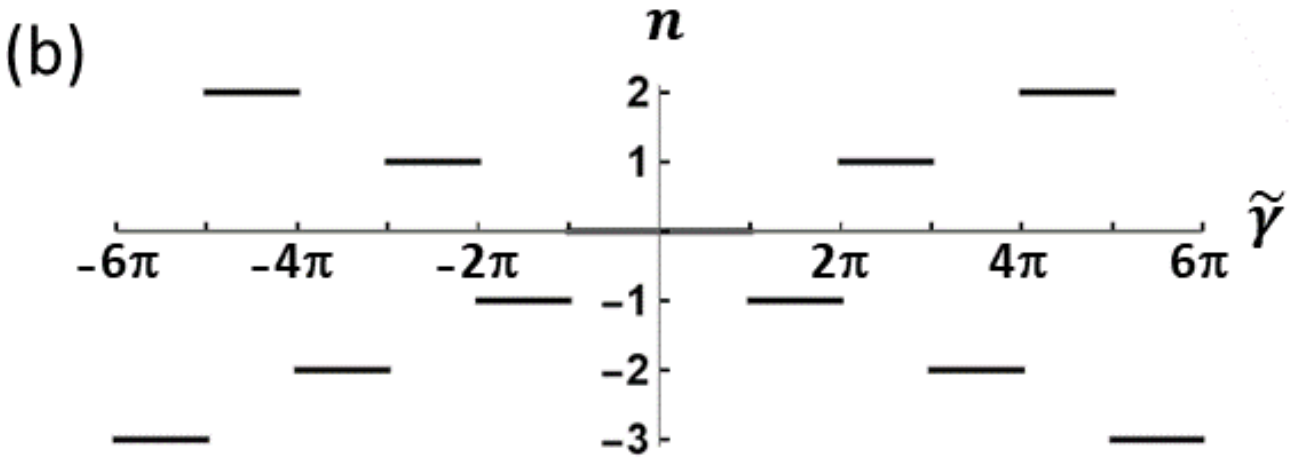

FIG 3. (a) Angular difference $\gamma$ between initial and final locations for geodesic travel around a sphere, as a function of the *total* angular displacement $\tilde{\gamma}$. (b) Resulting values of winding number $n$.

**D. Influence of the phase of prefactor terms in the propagator**

The exact propagator of Eq. (4) is composed of a sum of probability amplitude terms, $\exp(iI\alpha_n^2/2\hbar T)$, together with prefactors that multiply these terms. Of particular importance in determining the time dependence of the propagator is the phase of the prefactors. There are two contributions to this phase, one from the leading $\exp(i\hbar T/8I)$ prefactor and the other arising the phase of the terms that depend on winding number. The latter phase is determined by the $(-1)^n\alpha_n = (-1)^n(\alpha + 2\pi n)$ term, which has a sign of

$$(-1)^n \operatorname{sgn}(\alpha + 2\pi n) \tag{12}$$

where $\operatorname{sgn}(u)$ equals $+1$ if $u > 0$, $-1$ if $u < 0$, and 0 if $u = 0$. The phase of Eq. (12) changes by $\pm\pi$ for each increment of $n$ (except when $n$ changes from $-1$ to 0 in which case the $\alpha + 2\pi n$ term also changes sign). These phase jumps are shown in Fig. 2(a), and they lead to a simple but profound effect on the time dependence of the $S^2$ propagator and the associated eigenstates.

To demonstrate this influence of Eq. (12), we compare the $S^2$ propagator to the one for motion in $S^1$, evaluating the time dependence of each. For $S^1$, the propagator is

$$K^{(S^1)}(\phi_0, \phi_{\mathrm{f}}, T) = \sqrt{\frac{I}{2\pi\hbar i T}} \sum_{n=-\infty}^{\infty} \exp\left(\frac{iI(\phi_{\mathrm{f}} - \phi_0 + 2\pi n)^2}{2\hbar T}\right) \tag{13}$$

for motion between initial and locations $\phi_0$ and $\phi_{\mathrm{f}}$, respectively [2,18,20,31,32]. Its time dependence can be evaluated in a well-known manner using the Poisson summation formula [32],

$$\sum_{n=-\infty}^{\infty} f(2\pi n) = \sum_{k=-\infty}^{\infty} \frac{1}{2\pi} \int_{-\infty}^{\infty} dy\, f(y) \exp(-iky)\ , \tag{14}$$

which yields

$$K^{(S^1)}(\phi_0, \phi_{\mathrm{f}}, T) = \frac{1}{2\pi} \sum_{\ell=-\infty}^{\infty} e^{i\ell(\phi_{\mathrm{f}} - \phi_0)}\, e^{-i\hbar\ell^2 T/2I}. \tag{15}$$

From the exponential term on the far right-hand side, energy eigenvalues of $\hbar^2\ell^2/2I$ are obtained.

To compare this result to the time-dependence of the $S^2$ propagator, we evaluate that, for simplicity, at $\gamma = \pi$ (the treatment for arbitrary $\gamma$ is described in Appendix II). For that special case, and in order to achieve a form like the left-hand side Eq. (14), we write Eq. (4) as

$$K^{(S^2)}(\pi, T) = \pi\, e^{i\hbar T/8I} \left(\frac{I}{2\pi\hbar i T}\right)^{3/2} \sum_{n=-\infty}^{\infty} (-1)^{(2\pi n/2\pi)}\, (\pi + 2\pi n)\, e^{iI(\pi+2\pi n)^2/2\hbar T}. \quad (16)$$

Evaluation of this expression according to Eq. (14) is more complicated than for motion on $S^1$, but nonetheless it can be readily accomplished as demonstrated in Appendix II. We find that

$$K^{(S^2)}(\pi, T) = \frac{e^{i\hbar T/8I}}{2\pi} \sum_{\ell=0}^{\infty} (-1)^{\ell}\ (\ell + 1/2)\, e^{-i\hbar(\ell+1/2)^2 T/2I}\,. \quad (17)$$

This expression provides the same result as obtained if we write the propagator in its spectral representation (as a sum over eigenstates), as detailed in Appendix II.

Comparing the exponentials on the far right-hand sides of Eqs. (15) and (17), it is clear that the angular momenta involved in the respective eigenstates have shifted from $\hbar\ell$ for motion on a circle to $\hbar(\ell + 1/2)$ for motion on a sphere. This shift is due to the phase factor of Eq. (12) (in particular, the $(-1)^n$ part of that phase factor is primarily what leads to the shift, as discussed in Appendix II). This result of $\hbar(\ell + 1/2)$ for the angular momenta associated with the spherical harmonic eigenstates is a very important one, and it is the same as what we will find in Section III using stationary-phase analysis of dominant paths. Of course, to obtain an eigenvalue of the energy eigenstate, we must include effect of the $\exp(i\hbar T/8I)$ prefactor term in Eq. (17). Thus, the energy eigenvalues are seen to be composed of

$$\frac{\hbar^2 \ell(\ell+1)}{2I} = \frac{\hbar^2(\ell + 1/2)^2}{2I} - \frac{\hbar^2}{8I} \quad (18)$$

with the first term on the right arising from the dominant paths and the second term from the $\exp(i\hbar T/8I)$ prefactor. It is clear that this latter prefactor is absolutely essential in determining the resultant energy eigenvalue (especially for small values of $\ell$).

This result of dominant paths with angular momenta of $\hbar(\ell + 1/2)$ for motion in $S^2$ is actually a well-known one; it was obtained in the early analysis of Gutzwiller [33] and it is readily apparent from semiclassical analysis of the Schrödinger equation [34,35,36]. In any case, in Eq. (17) and in Section III we find that this result for the dominant paths is an *exact* one; the dominant angular momentum differs from what might be (incorrectly) inferred from the square root of an eigenvalue, $\hbar\sqrt{\ell(\ell+1)}$ due to presence of the $\exp(i\hbar T/8I)$ prefactor of the propagator, the origin of which we now consider.

### E. Explanation of the $\exp(i\hbar T/8I)$ prefactor

Whereas influence of the $\exp(i\hbar T/8I)$ prefactor in Eqs. (4) and (17) is straightforward to discern – it produces a phase that increases linearly with travel time, thereby affecting the energy eigenvalues in the manner shown in Eq. (18) – the fundamental origin of the prefactor is not so easy to establish. It has been called "mysterious" [37] and "confusing" [38], and it apparently was the source of considerable controversy in the path-integral field in the 1980s [39,40,41,42]. Such terms are associated with motion in curved spaces [43] (ones having a metric that depends on spatial location), and even though methods are now available that enable proper treatment of such spaces [4,5,44,45] , these "curvature terms" continue to play a large (and troublesome) role in many areas of path-integral application [46,47].

To understand the origin of the $\exp(i\hbar T/8I) = \exp(i\hbar T/8MR^2)$ prefactor for free-particle motion on a sphere, it is useful to compare it with what occurs for motion on a *pseudo-sphere*, a 2-dimensional hyperbolic space with negative curvature. In that case, the corresponding prefactor is $\exp(-i\hbar T/8MR^2)$ [45]. The exponent here has opposite sign compared to the one for the sphere, consistent with the change in sign of the *scalar curvature* $R_S$ for the two cases ($R_S = \pm d(d-1)/R^2$ for a $d$-dimensional sphere or pseudo-sphere, with the upper sign for the former and the lower sign for the latter). Qualitatively, the differing prefactors for the two cases can be understood as arising from fluctuations of the paths (away from a geodesic), with these fluctuations encountering differing types of spaces for motion on the sphere compared to the pseudo-sphere: As one moves some distance $r$ away from a given point in a 2-dimensional space, then for a flat space the area available (for points that fluctuate away from the given point) grows like $\int_0^r r' dr'$. In contrast, for the sphere the area grows more slowly, like $\int_0^r |\sin(r'/R)| dr'$, whereas for the pseudo-sphere the area grows more quickly, like $\int_0^r \sinh(r'/R)\, dr'$. Compared to the flat space, there are thusly fewer fluctuating paths available for the sphere and more for the pseudo-sphere.

It is this differing influence of fluctuations that gives rise to the differing prefactors for a sphere, a flat space, and a pseudo-sphere. Paths within a path integral are inherently stochastic [48,49], and DeWitt-Morette *et al.* have demonstrated how the associated fluctuations produce the prefactor term [50,51]. We consider this mechanism to provide the correct and most fundamental origin for prefactor terms such as $\exp(i\hbar T/8MR^2)$ in propagators associated with curved spaces. This origin of the prefactor holds true for all travel times, including short ones.

Nevertheless, we feel that it should be pointed out that this explanation for the prefactor is not one that is commonly stated (at least not explicitly) in the literature. Rather, the vast majority of work dealing with evaluation of curved-space path integrals discusses how a short-time expression for the propagator can be formulated, and then through successive application of that, how an expression can be obtained that is valid for times that are not necessarily small. Of course, this procedure is of the utmost importance in evaluating propagators – it is undoubtedly the most fundamental way of doing so. However, in this procedure, it is well known that there are many (equivalent) ways that prefactor terms within the *short-time* propagator can be included [4,52,53,54]. The inherent variability of the prefactor within the short-time propagator makes it easy to lose sight of the single, well-defined origin of the prefactor that occurs in the resultant, closed-form expression (valid for all travel times) of the propagator. This origin, again, is the stochastic nature of the paths and the influence of that on the resultant propagator when the metric of the space is also considered [50].

### F. Nonclassical paths of the $S^2$ propagator

Let us now consider the nonclassical paths that are explicitly contained in the probability amplitudes of the exact propagator of Eq. (4), *i.e.* those paths described by terms within the integral over $\alpha$ that have $\alpha \neq \gamma$. These paths are required such that the propagator provides an exact description of the dynamics. It is interesting to consider how explicit space-time curves can be formed that describe these nonclassical paths. We consider constant speed of the particle, so the problem reduces to forming a spatial curve with length determined by the action term of the propagator. Consider a single term of Eq. (4), for specific values of $\alpha$ and $n$. The action of each path is $I\alpha_n^2/2T$, so that with our assumption of constant speed, the corresponding angular length is $|\alpha_n| = |\alpha + 2\pi n|$. The integral over $\alpha$ in Eq. (4) extends between $\gamma$ and $\pi$, with the path having $\alpha = \gamma$ corresponding to a classical trajectory. Consider the situation with $n = 0$ and $\gamma = \pi/2$. Due to the integral over $\alpha$, we then end up with paths having angular lengths that vary between $\pi/2$ and $\pi$. Now consider the case of $n = -1$, in which case we have paths with angular length that vary between $\pi$ and $3\pi/2$. Combining these paths for the $n = 0$ and $-1$ terms, we end up with angular lengths that vary continuously between $\pi/2$ and $3\pi/2$. Figure 4 shows an example of such paths; each path can be represented as a segment of a circle having some radius $\rho \leq 1$, *i.e.* including small circles ($\rho < 1$) as well as great circles ($\rho = 1$). Each of these small-circle segments has tangent vector at the initial location that makes an angle $\beta$ with the geodesic path (curve (i)); we refer to $\beta$ as the *take-off* angle, as pictured in the inset of Fig. 4. This same angle also applies to the *approach* of the path to the geodesic at the final location. Appendix III provides the relationship between $\alpha$ and $\beta$ (also dependent on $n$) so that, if desired, the propagator can be expressed in terms of $\beta$ rather than $\alpha$.

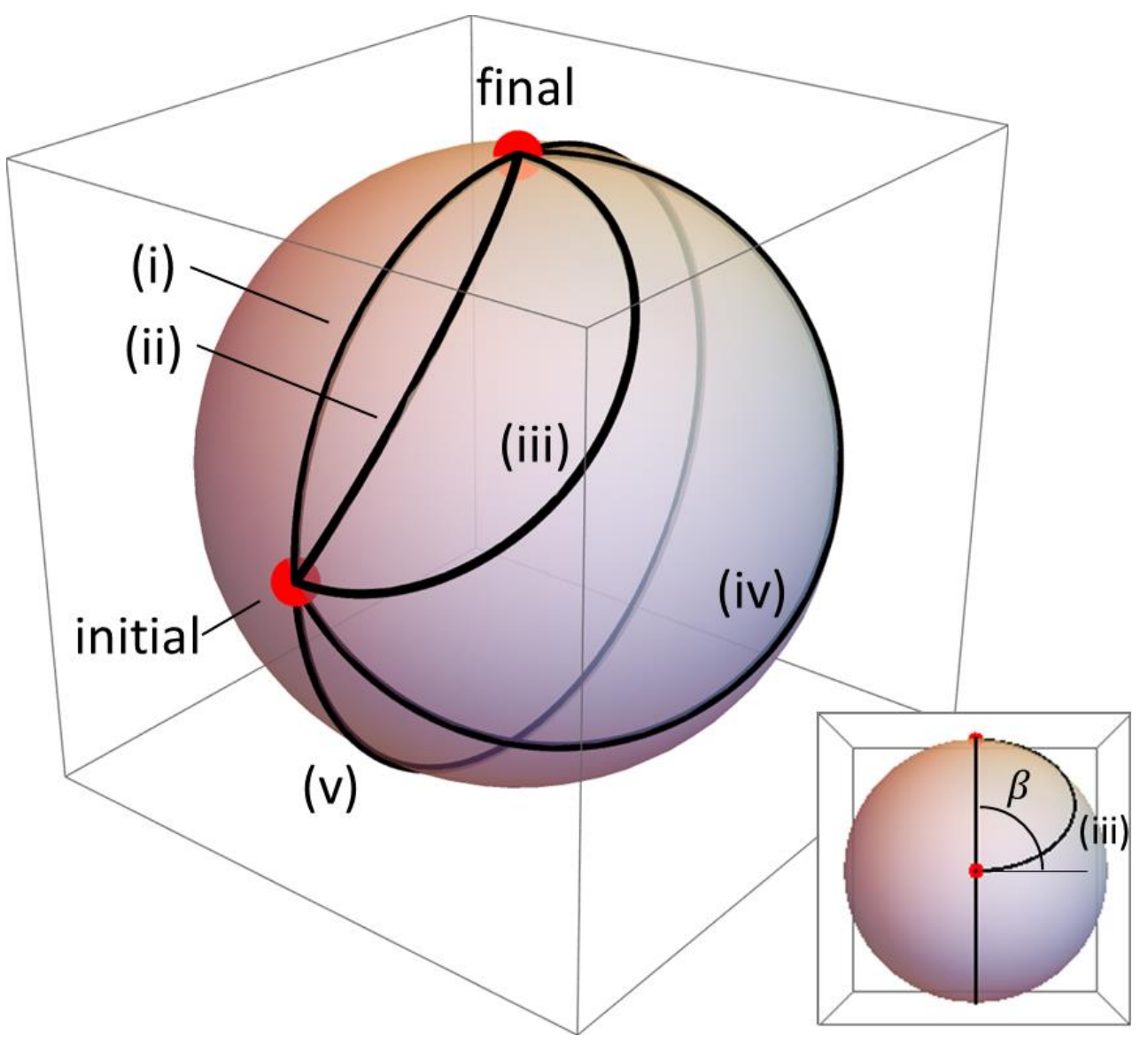

FIG 4. Circular paths extending between initial and final locations on a unit sphere, corresponding to winding numbers of $0$ and $-1$. Paths (i) and (v) are segments of a great circle, with angular lengths of $\pi/2$ and $3\pi/2$, respectively. Paths (ii) – (iv) are segments of small circles, with specified take-off angle $\beta$ from the initial location (as shown in the inset for path (iii)) being $\pi/4$, $\pi/2$, and $3\pi/4$, respectively. The angular length of path (iii) is $\pi$, whereas for (ii) and (iv) it is approximately $0.608\pi$ and $1.392\pi$, respectively.

This interpretation of the paths within the propagator in terms of small-circle motion also follows directly from the homotopy SO(3)/SO(2) that describes $S^2$, as described by Zee [55]; the product of two rotations, one producing curve (i) or (v) of Fig. 4 and the other being about an axis extending to the initial (or final) point, yields a rotation that produces curves (ii) – (iv). This type of small-circle motion can be extended such that it applies of other values of winding number, utilizing the method described in Appendix III. Curves are constructed as equal-length segments of small circles, pieced together with some torsion (twist) between the segments. Such curves are examples of *elastica* (elastic curves) [19,56,57,58,59,60]. They minimize a functional which is the sum of the squared curvature (bending energy) along the curve. It is not surprising that the nonclassical paths of the propagator are described in terms of *elastica*, since as mentioned by Langer and Singer [19], the fact that a homotopy exists for obtaining the propagator implies the existence of a new functional (involving bending energy in the case of *elastica*) for describing paths in the propagator.

## III. Path distributions
### A. Preliminaries

Let us now proceed with the derivation of the path distributions for eigenstates of orbital angular momentum. We follow the methodology of our prior work [17] which starts by writing an expression for the wave function in terms of the propagator,

$$Y_{\ell,m}(\theta_{\mathrm{f}},\phi_{\mathrm{f}}) = e^{iE_\ell T/\hbar}\int_0^{2\pi} d\phi_0 \int_0^{\pi} \sin\theta_0\, d\theta_0\, Y_{\ell,m}(\theta_0,\phi_0)\;\; K^{(S^2)}(\gamma(\theta_0,\phi_0,\theta_{\mathrm{f}},\phi_{\mathrm{f}}),T)\,, \quad (19)$$

with $\gamma(\theta_0,\phi_0,\theta_{\mathrm{f}},\phi_{\mathrm{f}})$ given by Eq. (3), $Y_{\ell,m}$ denotes a spherical harmonic having integer values of $\ell$ and $m$ satisfying $\ell \geq 0$ and $|m| \leq \ell$, and $E_\ell$ is the energy eigenvalue as listed in Eq. (18). To obtain path distributions, we proceed in analogy with the 1-dimensional analysis of our prior work [17], rewriting Eq. (19) such that one of the integrals in it ranges over an extended space $-\infty < \tilde{\gamma} < \infty$, and then further rewriting that integral in terms of a characteristic angular momentum $L_{\mathrm{c}}$. A simplified form of this procedure, which can be accomplished using closed-form analytic expressions but is restricted to the (approximate) semiclassical propagator and to a final location of $\theta_{\mathrm{f}} = 0$, is provided in Appendix IV. For the more general procedure that we describe here, it necessary to first rewrite Eq. (19) in terms of a rotated coordinate frame that is fixed to the final location of the particle $(\theta_{\mathrm{f}},\phi_{\mathrm{f}})$, as illustrated in Fig. 5. Basis vectors for Cartesian coordinates $(X,Y,Z)$ in this new frame are pictured there, and note in particular the choice of the $X$-axis orientation (such that the $XZ$-plane has a fixed value of $\phi$) which determines $\Phi_0$ values for initial locations. Since the new coordinates are fixed to the final location, we have $\Theta_0 = \gamma$.

Using coordinates in this new frame, Eq. (19) can be written as

$$Y_{\ell,m}(\theta_{\mathrm{f}},\phi_{\mathrm{f}}) = e^{iE_\ell T/\hbar}\int_0^{2\pi} d\Phi_0 \int_0^{\pi} \sin\gamma\, d\gamma\, Y_{\ell,m}(\theta_0,\phi_0)\;\; K^{(S^2)}(\gamma,T) \quad (20)$$

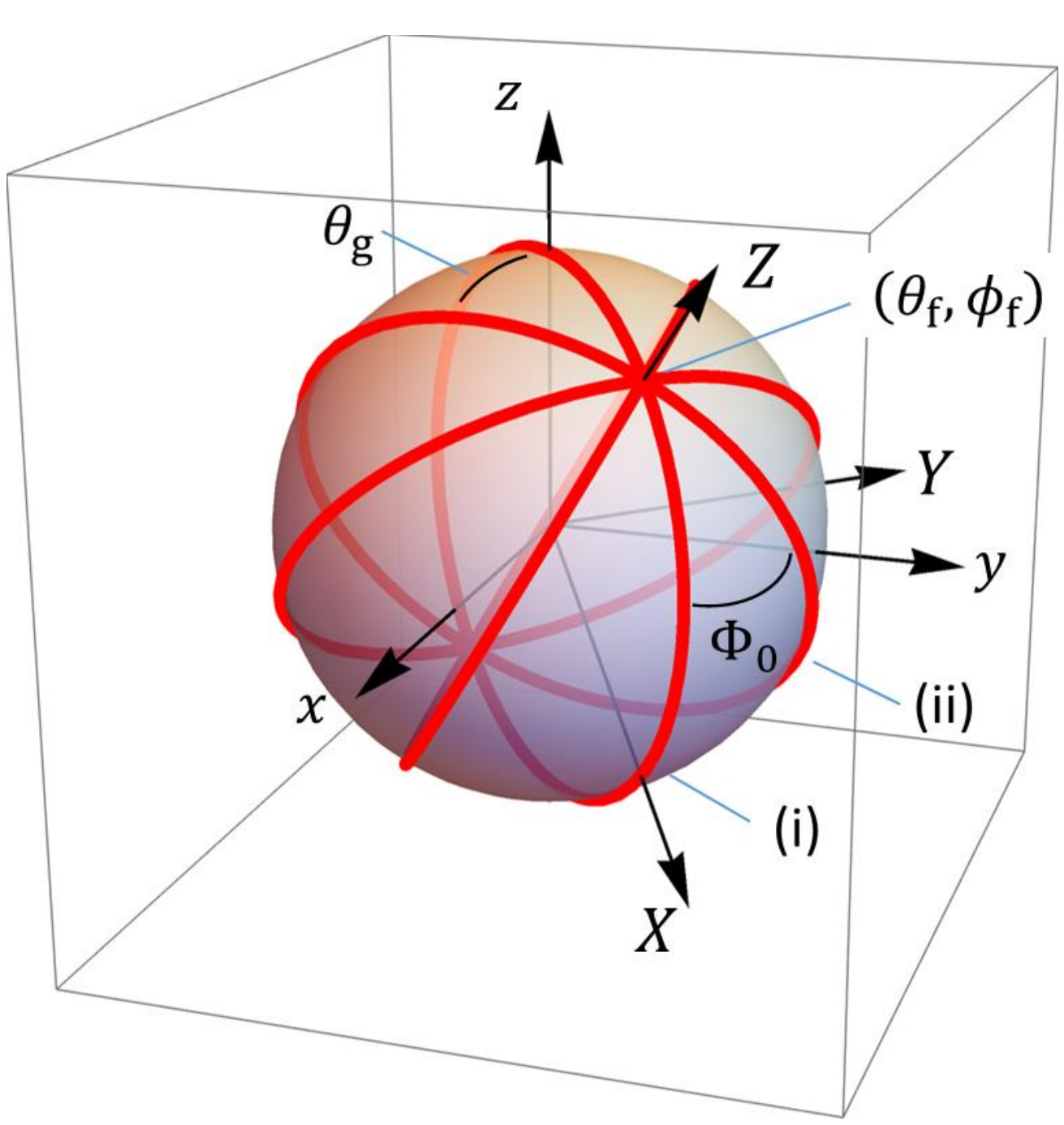

FIG 5. Great circles arrayed about a final location of a particle at $(\theta_{\mathrm{f}},\phi_{\mathrm{f}})$. A rotated coordinate frame $(X,Y,Z)$ is indicated, aligned to the great circles, with $(\Theta_{\mathrm{f}},\Phi_{\mathrm{f}}) = (0,0)$ in the rotated frame. Two great circles are indicated by (i) and (ii). The angle between them is labeled $\Phi_0$, whereas the angle between the plane containing great-circle (ii) and the $z$-axis is labeled $\theta_{\mathrm{g}}$.

where we must specify how $(\theta_0, \phi_0)$ (initial location expressed in the original frame) can be obtained from $(\Theta_0, \Phi_0) = (\gamma, \Phi_0)$ (initial location expressed in the new frame). The matrix for this passive transformation is identical to the matrix for actively rotating the $(x, y, z)$ coordinate axes to the $(X, Y, Z)$ ones, *i.e.* first rotating about the *y*-axis by $\theta_\mathrm{f}$ and then rotating about the *z*-axis by $\phi_\mathrm{f}$ (these angles being members of an Euler *z*-*y*-*z* sequence [61]). Multiplying the $(X, Y, Z)^T = (\cos\Phi\sin\Theta, \sin\Phi\sin\Theta, \cos\Theta)^T$ vector by this matrix (with $T$ denoting transpose, not to be confused with travel time which is never used as a superscript), and equating the result to $(x, y, z)^T = (\cos\phi\sin\theta, \sin\phi\sin\theta, \cos\theta)^T$, yields

$$\theta(\Theta, \Phi) = \cos^{-1}[-\sin\theta_\mathrm{f}\cos\Phi\sin\Theta + \cos\theta_\mathrm{f}\cos\Theta] \tag{21a}$$

$$\begin{aligned}\phi(\Theta, \Phi) = \tan^{-1}[&-\sin\phi_\mathrm{f}\sin\Phi\sin\Theta + \cos\phi_\mathrm{f}(\cos\theta_\mathrm{f}\cos\Phi\sin\Theta + \sin\theta_\mathrm{f}\cos\Theta), \\ &\cos\phi_\mathrm{f}\sin\Phi\sin\Theta + \sin\phi_\mathrm{f}(\cos\theta_\mathrm{f}\cos\Phi\sin\Theta + \sin\theta_\mathrm{f}\cos\Theta)]\end{aligned} \tag{21b}$$

where it is the $(\Theta, \Phi)$ values that are being transformed, with $\theta_\mathrm{f}$ and $\phi_\mathrm{f}$ being parameters for this transformation. In Eq. (21a), the principal branch of $\cos^{-1}$ is utilized (ranging between 0 and $\pi$), whereas in Eq. (21b), $\tan^{-1}(u, v)$ refers to the value of $\tan^{-1}(v/u)$ which takes into account the quadrant that the point $(u, v)$ is in, and values of $\tan^{-1}$ ranging between 0 to $2\pi$ are utilized. Using Eq. (21), the values of $\theta_0$ and $\phi_0$ in the argument of the spherical harmonic within the integrand of Eq. (20) are given by

$$\theta_0 = \theta(\gamma, \Phi_0) \tag{22a}$$

$$\phi_0 = \phi(\gamma, \Phi_0)\ . \tag{22b}$$

Inserting the exact propagator of Eq. (4) into Eq. (20), then following our prior work [17] we must rewrite the sum over $n$ together with the integral over $\gamma$ in terms of an integral over an extended angle $\tilde{\gamma}$. For ease of explanation, we initially consider a range of just $0 \le \tilde{\gamma} < \infty$. Utilizing Eq. (11) which provides values of $n$ and $\gamma$ in terms of $\tilde{\gamma}$, we find that Eq. (20) can be rewritten as

$$\begin{aligned} Y_{\ell,m}(\theta_\mathrm{f}, \phi_\mathrm{f}) = e^{iE_\ell T/\hbar} \int_0^{2\pi} d\Phi_0 \int_0^{\infty} |\sin\tilde{\gamma}|\, d\tilde{\gamma}\; Y_{\ell,m}(\theta_0, \phi_0) \\ \times \sqrt{2}\, e^{i\hbar T/8I} \left(\frac{I}{2\pi\hbar i T}\right)^{3/2} (-1)^n \int_\gamma^\pi \frac{d\alpha\, \alpha_n\, e^{iI\alpha_n^2/2\hbar T}}{(\cos\gamma - \cos\alpha)^{1/2}}\ . \end{aligned} \tag{23}$$

We then must further rewrite Eq. (23) using an extended range $-\infty < \tilde{\gamma} < \infty$ (such that both positive and negative values of $L_\mathrm{c}$ are included, *i.e.* when we transform the integral to this variable). Equation (11) works well for both positive and negative values of $\tilde{\gamma}$, as already mentioned in Section II, and similarly we find that the rotational transformation of Eq. (21) also works fine for both positive and negative $\Theta$ values of arbitrary magnitude (with $\Theta = \Theta_0 = \gamma$ when we utilize Eq. (21) in Eq. (22)). Thus, in place of Eq. (22) we use

$$\theta_0 = \theta(\tilde{\gamma}, \Phi_0) \tag{24a}$$

$$\phi_0 = \phi(\tilde{\gamma}, \Phi_0)\ . \tag{24b}$$

As a check on the validity of Eqs. (21) and (24), the great-circle curves shown in Fig. 5 were obtained using precisely these expressions (and these curves correctly form the great circles).

Turning now to reconsider the integration ranges of Eq. (23), in order for $\tilde{\gamma}$ to take on both positive and negative values we restrict the range of $\Phi_0$ in the integral to be $0 \leq \Phi_0 < \pi$. Re-expressing the integral in this manner is not a problem, although a small detail arises with this choice of a restricted range for $\Phi_0$: If we consider a great-circle path in Fig. 5 with $\Phi_0 \approx \pi/2$ and with positive polar angle $\Theta_0 = \gamma$ between initial and final locations, then as a particle moves from the initial to the final point, the associated angular momentum vector points *downwards*, with *negative* z-component (and similarly when we extend $\gamma$ to $\tilde{\gamma}$). It is convenient to associate this motion with a negative (rather than positive) angular momentum $L_\mathrm{c}$. Thus, we rewrite the integral of Eq. (23) utilizing a scalar characteristic angular momentum of

$$L_\mathrm{c} = -\frac{I\tilde{\gamma}}{T} \tag{25}$$

to define the classical angular momentum that is determined by the endpoints of a path. (We could alternatively employ $\pi \leq \Phi_0 < 2\pi$ along with $L_\mathrm{c} = +I\tilde{\gamma}/T$, but we find it more convenient to use $0 \leq \Phi_0 < \pi$ and Eq. (25) because of simplification in subsequent expressions involving $\Phi_0$).

With these considerations, the form of Eq. (23) that is written with $\tilde{\gamma}$ values extending over $-\infty < \tilde{\gamma} < \infty$ is found to be

$$\begin{aligned} Y_{\ell,m}(\theta_\mathrm{f},\phi_\mathrm{f}) = \sqrt{2}\, e^{i\hbar(\ell+1/2)^2 T/2I} \left(\frac{I}{2\pi\hbar i T}\right)^{3/2} \int_0^\pi d\Phi_0 \int_{-\infty}^{\infty} |\sin\tilde{\gamma}|\, d\tilde{\gamma} \\ \times Y_{\ell,m}(\theta_0,\phi_0)(-1)^n \int_\gamma^\pi \frac{d\alpha\, \alpha_n\, e^{iI\alpha_n^2/2\hbar T}}{(\cos\gamma - \cos\alpha)^{1/2}} \end{aligned} \tag{26}$$

with $(\theta_0, \phi_0)$ obtained using Eq. (24) (which refers back to Eq. (21)), and where $n$ and $\gamma$ are obtained using Eq. (11). The right-hand side of Eq. (26) is then trivially rewritten using the integration variable of $L_\mathrm{c}$ of Eq. (25), from which an expression for $\Delta\mathcal{J}_{\ell,m}(L_\mathrm{c},\Phi_0,\theta_\mathrm{f},\phi_\mathrm{f},T)$ is obtained as [17],

$$\begin{aligned} \Delta\mathcal{J}_{\ell,m}(L_\mathrm{c},\Phi_0,\theta_\mathrm{f},\phi_\mathrm{f},T) = \frac{e^{i\hbar(\ell+1/2)^2 T/2I}}{\sqrt{2}\,\pi\hbar i} \left(\frac{I}{2\pi\hbar i T}\right)^{1/2} \int_{L_\mathrm{c}-\sqrt{\hbar I/T}}^{L_\mathrm{c}+\sqrt{\hbar I/T}} |\sin L'T/I|\, dL' \\ \times Y_{\ell,m}(\theta_0,\phi_0)(-1)^n \int_\gamma^\pi \frac{d\alpha\, \alpha_n\, e^{iI\alpha_n^2/2\hbar T}}{(\cos\gamma - \cos\alpha)^{1/2}} \end{aligned} \tag{27}$$

where, again, $n$, $\gamma$, $\theta_0$, and $\phi_0$ are obtained using Eqs. (11), (21), and (24), but now in those equations, $\tilde{\gamma}$ is obtained from $\tilde{\gamma} = -L'T/I$.

Using Eq. (27), the wave function $Y_{\ell,m}(\theta_\mathrm{f},\phi_\mathrm{f})$ is obtained according to [17]

$$Y_{\ell,m}(\theta_\mathrm{f},\phi_\mathrm{f}) = \int_0^\pi d\Phi_0 \sum_{\nu=-\infty}^{\infty} \Delta\mathcal{J}_{\ell,m}(L_\nu,\Phi_0,\theta_\mathrm{f},\phi_\mathrm{f},T) \tag{28a}$$

with

$$L_\nu = 2\nu \sqrt{\frac{\hbar I}{T}} + \delta L \tag{28b}$$

and where $\delta L$ is any value of angular momentum. Averaging over $\delta L$ (for which it is sufficient to consider values varying between 0 and $2\sqrt{\hbar I/T}$ [17]) then yields

$$Y_{\ell,m}(\theta_\mathrm{f},\phi_\mathrm{f}) = \int_0^\pi d\Phi_0 \int_{-\infty}^{\infty} dL_\mathrm{c}\, \frac{\Delta\mathcal{J}_{\ell,m}(L_\mathrm{c},\Phi_0,\theta_\mathrm{f},\phi_\mathrm{f},T)}{2\sqrt{\hbar I/T}} \,. \tag{29}$$

Finally, we multiply both sides of Eq. (29) by $Y_{\ell,m}^*(\theta_\mathrm{f},\phi_\mathrm{f})$ and integrate over $\theta_\mathrm{f}$ and $\phi_\mathrm{f}$, forming

$$\int_0^{2\pi} d\phi_\mathrm{f} \int_0^\pi \sin\theta_\mathrm{f}\, d\theta_\mathrm{f} \int_0^\pi d\Phi_0 \int_{-\infty}^{\infty} dL_\mathrm{c}\, \frac{Y_{\ell,m}^*(\theta_\mathrm{f},\phi_\mathrm{f})\,\Delta\mathcal{J}_{\ell,m}(L_\mathrm{c},\Phi_0,\theta_\mathrm{f},\phi_\mathrm{f},T)}{2\sqrt{\hbar I/T}} = 1 \,. \tag{30}$$

Still following prior work [17], we could then switch the order of the integrals over $\Phi_0$ and $L_\mathrm{c}$ with those over $\phi_\mathrm{f}$ and $\theta_\mathrm{f}$, and thereby isolate a term

$$\int_0^{2\pi} d\phi_\mathrm{f} \int_0^\pi \sin\theta_\mathrm{f}\, d\theta_\mathrm{f}\, \frac{Y_{\ell,m}^*(\theta_\mathrm{f},\phi_\mathrm{f})\,\Delta\mathcal{J}_{\ell,m}(L_\mathrm{c},\Phi_0,\theta_\mathrm{f},\phi_\mathrm{f},T)}{2\sqrt{\hbar I/T}} \,. \tag{31}$$

This term, as a function of $L_\mathrm{c}$ and $\Phi_0$, provides a bi-variate distribution which describes the contribution that paths which have endpoints related by both the classical angular momentum $L_\mathrm{c}$ and the azimuthal angle $\Phi_0$ makes towards forming the $Y_{\ell,m}$ wave function. This sort of distribution is fine for examining the $L_\mathrm{c}$-dependence of how paths contribute towards forming the wave functions, but for the $\Phi_0$-dependence some additional manipulations are necessary, as we will now discuss in Section III(B). In any case, regarding the detailed computations needed to obtain path distributions for describing eigenstates of orbital angular momentum, the definition of $\Delta\mathcal{J}_{\ell,m}(L_\mathrm{c},\Phi_0,\theta_\mathrm{f},\phi_\mathrm{f},T)$ in Eq. (27), referring back to Eqs. (4b), (11), (21), and (24), is in its final form, suitable for evaluation.

**B. Results**

The path distribution indicated in Eq. (31) involves two variables, $L_\mathrm{c}$ and $\Phi_0$. However, the manner in which the $L_\mathrm{c}$-dependence of $\Delta\mathcal{J}_{\ell,m}(L_\mathrm{c},\Phi_0,\theta_\mathrm{f},\phi_\mathrm{f},T)$ leads to the wave functions is practically independent of $\Phi_0$. Hence, we integrate over the latter variable, forming the distribution

$$\mathcal{P}_{\ell,m}^{(1)}(L_\mathrm{c},T) \equiv \int_0^\pi d\Phi_0 \int_0^{2\pi} d\phi_\mathrm{f} \int_0^\pi \sin\theta_\mathrm{f}\, d\theta_\mathrm{f}\, \frac{Y_{\ell,m}^*(\theta_\mathrm{f},\phi_\mathrm{f})\,\Delta\mathcal{J}_{\ell,m}(L_\mathrm{c},\Phi_0,\theta_\mathrm{f},\phi_\mathrm{f},T)}{2\sqrt{\hbar I/T}} \tag{32a}$$

with

$$\int_{-\infty}^{\infty} dL_\mathrm{c}\, \mathcal{P}_{\ell,m}^{(1)}(L_\mathrm{c},T) = 1 \,. \tag{32b}$$

In Fig. 6 we plot this distribution, for various values of $\ell$ and $m$ and for a travel time of $T = 32\pi I/\hbar$. We evaluate the integrals in Eqs. (27) and (32a) using the numerical method described in Appendix V.

Peaks at $L_\mathrm{c} = \pm\hbar(\ell + 1/2)$ are clearly evident in Fig. 6; it is these two types of paths, having endpoints related by these values of characteristic angular momenta, that primarily constitute the spherical harmonic wave functions. However, for $m \neq 0$ there are small secondary peaks seen in the distributions of Fig. 6. These secondary peaks occur at values of $L_\mathrm{c}$ separated from the primary peaks by integer spacings of $\pm\hbar$. The peak values of these secondary features alternate in sign as $L_\mathrm{c}$ varies away from the primary peak. The paths associated with these secondary peaks play a significant role in determining the numerical convergence for the angular dependence of the path distributions, as further discussed in Appendix V.

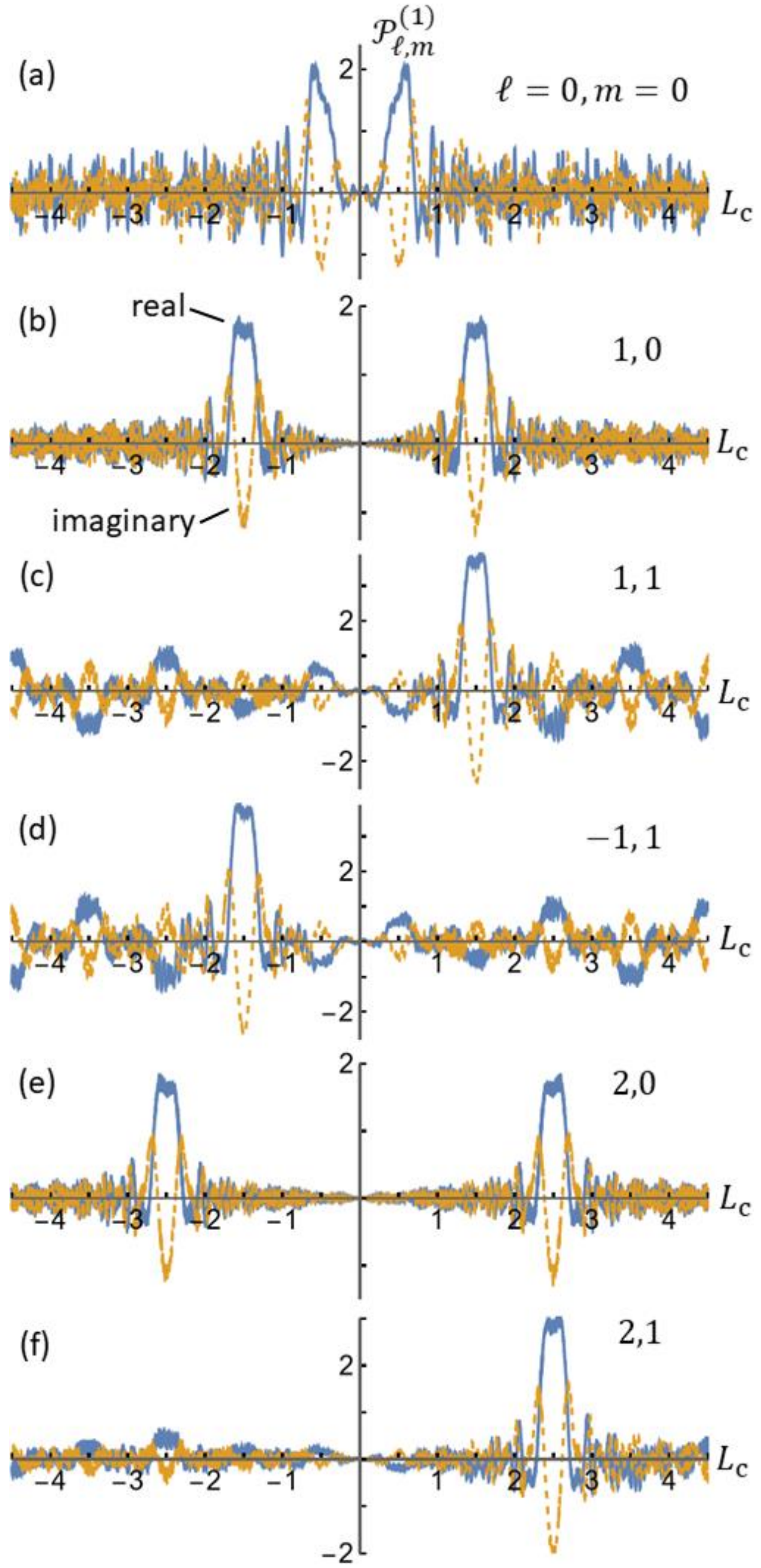

FIG 6. Path distributions $\mathcal{P}_{\ell,m}^{(1)}(L_\mathrm{c}, T)$ as given by Eq. (32), for integer values of $\ell$ and $m$, as a function of the scalar characteristic angular momentum $L_\mathrm{c}$ that is determined by endpoints of a path. Real and imaginary parts of the distributions are shown by solid and dotted lines, as indicated. In all cases, peaks are obtained at one or both of the $L_\mathrm{c}$ values equal to $\pm\hbar(\ell + 1/2)$. Computations are performed for a travel time of $T = 32\pi I/\hbar$ (units of $\hbar = I = 1$ are used in the plots).

The relative intensity of the two primary peaks that occur in Fig. 6 for a given $\ell$ value is seen to vary with $m$, but the $L_\mathrm{c}$ values of the peak positions do not. We have also computed distributions for a travel time of $3200\pi I/\hbar$, with those results showing $\mathcal{P}^{(1)}_{\ell,m}$ distributions having peaks at the same locations as in Fig. 6 but with the width of the peaks decreasing as $1/\sqrt{T}$. This sharpening of the peaks is somewhat similar to what occurs for free-particle motion in $\mathbb{R}^1$ [17] although the nonclassical paths of the $S^2$ propagator as discussed in Section II(F) give the peaks in Fig. 6 (especially the ones for the $\ell = 0$, $m = 0$ state) some additional breadth that does not occur for motion in $\mathbb{R}^1$. The real part of the peaks in Fig. 6 are seen to approximately have the form of a Gaussian whereas the imaginary part resembles a second derivative of such a Gaussian. Since the integrated value of the distribution is unity, Eq. (32b), then in the $T \to \infty$ limit we can, for practical purposes, simply utilize the real part of the distribution for describing its properties [17].

To analyze the angular dependence of the path distributions, one method would be to utilize the $\Phi_0$-dependence of $\Delta\mathcal{I}_{\ell,m}(L_c, \Phi_0, \theta_f, \phi_f, T)$; from Fig. 5 it is clear that $\Phi_0$ can be used to characterize the orientation of a vector that is perpendicular to the plane that contains the geodesic which connects the endpoints for the travel along any given path. However, this angle is not the best one to use for this purpose. A much better means of characterizing this vector is to use its polar angle relative to the $z$-axis; we denote this angle by $\theta_{\mathbf{L}_c}$. To obtain it, we first consider the angle between the $z$-axis and the plane containing the geodesic, *i.e.* the polar angle $\theta_g$ as shown in Fig. 5. The relationship between $\theta_g$ and $\Phi_0$ is obtained by examining the derivative with respect to $\Theta$ of Eq. (21a), setting this equal to 0 in order to obtain the minimum value of $\theta$ for a point along the geodesic. This value corresponds to $\theta_g$, which we find to be

$$\cos\theta_g = |\cos\theta_f|\,[1 + \cos^2\Phi_0 \tan^2\theta_f]^{1/2} \,. \tag{33}$$

The quantity on the right-hand side here is positive, so using the principal branch of $\cos^{-1}$ in order to obtain $\theta_g$ yields $0 \le \theta_g \le \pi/2$. When the final location is near $\theta_f = \pi/2$, there is little difference between using $\theta_g$ or $\Phi_0$ to characterize a path. However, for $\theta_f$ near 0, the associated planes of the great-circle paths arrayed about this final location all have $\theta_g = 0$ but their $\Phi_0$ values range from 0 to $\pi$. These paths with varying $\Phi_0$ all contribute identically to the wave function in this case, so a distribution that employs $\theta_g$ rather than $\Phi_0$ is clearly desirable.

In this manner, we can replace the integral over $\Phi_0$ in Eq. (30) with one that extends over $\theta_g$,

$$\int_0^{2\pi} d\phi_f \left( \int_0^{\pi/2} d\theta_f \int_{\pi-\theta_f}^{\pi} \cos\theta_g \, d\theta_g + \int_{\pi/2}^{\pi} d\theta_f \int_{\theta_f}^{\pi} \cos\theta_g \, d\theta_g \right) \int_{-\infty}^{\infty} dL_c \sin\theta_f \sin\theta_g$$
$$\times \frac{Y^*_{\ell,m}(\theta_f,\phi_f) \Big( \Delta\mathcal{I}_{\ell,m}(L_c,\Phi_0,\theta_f,\phi_f,T) + \Delta\mathcal{I}_{\ell,m}(L_c,\pi-\Phi_0,\theta_f,\phi_f,T) \Big)}{2\cos^2\theta_f \sqrt{\hbar I \big(\cos^2\theta_g/\cos^2\theta_f - 1\big)\big(\tan^2\theta_f - \cos^2\theta_g/\cos^2\theta_f + 1\big)/T}} = 1 \tag{34}$$

with $\Phi_0$ obtained from Eq. (33) as $\tan^{-1}\left\{ \left( \left[ \tan^2\theta_f / \left( \cos^2\theta_g/\cos^2\theta_f - 1 \right) \right] - 1 \right)^{1/2} \right\}$, using the principal branch of $\tan^{-1}$. Interchange of the integrals over $\theta_g$ and $L_c$ in Eq. (34) with those over $\phi_f$ and $\theta_f$ then leads to the bi-variate distribution

$$\mathcal{P}_{\ell,m}\big(L_c, \theta_{\mathbf{L}_c}, T\big) \equiv \int_0^{2\pi} d\phi_f \int_{\theta_g}^{\pi-\theta_g} d\theta_f \, \sin\theta_f \sin\theta_g$$

$$\times \frac{Y^*_{\ell,m}(\theta_\mathrm{f},\phi_\mathrm{f})\left(\Delta\mathcal{J}_{\ell,m}(L_\mathrm{c},\Phi_0,\theta_\mathrm{f},\phi_\mathrm{f},T)+\ \Delta\mathcal{J}_{\ell,m}(L_\mathrm{c},\pi-\Phi_0,\theta_\mathrm{f},\phi_\mathrm{f},T)\right)}{\cos^2\theta_\mathrm{f}\sqrt{\hbar I\left(\cos^2\theta_\mathrm{g}/\cos^2\theta_\mathrm{f}-1\right)\left(\tan^2\theta_\mathrm{f}-\cos^2\theta_\mathrm{g}/\cos^2\theta_\mathrm{f}+1\right)/T}}\ . \quad (35a)$$

We have expressed the angular argument of this distribution in terms of the angle $\theta_{\mathbf{L}_\mathrm{c}}$ mentioned above (and already introduced in Section I), rather than $\theta_\mathrm{g}$, with the former being the polar angle of the $\mathbf{L}_\mathrm{c}$ vector perpendicular to the plane of the geodesic; the two angles are related by $\theta_{\mathbf{L}_\mathrm{c}} \equiv \pi/2-\theta_\mathrm{g}$ so that $\sin\theta_{\mathbf{L}_\mathrm{c}}=\cos\theta_\mathrm{g}$ and $0\le\theta_{\mathbf{L}_\mathrm{c}}\le\pi/2$. In terms of $\theta_{\mathbf{L}_\mathrm{c}}$, the $\cos\theta_\mathrm{g}$ term at the front of the integrand in Eq. (34) becomes $\sin\theta_{\mathbf{L}_\mathrm{c}}$, which serves as a metric term for the distribution of Eq. (35a) such that

$$\int_0^{\pi/2}\sin\theta_{\mathbf{L}_\mathrm{c}}\,d\theta_{\mathbf{L}_\mathrm{c}}\int_{-\infty}^{\infty}dL_\mathrm{c}\,\mathcal{P}_{\ell,m}\left(L_\mathrm{c},\theta_{\mathbf{L}_\mathrm{c}},T\right)=1\,. \quad (35b)$$

We furthermore consider distributions formed by integrating $\mathcal{P}_{\ell,m}\left(L_\mathrm{c},\theta_{\mathbf{L}_\mathrm{c}},T\right)$ over its dependence on the scalar value of $\mathbf{L}_\mathrm{c}$, defining separate distributions for the contributions from $L_\mathrm{c}<0$ and $L_\mathrm{c}>0$,

$$\mathcal{P}^{(2+)}_{\ell,m}\left(\theta_{\mathbf{L}_\mathrm{c}},T\right)\equiv\int_0^{\infty}dL_\mathrm{c}\,\mathcal{P}_{\ell,m}\left(L_\mathrm{c},\theta_{\mathbf{L}_\mathrm{c}},T\right)\ , \quad (36a)$$

$$\mathcal{P}^{(2-)}_{\ell,m}\left(\theta_{\mathbf{L}_\mathrm{c}},T\right)\equiv\int_{-\infty}^{0}dL_\mathrm{c}\,\mathcal{P}_{\ell,m}\left(L_\mathrm{c},\theta_{\mathbf{L}_\mathrm{c}},T\right)\ . \quad (36b)$$

Finally, we combine these two distributions into a single one, now defined over an extended range of $\theta_{\mathbf{L}_\mathrm{c}}$ angles, $0\le\theta_{\mathbf{L}_\mathrm{c}}\le\pi$,

$$\mathcal{P}^{(2)}_{\ell,m}\left(\theta_{\mathbf{L}_\mathrm{c}},T\right)\equiv\begin{cases}\mathcal{P}^{(2+)}_{\ell,m}\left(\theta_{\mathbf{L}_\mathrm{c}},T\right)\,, & 0\le\theta_{\mathbf{L}_\mathrm{c}}\le\pi/2\\ \mathcal{P}^{(2-)}_{\ell,m}\left(\pi-\theta_{\mathbf{L}_\mathrm{c}},T\right)\,, & \pi/2<\theta_{\mathbf{L}_\mathrm{c}}\le\pi\end{cases} \quad (37)$$

(with $\mathcal{P}^{(2+)}_{\ell,m}(\pi/2\,,T)=\mathcal{P}^{(2-)}_{\ell,m}(\pi/2\,,T)$). Using Eq. (35b), we then have the result for $\mathcal{P}^{(2)}_{\ell,m}\left(\theta_{\mathbf{L}_\mathrm{c}},T\right)$ that

$$\int_0^{\pi}\sin\theta_{\mathbf{L}_\mathrm{c}}\,\mathcal{P}^{(2)}_{\ell,m}\left(\theta_{\mathbf{L}_\mathrm{c}},T\right)d\theta_{\mathbf{L}_\mathrm{c}}=1\,. \quad (38)$$

It is clear from this expression that the $\mathcal{P}^{(2)}_{\ell,m}$ distributions can be rotated about a vertical axis in $\mathbf{L}_\mathrm{c}$-space in order to fully visualize their full form. This type of display is what was already presented in Fig. 1(c).

Figure 7 displays results for the distribution $\mathcal{P}^{(2)}_{\ell,m}\left(\theta_{\mathbf{L}_\mathrm{c}},T\right)$, shown on a polar plot utilizing the angle $\theta_{\mathbf{L}_\mathrm{c}}$ and with the radial distance of a point from the origin being proportional $\mathcal{P}^{(2)}_{\ell,m}\left(\theta_{\mathbf{L}_\mathrm{c}},T\right)$. (The distributions actually have imaginary components, as further discussed below, but these are very small and so we ignore them for the present discussion). With values of $\theta_{\mathbf{L}_\mathrm{c}}$ ranging only from 0 to $\pi$, the plots are repeated by reflection about the vertical axis (more precisely, the distributions are valid for all values of azimuthal angle, *i.e.* all possible rotations about the vertical axis, with Fig. 7 showing results for azimuthal angles of 0 and $\pi$). The computational results in Fig. 7 were obtained using numerical integration in the same manner as for the distributions of Fig. 6, with details provided in Appendix V.

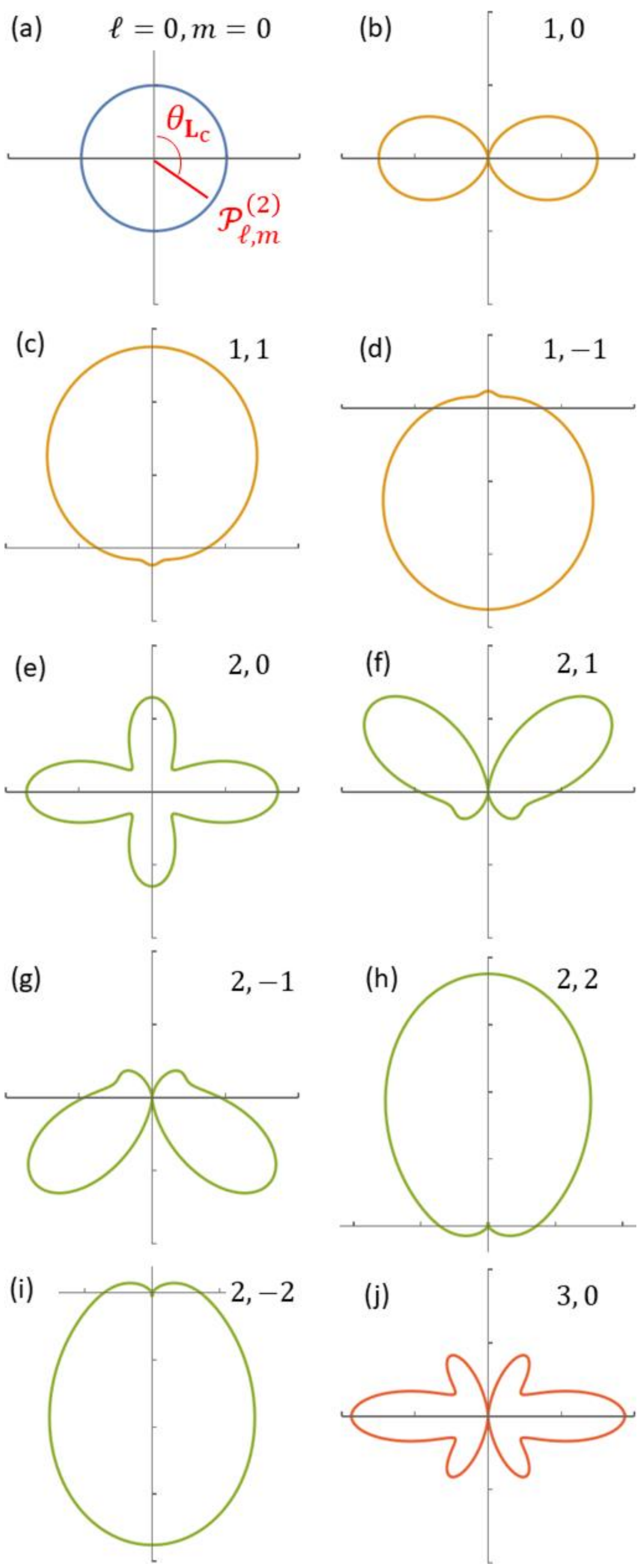

FIG 7. Polar plots for the real part of the $\mathcal{P}^{(2)}_{\ell,m}(\theta_{\mathbf{L_c}}, T)$ as given by Eq. (37), for a travel time of $T = 32\pi I/\hbar$ (units of $\hbar = I = 1$ are used in the plots). In these polar plots, $\theta_{\mathbf{L_c}}$ is an angle measured with respect to the vertical axis ($z$-axis), as shown in (a), and the radial distance to the plotted lines is equal to the real part of $\mathcal{P}^{(2)}_{\ell,m}(\theta_{\mathbf{L_c}}, T)$. Tick marks along the horizontal and vertical axes of all plots correspond to a values of this quantity at increments of 0.5 (values of $\mathcal{P}^{(2)}_{0,0}(\theta_{\mathbf{L_c}}, T)$ in panel (a) are very close to 0.5, so they overlap the tick marks on the axes at 0.5). The $\ell$ and $m$ values for each plot are listed.

The $\mathcal{P}^{(2)}_{\ell,m}(\theta_{\mathbf{L_c}}, T)$ distributions correspond to the polar *direction* of a characteristic $\mathbf{L_c}$ vector as determined by the endpoints of a path, with this vector predominantly having *magnitude* of $\hbar(\ell + 1/2)$ for a state with quantum number $\ell$. The distributions are consistent with what one might expect based on qualitative reasoning, if one thinks in terms of an eigenstate being composed of a collection of great-circle orbits distributed with various $\theta_{\mathbf{L}_c}$ values for each state (as in Fig. 1(c)). For the $\ell = 0, m = 0$ eigenstate, a spherically symmetric distribution is found, indicative of the fact that paths with endpoints determined by classical angular momentum vectors extending in all possible directions all contribute equally to the wave function. For states with $\ell > 0$ but $m = 0$, we see in Fig. 7 that angular momenta vectors extending in the horizontal direction are favored. However, for $\ell > 1$ and $m = 0$, addition features (lobes) of the distribution also occur at angles of $\pi/\ell$ relative to that horizontal plane. For $\ell > 0$ and $m \neq 0$, the horizontal lobes of the $m = 0$ distributions corresponding to the same $\ell$ become tilted upwards or downwards (depending on the sign of $m$), with this tilt becoming larger as $|m|$ increases. For the case of $|m| = \ell \neq 0$, the distributions are peaked along the vertical axis, with these peaks becoming sharper as $|m|$ increases. Some additional $\mathcal{P}_{\ell,m}^{(2)}(\theta_{\mathbf{L}_c}, T)$ distributions, for $\ell$ and/or $m$ values not shown in Fig. 7, are presented in Appendix V.

The values of the $\mathcal{P}_{\ell,m}^{(2)}(\theta_{\mathbf{L}_c}, T)$ distributions obtained here are predominantly real. Considering the ratio of the average magnitude of the imaginary parts to the average magnitude of the real parts for each distribution, and then averaging over the distributions, we find an average ratio of 0.008 for the results of Fig. 7. We find that these imaginary parts become significantly smaller for longer travel times (and also when the range of the $L_c$ integration used in the numerical computations, discussed in Appendix V, is increased). It thus appears, at least within our numerical accuracy, that the imaginary parts of the $\mathcal{P}_{\ell,m}^{(2)}$ distributions approach zero as $T \to \infty$. Concerning the sign of the real part of the $\mathcal{P}_{\ell,m}^{(2)}$ distributions, all of the results in Fig. 7 have positive real parts. Although there is nothing in our general methodology for path distributions that prohibits negative values, for all of the situations we have examined in the past the distributions turn out to be (or asymptotically can be taken to be) real and non-negative, thereby providing a convenient measure of how paths contribute to an eigenstate [17]. The same situation appears to hold for the present case of orbital angular momentum.

An important aspect of the results of Fig. 7 is that, if we sum up the distributions over all $m$ values for a given $\ell$, then for each value of $\theta_{\mathbf{L}_c}$ we find a constant result for this sum. That is to say, within our numerical accuracy, we find that

$$\sum_{m=-\ell}^{\ell} \mathcal{P}_{\ell,m}^{(2)}(\theta_{\mathbf{L}_c}, T) = \text{constant} \ . \tag{39}$$

Given this relationship, then the value of the constant can be easily determined by multiplying both sides of the expression by $\sin\theta_{\mathbf{L}_c}$, integrating over $\theta_{\mathbf{L}_c}$, and using Eq. (38), thus yielding a value of $\ell + 1/2$. We find that Eq. (38) is satisfied by our computed distributions in Fig. 7 to within an average absolute error of 0.0015, while Eq. (39) (moving the constant $\ell + 1/2$ term over to the left-hand side as a denominator) is satisfied on average to within 0.0031. Hence, not only have we managed obtain the distributions $\mathcal{P}_{\ell,m}^{(2)}(\theta_{\mathbf{L}_c}, T)$ that are almost entirely real and non-negative, but additionally, we find that when we sum up the distributions over $m$, for each value of $\ell$, we obtain a spherically symmetric result. In this manner, we can describe the overall composition of an eigenstate of orbital angular momentum in terms of paths. Although the derivation of the distributions in Fig. 7 is admittedly rather complicated, we feel that the results themselves are qualitatively quite simple to understand, in that they provide a description that is no more complicated than that of the vector model (but with our distributions being exact).

## IV. Discussion

Some of the aspects of the $\mathcal{P}_{\ell,m}^{(2)}\left(\theta_{\mathbf{L}_{\mathrm{c}}}, T\right)$ distributions of Fig. 7 bear qualitative resemblance to the well-known "vector model" for orbital angular momentum [13,14,15], as already mentioned in connection with Fig. 1. However, as discussed there, the vector model is substantially in error in its assumption that a delta-function distribution (nonzero at only a single value of $\theta_{\mathbf{L}_{\mathrm{c}}}$) is sufficient for describing each eigenstate. Rather, we conclude from our path distributions that a broad distribution of $\theta_{\mathbf{L}_{\mathrm{c}}}$-values is required in order to characterize each eigenstate. An additional error of the vector model is the claim that a classical angular momentum of $\hbar\sqrt{\ell(\ell+1)}$ can be associated with motion within an eigenstate with orbital quantum number of $\ell$. On the contrary, as discussed in Sections II(D) and III(B), the actual angular momentum associated with dominant paths is $\hbar(\ell+1/2)$. It is only by inclusion of fluctuations of the paths and the associated $\exp(i\hbar T/8I)$ prefactor of the propagator that the eigenvalues for angular momentum squared of $\hbar^2\ell(\ell+1)$ are produced, as in Eq. (18).

Focusing on *s*-states in particular, with quantum number of $\ell=0$, in this case the vector model yields no description whatsoever (since it assigns the classical angular momentum of the associated paths to be 0). In contrast, in the path-distribution method we obtain a spherically symmetric distribution (Fig. 7(a)) with classical angular momentum value of $|L_{\mathrm{c}}|=\hbar/2$. The correctness of this result relies upon the analysis of DeWitt-Morette *et al.* [50,51], discussed in Section II(D), where it is demonstrated how the $\exp(i\hbar T/8I)$ prefactor arises from fluctuations in the motion. It is the combined effect of those fluctuations together with their dependence upon spatial location (due to the varying metric of the space) that leads to the prefactor. It is important to understand that, for both flat or curved spaces, fluctuations will always occur relative to the paths that are explicitly contained within a closed-form expression of the propagator. These fluctuations then contribute prefactor(s) to the probability-amplitude terms within the propagator. In the simple case of motion in a flat space for which a semiclassical propagator provides an exact description, then these prefactor(s) simply consist of the square root of a VPM determinant(s) as mentioned in Section II(B). However, in a curved space, the fluctuations lead, additionally, to a prefactor such as $\exp(\pm i\hbar T/8I)$ (or possibly with a different numerical factor in the exponent than $1/8$, depending on the curvature of the space). For the case of motion on a positive-curvature sphere in particular, the resulting fluctuations then act to *decrease* the energy eigenvalues for the motion, in accordance with Eq. (18), compared to what would otherwise be inferred from the classical energy of the dominant paths alone.

We also comment on how the results obtained herein can be compared with those from the "azimuthal decomposition" of the propagator for motion in $S^2$. This decomposition is described in detail by Kleinert [3]; it relies upon the transformation of the curved-space $S^2$ problem into an equivalent one for a flat space, using a well-established method [62]. The resulting propagators, one for each value of quantum number $m$, are formed by utilizing wave functions ($e^{im\phi}$ Fourier components) for the $\phi$-portion of the motion along with explicit $\theta(t)$ paths for the $\theta$-portion. For values of $m\neq 0$, a propagator is produced that is identical to one that arises from a 1-dimensional Pöschl-Teller potential.

In principle, the fact that the azimuthal decomposition is a description that involves *both* wave functions and paths prevents it from being directly comparable to the path-distribution description developed here. Nevertheless, in an effort to make some sort of comparison – even a qualitative one – we have performed an approximate type of semiclassical analysis of the azimuthally decomposed propagators in which full $(\theta(t), \phi(t))$ paths are constructed. We describe this analysis in Appendix VI, comparing those paths with the distributions found in Fig. 7. Significant differences are found, but these are attributed to the fact that the results of Fig. 7 are based upon paths having constant value for the *total* angular momentum along a path, whereas paths of the azimuthally decomposed propagators have constant *z-component* of the angular momentum. The descriptions are thus seen as being complementary. (Regarding the $\exp(i\hbar T/8I)$ prefactor in Eq. (4), the *same* prefactor occurs in the azimuthal decomposition [3], so in this sense the two types of propagators are similar).

Finally, we comment on the possibility of extending our analysis to the case of *spin* angular momentum. On a very simple level, the vector model is commonly employed for describing *both* orbital and spin angular momentum [13,14,15,63], and so it is natural to ask if the methodology of the present work might yield some sort of path distributions analogous to those of Figs. 6 and 7 for the case of spin, *i.e.*, half-integral value of spin in particular. In terms of homotopy, half-integral values of angular momentum are described by a very similar propagator as Eq. (4), except that a homotopy term of $(+1)^n$ occurs in this revised propagator rather than $(-1)^n$ term of Eq. (4) [2,12,20]. The associated spectral representation can then be expressed in terms of the Wigner *D*-matrices, written as a function of 3 angles, *i.e.* the Euler angles describing the rotation of a spherical top [2,12,20]. We consider it quite likely that path distributions for this 3-dimensional problem can be deduced using our methodology, although these would take the form of tri-variate distributions. Reduction to a bi-variate form, and furthermore to 1-dimensional distributions, may well be possible in a similar manner as for the $S^2$ problem of integer angular momentum (Eqs. (40), (44), and (45)). The double-valued spatial wave functions developed by both Pauli and Merzbacher [64,65] might well provide a route for achieving this reduction. In any case, detailed considerations are necessary for any such analysis, so we do not further pursue it here.

## V. Summary

In this work, we have produced path distributions for the problem of orbital angular momentum, utilizing a path-integral method consisting of a generalization of stationary-phase analysis. The analysis is exact, with results obtained for all value of travel time, but they are most easily displayed for large travel-times in which case we find real-valued, non-negative results. The distributions thusly form a well-defined *measure* [5] of how paths contribute towards forming a given wave function. We similarly found real, non-negative measures in our prior work involving various 1-dimensional problems [17]; we do not understand *why* such distributions are obtained in our analysis (since in general they could be negative and/or complex-valued), but in any case it is a very convenient result for the purpose of understanding how paths contribute towards eigenstates.

For an eigenstate of orbital angular momentum having quantum number $\ell$, we find that paths which dominantly contribute towards the state have starting and ending locations which differ in accordance with a classical angular momentum of magnitude $\hbar(\ell + 1/2)$. This result, although it has been known in path-integral theory since the early work of Gutzwiller [33] (and from semiclassical analysis in general [34,35,36]), is not one that is mentioned, to the best of the author's knowledge, in any quantum mechanics curricula for undergraduate courses. However, we feel that it may well be suitable for discussion in such courses, since it is an *exact* result that is so distinctly different than what occurs for motion around a circle (with dominant paths have starting and ending locations which differ in accordance with an angular momentum of $\hbar\ell$).

This distinction between motion around a sphere as compared to a circle is a direct consequence of a phase factor in the former such that paths which have lengths that consecutively increase by $2\pi$ (*i.e.* winding numbers that increase by an integer) sum together with *alternating* sign for motion on the sphere but with the *same* sign for motion on the circle, as discussed in Section II(D) and Appendix II. However, this alternation in sign of the *n*-dependent prefactor in the propagator for motion in $S^2$ is only half the story in terms of producing the eigenvalues of the eigenstates. The other half involves the time-dependent prefactor $\exp(i\hbar T/8I)$ that occurs for $S^2$ (but not for $S^1$, since a circle is not considered to be a curved space owing to the fact that it has a constant metric). As discussed by previous authors [50,51] and in Section II(E), this prefactor can be attributed to fluctuations of paths away from the ones that the closed-form expression of the propagator is explicitly written in terms of. Whereas such fluctuations do not yield a time-dependence prefactor term for motion in a flat space, they *do* yield this sort of prefactor for motion in a curved space.

Turning to the orientational dependence of the path distributions, as displayed in the $\mathcal{P}^{(2)}_{\ell,m}(\theta_{\mathbf{L}_c}, T)$ distributions of Fig. 7, some of that dependence bears qualitative resemblance to the well-known "vector model" for orbital angular momentum [13,14,15] as discussed in connection with Fig. 1 and in Section III(B). However, the vector model is substantially in error in its assumption that a delta-function distribution (nonzero at only a single value of $\theta_{\mathbf{L}_c}$) is sufficient for describing each eigenstate. Rather, we conclude from our path distributions that a broad distribution of $\theta_{\mathbf{L}_c}$-values is required in order to characterize each eigenstate. An additional error of the vector model is the claim that a classical angular momentum of $\hbar\sqrt{\ell(\ell+1)}$ can be associated with motion within an eigenstate with orbital quantum number of $\ell$. On the contrary, as just mentioned, the actual angular momentum associated with dominant paths is $\hbar(\ell+1/2)$. It is only by inclusion of fluctuations of the paths and the associated $\exp(i\hbar T/8I)$ prefactor of the propagator that the eigenvalues for angular momentum squared of $\hbar^2\ell(\ell+1)$ are produced. Focusing on *s*-states in particular, with quantum number of $\ell = 0$, in this case the vector model yields no description whatsoever (since it assigns the classical angular momentum of the associated paths to be 0). In contrast, in the path-distribution method we obtain a spherically symmetric distribution (Fig. 7(a)) with classical angular momentum value of $|L_c| = \hbar/2$.

**Acknowledgements**

The author is very grateful to Riccardo Penco and Michael Widom for numerous discussions. Thanks also go to Huili (Grace) Xing for encouragement and comments. This work was supported by the National Science Foundation, grant DMR-1809145.

## Appendix I

In Section II(B), in comparing the semiclassical and the exact forms of the $S^2$ propagator, a difference in their phase discontinuities was found, amounting to $\pm\pi/2$ for each $\pi$ change in $|\gamma + 2\pi n|$. Here, we demonstrate that this difference arises from the fact that the former is actually valid only for large values of angular momentum, with the exact propagator producing precisely this phase difference in the limit of large angular momentum. To illustrate this point, we consider the difference in phase between the terms of Eqs. (9b) and (10b), *not including* the phase jumps arising from the $(-1)^n(\alpha + 2\pi n)$ term for the exact form and the $\exp(-i\pi\lfloor|\gamma_n|/\pi\rfloor/2)$ term of the semiclassical form. This difference then should reflect the *additional* phase that arises in the exact propagator (through interference of its probability amplitude terms) that leads to the difference between the two propagators of $\pm\pi/2$ at integer values of $|\gamma + 2\pi n|/\pi$. This phase difference is shown in Fig. 8, and as $|\gamma + 2\pi n|$ increases the expected behavior is indeed found. This phase difference can be understood in analogy with the $\pi/2$ phase shift that occurs in the phase of a focused electromagnetic wave as one moves spatially through the focal point [29]. Importantly, a further phase difference between the semiclassical and exact propagators arises from the $\exp(i\hbar T/8I)$ prefactor that is present in the latter but not the former, as discussed in Section II(E).

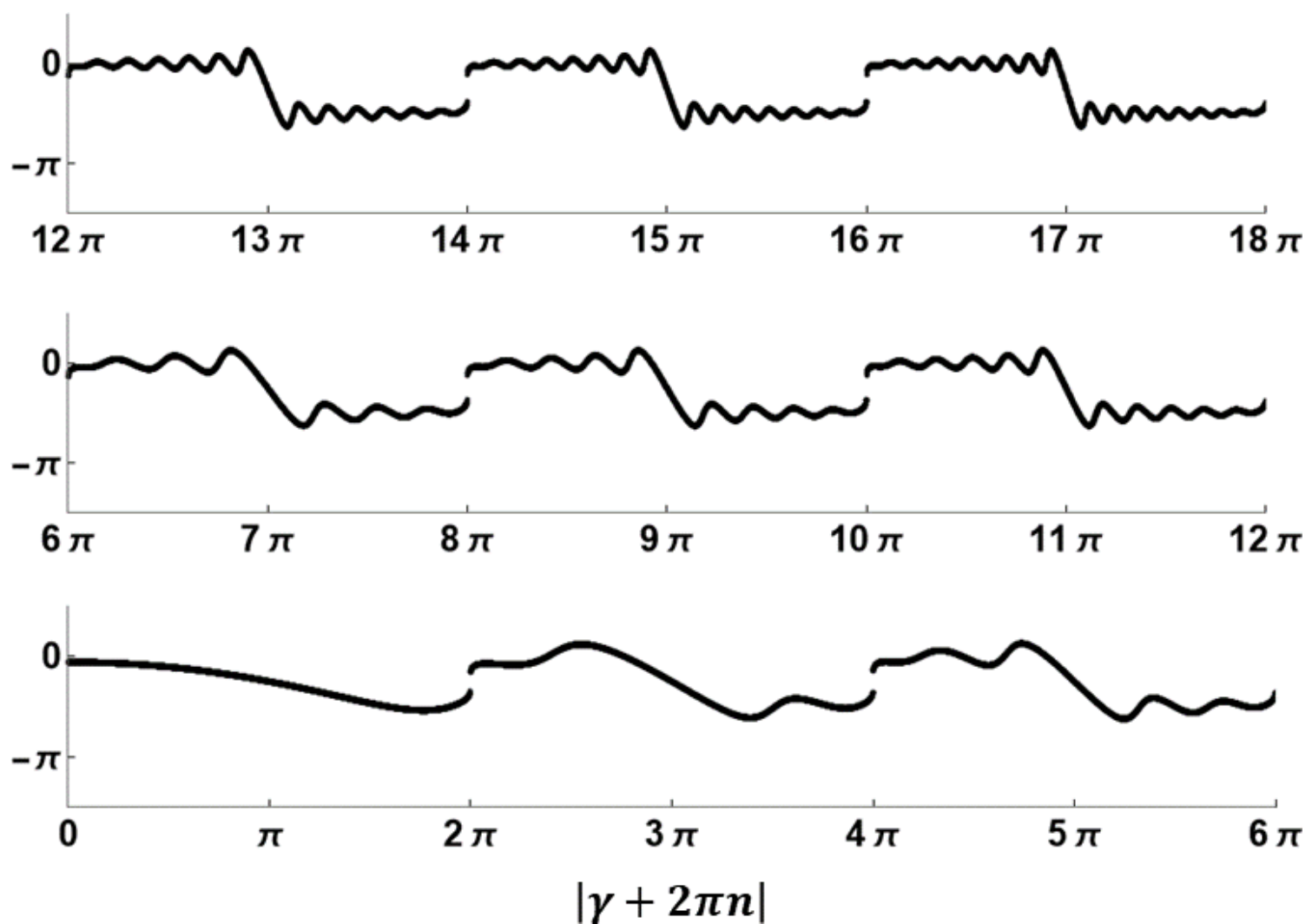

FIG 8. Phase difference between exact and semiclassical propagators, not including the discontinuous phase jumps of each. Results are for a travel time of $T = 3$. The panels have differing ranges of $|\gamma + 2\pi n|$ values, as indicated on the respective horizontal axes.

## Appendix II

In Section II(D), the result of Poisson summation of the $S^2$ propagator of Eq. (4) is described, as a means of obtain its time dependence which in turn provides the energy eigenvalues for the problem. Details of this evaluation are provided here. We first consider the special case of an angular difference between initial and final locations of $\gamma = \pi$. Then, starting from Eq. (16), the function $f$ on the left-hand side of Eq. (14) is identified as

$$f(y) = \pi\, e^{i\hbar T/8I} \left(\frac{I}{2\pi\hbar i T}\right)^{3/2} (-1)^{(y/2\pi)} (\pi + y)\, e^{iI(\pi+y)^2/2\hbar T} \quad . \tag{40}$$

Proceeding with the evaluation of the right-hand side of Eq. (14),

$$\sum_{k=-\infty}^{\infty} \frac{1}{2\pi} \int_{-\infty}^{\infty} dy\, f(y) \exp(-iky)$$

$$= \frac{e^{i\hbar T/8I}}{2} \left(\frac{I}{2\pi\hbar i T}\right)^{3/2} \sum_{k=-\infty}^{\infty} \int_{-\infty}^{\infty} dy\, (-1)^{(y/2\pi)} (\pi + y)\, e^{iI(\pi+y)^2/2\hbar T} e^{-iky} \tag{41a}$$

$$= -\frac{i\, e^{i\hbar T/8I}}{2} \left(\frac{I}{2\pi\hbar i T}\right)^{3/2} \sum_{k=-\infty}^{\infty} e^{ik\pi} \int_{-\infty}^{\infty} dz\, e^{iz/2} z\, e^{iIz^2/2\hbar T} e^{-ikz} \tag{41b}$$

$$= -\frac{i\, e^{i\hbar T/8I}}{4\pi} \sum_{k=-\infty}^{\infty} (-1)^k \left(k - \frac{1}{2}\right) \exp\left[-\frac{i\hbar T}{2I}\left(k - \frac{1}{2}\right)^2\right] \tag{41c}$$

$$= \frac{e^{i\hbar T/8I}}{4\pi} \sum_{\ell=-\infty}^{\infty} (-1)^\ell \left(\ell + \frac{1}{2}\right) \exp\left[-\frac{i\hbar T}{2I}\left(\ell + \frac{1}{2}\right)^2\right] \tag{41d}$$

$$= \frac{e^{i\hbar T/8I}}{2\pi} \sum_{\ell=0}^{\infty} (-1)^\ell \left(\ell + \frac{1}{2}\right) \exp\left[-\frac{i\hbar T}{2I}\left(\ell + \frac{1}{2}\right)^2\right] \tag{41e}$$

where in going from Eqs. (41a) and (41b) we introduce the integration variable $z = y + \pi$, and in going from (41c) to (41d) we introduce the summation variable $\ell = k - 1$.

We compare the result in Eq. (41e) with what is obtained from the spectrally resolved form of the propagator,

$$K^{(S^2)}(\theta_0, \phi_0, \theta_f, \phi_f, T) = \sum_{\ell,m} Y^*_{\ell,m}(\theta_0, \phi_0) Y_{\ell,m}(\theta_f, \phi_f) e^{-iE_\ell T/\hbar} \tag{42}$$

with $E_\ell = \hbar^2 \ell(\ell+1)/2I$. An angular difference of $\gamma = \pi$ corresponds to inversion of space, with the spherical harmonic then having parity of $(-1)^\ell$. Hence, in Eq. (42) we can write $Y_{\ell,m}(\theta_f, \phi_f) = Y_{\ell,m}(\pi - \theta_0, \pi + \phi_f) = (-1)^\ell\, Y_{\ell,m}(\theta_0, \phi_f)$. Summing the resultant product of spherical harmonics in Eq. (42) over $m$ then yields the well-known result of $(2\ell + 1)/4\pi$, thereby establishing agreement between Eqs. (41e) and (42).

Returning for a moment to Eq. (41b), it is the evaluation of the integral there that leads to the $\exp[-i\hbar T(k - 1/2)^2/2I]$ in Eq. (41c) which then produces the $\exp[-i\hbar T(\ell + 1/2)^2/2I]$ term in Eqs. (41d) and (41e). This latter term, with its angular momenta of $\hbar(\ell + 1/2)$ that are shifted compared to the $\hbar\ell$ for motion on a circle, is primarily produced by the $(-1)^n$ term in Eq. (14), which then produces the $e^{iz/2}$ term in Eq. (41b). That is to say, if we modify the integrand of Eq. (41b) by leaving out the lone $z$ term there, or by replacing the $z$ term by $\mathrm{sgn}(z)$, in both cases evaluation of the integral still produces a result that includes a term of $\exp[-i\hbar T(k - 1/2)^2/2I]$. Of course, additional time dependence is also produced in such an evaluation, so for quantitative purposes all of the terms in the integrand of Eq. (41b) are essential. Nevertheless, it is the $(-1)^n$ term of Eq. (14) – which can be viewed as a homotopy term within the propagator of Eq. (4) – that is the primary cause of the shift in the angular momenta.

The Poisson summation procedure for a general value of $\gamma$ is only modestly more complicated than that of Eq. (41). The $f(y)$ function then contains the full integral over $\alpha$ from the propagator, Eq. (4). The corresponding integral over $y$ (analogous to Eq. (41b)) of that integrand can be performed directly, with the remaining $1/(\cos\gamma - \cos\alpha)^{1/2}$ term from the propagator then being a multiplier of that result. The integral over $\alpha$ thus formed can be identified as a Laplace-Mehler integral [66], which is given by a Legendre polynomial, and the spectral representation of the propagator easily follows from that.

**Appendix III**

In Section II(F), constant-speed motion on *small* circles is utilized for explaining the angular curve length $|\alpha_n|$ within the action terms of nonclassical paths contained in the exact $S^2$ propagator, for the case of winding numbers of $n = 0$ and $-1$. However, to fully understand all paths, we must somehow extend this small-circle type of motion so that we can obtain angular lengths of $|\alpha_n|$ for other values of $n$ as well. This can be accomplished as shown in Fig. 9 for the case of $n = 1$ and $-2$, again taking $\gamma = \pi/2$ as in Fig. 4. The resultant angular lengths $|\alpha_n| = |\alpha + 2\pi n|$ range from $5\pi/2$ to $7\pi/2$, with the curves shown in Fig. 9 having this same range of lengths. The curves are constructed as segments of small circles pieced together with some torsion (twist) between the segments. Red filled circles in Fig. 9 mark the starting and ending locations of the travel, whereas the black filled circles mark inflection points where the *signed geodesic curvature* changes sign (for a curve constrained to lie on any specified surface, its total curvature can be separated into normal and geodesic components, the former arising from the curvature of the surface and the latter indicative of the curvature within the manifold itself [58]). Hence, there is nonzero torsion applied to the circular segments at these points, with the derivative of the direction of the tangent to the curve changing discontinuously although the total curvature remains constant. We will refer to these inflection points as *torsion points*. The curves shown in Figs. 4 and 9 are constant-curvature spherical curves, and in particular they are examples of *elastica* (elastic curves).

*Elastica* were studied early-on by both James and Daniel Bernoulli and by Euler (see Truesdell for a fascinating historical account [56]), as well as by Birkhoff [57], and more recently by Langer and Singer [19] and others [58,59,60]. They minimize a functional which is the sum of the squared curvature (bending energy) along the curve. For the curves described in the present work, lying on a sphere, they all consist of segments of circles and this functional refers to their total curvature, which is simply the inverse of the radius of curvature of each circular segment.

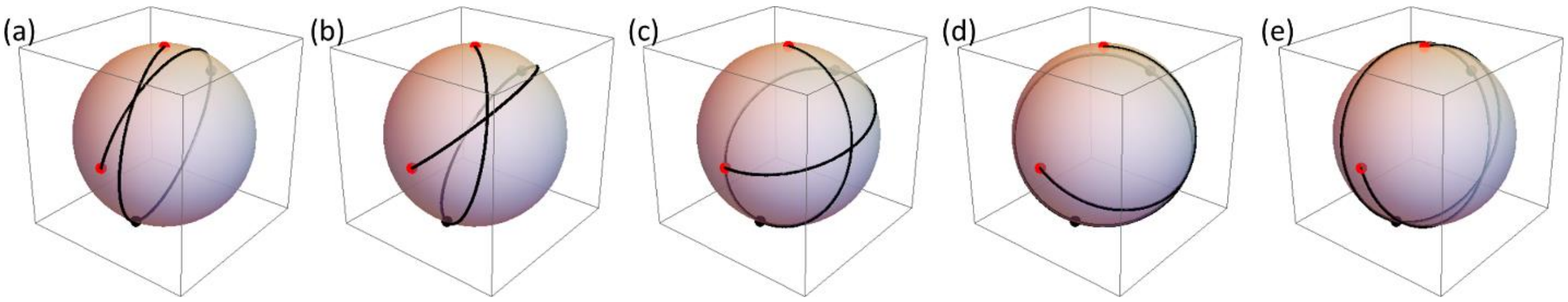

FIG 9. Paths composed of segments of circles extending between the same initial and final locations on a unit sphere as in Fig. 4, corresponding here to winding numbers of 1 for (a) – (c) and $-2$ for (d) and (e). The paths have take-off angles of $\beta = 0.1\pi$, $0.3\pi$, $0.5\pi$, $0.7\pi$, and $0.9\pi$, for (a) – (e), respectively, and their total lengths are approximately $2.515\pi$, $2.642\pi$, $2.898\pi$, $3.225\pi$, and $3.465\pi$.

To describe our general method for constructing *elastica* with angular lengths that match the $|\alpha + 2\pi n|$ values of Eq. (4), we start by defining a new type of winding number, $n'$, in terms of the winding number $n$ from Eq. (11b),

$$n' \equiv \lfloor |n + 1/2| \rfloor \tag{43a}$$

$$= \left\lfloor \frac{|\tilde{\gamma}|}{2\pi} \right\rfloor \tag{43b}$$

where $|\tilde{\gamma}| = |\gamma + 2\pi n|$ as discussed in connection with Eq. (11). The quantity $n'$ can be viewed as a *pair-wise* winding number, since for the pair $n = 0$ or $-1$ we have $n' = 0$, whereas for $n = 1$ or $-2$ we have $n' = 1$, *etc.* Now, for a set of *elastica* paths associated with a great-circle length between initial and final points of $|\tilde{\gamma}|$, the number of torsion points is taken to be

$$N_\tau = 2n' \tag{44}$$

so that, *e.g.*, for Fig. 4 we have $N_\tau = 0$ and for Fig. 9 we have $N_\tau = 2$. Figure 10 shows another set of elastica curves, for the case of $n = 2$ or $-3$ so that $n' = 2$ and $N_\tau = 4$. In all cases, an *elastica* path is drawn from some starting point, extends through the series of torsion points, and eventually reaches the final point. The initial and final points and all torsion points are placed in a common plane that we will denote the *torsion plane*. The angle separating one such point from the next is taken to be

$$\Gamma = \frac{|\tilde{\gamma}|}{N_\tau + 1} = \frac{|\tilde{\gamma}|}{2n' + 1} \tag{45}$$

resulting in $\Gamma \leq \pi$ for $n \geq 0$ (so that $|\tilde{\gamma}| = |\gamma + 2\pi n|$ is between 0 and $\pi$, or $2\pi$ and $3\pi$, *etc.*) and $\Gamma \geq \pi$ for $n < 0$ (so that $|\tilde{\gamma}|$ is between $\pi$ and $2\pi$, or $3\pi$ and $4\pi$, *etc.*), and where $\Gamma \rightarrow \pi$ for $n \rightarrow \pm\infty$. The linear distance (chord length) between neighboring torsion points is $2\sin(\Gamma/2)$.

We consider a circular segment leaving the initial location. We define a *take-off angle* $\beta$ for this curve as the angle it subtends (when it first leaves the initial location) from the shortest geodesic that connects the initial and final locations, as pictured in the inset of Fig. 4. The *elastica* curve is drawn to the first torsion point, twisted, drawn to the second torsion point, twisted again by the same amount, and this procedure is repeated until the curve leaves the final torsion point and extends to the final location. For each of these segments, the angle between the plane that the curve lies in and the torsion plane remains constant at $\beta$ (hence, the *approach angle* with which the curve approaches the final location, relative to the geodesic, is also equal to $\beta$).

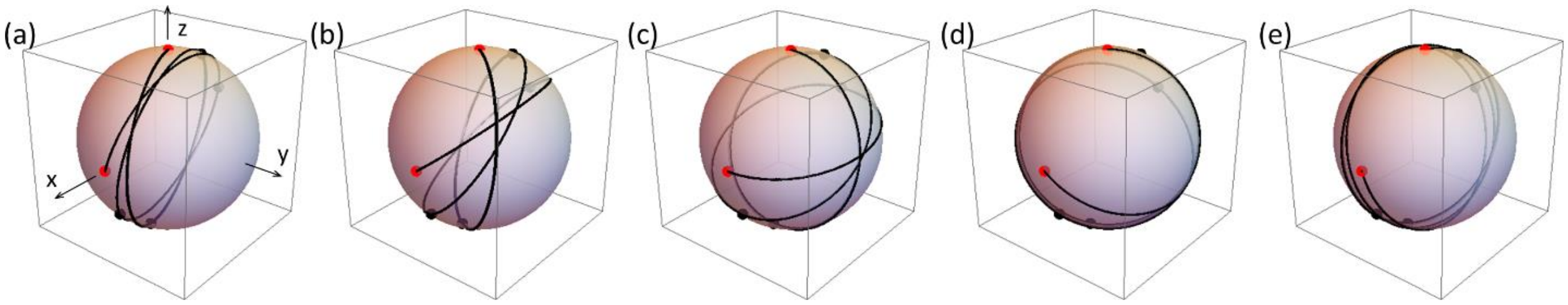

FIG 10. Similar caption as for Fig. 9, but with the paths here corresponding to winding numbers of 2 for (a) – (c) and and $-3$ for (d) and (e). Total path lengths are approximately $4.519\pi$, $4.667\pi$, $4.938\pi$, $5.253\pi$, and $5.470\pi$, respectively.

In terms of detailed geometry of these circular segment, we find that the radius of curvature of each segment is given by

$$\rho = \sqrt{\sin^2(\Gamma/2) + \cos^2\beta \cos^2(\Gamma/2)} \tag{46}$$

and the internal angle $\nu$ subtended by each segment is

$$\nu = 2\cos^{-1}[\mathrm{sgn}(n + 1/2) \cos\beta \cos(\Gamma/2)/\rho] \,, \tag{47}$$

using the principal value of $\cos^{-1}$. The corresponding arc length of each segment is $\rho\nu$. For the unit sphere we are working with, this arc length also equals the total angle subtended by a vector (extending from the origin to the *elastica*) as its end point travels along the *elastica*. Our method for plotting the *elastica* curves utilizes a sequence of rotational transformations in order to bring the circular segments, each with radius $\rho$ and internal angle $\nu$, into the appropriate orientation such that they join up end to end and form the complete curve.

The take-off angle $\beta$ provides a convenient means of labeling the curves. There is some critical value of $\beta$, which we will denote as $\beta_c$ with $\beta_c > \pi/2$, for which $\rho\nu = \pi$ and the value of winding number switches from $n = n' \geq 0$ to $n = -n' - 1 < 0$. The value of $\beta_c$ can be determined by first noting that the total angular length of each curve is given by $(N_\tau + 1)\rho\nu$, so that

$$|\alpha + 2\pi n| = 2(N_\tau + 1)\, \rho \cos^{-1}[\mathrm{sgn}(n + 1/2) \cos\beta \cos(\Gamma/2)/\rho] \tag{48}$$

with $N_\tau$, $\Gamma$, and $\rho$ given by Eqs. (44), (45) and (46). We then equate Eq. (48) to the upper limit of the $|\alpha + 2\pi n|$ angular length in Eq. (4), *i.e.* when $\alpha = \pi$, and consider the case of $n \geq 0$ so that

$$|\pi + 2\pi n'| = 2(N_\tau + 1)\, \rho \cos^{-1}[\cos\beta_c \cos(\Gamma/2)/\rho] \tag{49}$$

with $N_\tau$, $\Gamma$, and $\rho$ again given by Eqs. (44), (45), and (46) (with $\beta = \beta_c$ in the latter). Equation (49) is easily solved numerically (iteratively), using an initial guess for $\beta_c$ of $\pi/2$. Caution should be employed when using Eqs. (46) – (48) since, again, they are only applicable if $n \geq 0$ for $\beta \leq \beta_c$ or $n < 0$ for $\beta > \beta_c$ (when $\beta = \beta_c$ then either value of $n$ can be used), with the value of $n$ determined from $\tilde{\gamma}$ by Eq. (11b) and $\tilde{\gamma}$ also determining $N_\tau$ and $\Gamma$ by Eqs. (43) – (45).

As just described in Eqs. (43) – (49), each *elastica* path is characterized by a winding number $n$, starting and ending locations (separated by angle $\gamma$ with $0 \leq \gamma \leq \pi$), and the takeoff angle $\beta$ (which equals the approach angle). It is of interest to consider how an arbitrary path can be characterized as being a distortion of one of the *elastica* paths. We can always define a torsion plane for a path as being the plane that contains the initial and final locations along with the origin of the sphere (if the initial and final locations are either coincident or antipodal, then we can choose any one of the many planes that contain them and the origin). Let us refer to the normal to the torsion plane as the "$y$" direction, with coordinates $x$ and $z$ being in the torsion plane as pictured in Fig. 10(a). Then, for our arbitrary curve, we compute a non-negative winding number in this plane by a line integral over the curve,

$$w_\tau = \frac{1}{2\pi} \left| \int_C \frac{x dz - z dx}{x^2 + z^2} \right| \tag{50}$$

and we then obtain an integer value from this by $\lfloor w_\tau \rfloor$. For all of the *elastica*, we find that $\lfloor w_\tau \rfloor = n'$, so we use this same relationship in order to determine an $n'$ value for an arbitrary curve. Additionally, we require a $\beta$ value for the arbitrary curve. We take that to be the *average* of the take-off and approach angles (with the positive sense of the angles being as illustrated in Figs. 4, 9, and 10). Thus, together with the minimum geodesic separation ($\gamma$), we have all the values needed to characterize the arbitrary curve, except that we still must distinguish between the two possible $n$ values of $n'$ and $-n'-1$. For this purpose, we examine the fractional part of $w_\tau$. For all the *elastica*, we find that if this fractional part is less than 0.5 when the winding number for the curve is $n = n'$, whereas if it is greater than 0.5 when $n = -n' - 1$, and so we use the same criterion for characterizing the arbitrary curve.

**Appendix IV**

In Section III, path distributions for motion in $S^2$ are fully determined, utilizing the exact form of the propagator as given in Eq. (4). This general solution for the path distributions relies on numerical evaluation of several nested integrals. Here, we demonstrate how an approximate, special case of that solution can be solved using simple analytic expressions. In particular, we utilize the semiclassical form of the propagator, Eq. (8), and we evaluate the expression for the wave function, Eq. (19), only at the particular final location of $\theta_\mathrm{f} = 0$. At this location we have $\gamma = \theta_0$ and only states with $m = 0$ contribute to Eq. (19) since $Y_{\ell,m}(0, \phi_\mathrm{f}) = 0$ unless $m = 0$. Using $Y_{\ell,0}(\theta,\phi) = \sqrt{(2\ell+1)/4\pi}\, P_\ell(\cos\theta)$ for Legendre polynomial $P_\ell$, and with $P_\ell(1) = 1$, Eq. (19) reduces to

$$1 = 2\pi e^{iE_\ell T/\hbar} \int_0^\pi \sin\gamma \, d\gamma \, P_\ell(\cos\gamma) \;\; K^{(S^2)}(\gamma, T) \tag{51}$$

where the equality expressed here is valid only if an exact form of the $K^{(S^2)}(\gamma, T)$ propagator is utilized. As we will see, this equality is violated when we utilize the semiclassical propagator, although nonetheless the two sides of Eq. (51) are found to be approximately in agreement.

We insert the semiclassical propagator into Eq. (51) and rewrite the integrand utilizing an extended range for $\gamma$, such that the method of stationary phase [31,67] can be applied. We refer to the variable for this extended range as $\tilde{\gamma}$, as in Fig. 3(a), with $\tilde{\gamma} = |\gamma_n| = |\gamma + 2\pi n|$ with the sequence of winding numbers $n = 0, -1, 1, -2, 2, \ldots..$ (For this simplified problem it is sufficient to use $0 \le \tilde{\gamma} < \infty$, although for the general problem treated in Section III we had to use $-\infty < \tilde{\gamma} < \infty$). With $\tilde{\gamma} = |\gamma_n|$, stationary-phase analysis of Eq. (51) is easily accomplished since all of the $\gamma$ dependence of the semiclassical propagator occurs only in terms of $|\gamma_n|$. We thus have

$$2\pi e^{iE_\ell T/\hbar} \int_0^\pi \sin\gamma \, d\gamma \, P_\ell(\cos\gamma) \;\; K_\mathrm{sc}^{(S^2)}(\gamma, T)$$

$$= \frac{I \, e^{iE_\ell T/\hbar}}{\hbar i T} \int_0^\pi \sin\gamma \, d\gamma \, P_\ell(\cos\gamma) \sum_{n=-\infty}^{\infty} \left| \frac{\gamma_n}{\sin\gamma_n} \right|^{1/2} e^{-i\pi \lfloor |\gamma_n|/\pi \rfloor /2} \; e^{iI\gamma_n^2/2\hbar T} \tag{52a}$$

$$= \frac{I \, e^{iE_\ell T/\hbar}}{\hbar i T} \int_0^\infty |\sin\tilde{\gamma}|^{1/2} d\tilde{\gamma} \, P_\ell(\cos\tilde{\gamma}) \;\; \tilde{\gamma}^{1/2} e^{-i\pi\lfloor \tilde{\gamma}/\pi \rfloor/2} \; e^{iI\tilde{\gamma}^2/2\hbar T} \tag{52b}$$

where in going from Eq. (52a) to (52b) we evaluate $\sin\gamma = |\sin|\gamma_n||$ and $\cos\gamma = \cos|\gamma_n|$ for $0 \le \gamma \le \pi$ and integer $n$, and we rewrite the integral in terms of the variable $|\gamma_n|$. In Eq. (52b) we then relabel that variable as simply $\tilde{\gamma}$. To evaluate Eq. (52b), we first write the Legendre polynomial as

$$P_\ell(\cos\tilde{\gamma}) = \sum_{\substack{k=0 \\ \{\ell+k,\text{even}\}}}^{\ell} (-1)^{\ell+k} \frac{2-\delta_{k0}}{4^\ell} \binom{\ell-k}{(\ell-k)/2} \binom{\ell+k}{(\ell+k)/2} \cos k\tilde{\gamma} \tag{53}$$

(a form provided by Matt Majic in https://mathoverflow.net/questions/277155, and which, as indicated there, can be easily confirmed by direct evaluation). In Eq. (53), the sum runs over even values of $\ell + k$, $\delta_{k0}$ is the Kronecker $\delta$-function, and $\binom{u}{w}$ refers to a binomial coefficient (*i.e.* a generalized binomial coefficient for the case of half integer values of the argument $w$). For the $|\sin\tilde{\gamma}|^{1/2}\, e^{-i\pi\lfloor\tilde{\gamma}/\pi\rfloor/2}$ term in the integrand of Eq. (52b), it has periodicity of $4\pi$ so it can be expanded as a Fourier series $\sum_\nu a_{\nu,\text{sc}} \exp(i\nu\tilde{\gamma}/2)$, with

$$a_{\nu,\text{sc}} = \frac{1}{4\pi}\int_0^{4\pi} |\sin\tilde{\gamma}|^{1/2}\, e^{-i\pi\lfloor\tilde{\gamma}/\pi\rfloor/2}\, e^{-i\nu\tilde{\gamma}/2}\, d\tilde{\gamma} \tag{54a}$$

$$= \begin{cases} \dfrac{-(1+i)\,\Gamma[(\nu+5)/4 - 3/2]}{4\sqrt{\pi}\,\Gamma[(\nu+5)/4]} & \text{if} \quad \nu = -1, 3, 7, 11, \dots \\ 0 & \text{otherwise} \ . \end{cases} \tag{54b}$$

We relabel these coefficients as $a_{\mu,\text{sc}}$ using integer $\mu$, with $\nu = 4\mu - 1$ such that $\mu$ extends over all integer values $\geq 0$.

We insert Eqs. (53) and (54b) into Eq. (52b), yielding

$$2\pi e^{iE_\ell T/\hbar} \int_0^\pi \sin\gamma\, d\gamma\, P_\ell(\cos\gamma)\, K_{\text{sc}}^{(S^2)}(\gamma, T)$$

$$= \frac{I\, e^{iE_\ell T/\hbar}}{\hbar i T} \sum_{\substack{k=0 \\ \{\ell+k,\text{even}\}}}^{\ell} (-1)^{\ell+k} \frac{2-\delta_{k0}}{4^\ell} \binom{\ell-k}{(\ell-k)/2} \binom{\ell+k}{(\ell+k)/2}$$

$$\times \sum_{\mu=0}^{\infty} a_{\mu,\text{sc}} \int_0^\infty \cos(k\tilde{\gamma})\, \tilde{\gamma}^{1/2}\, e^{i(4\mu-1)\tilde{\gamma}/2}\, e^{i\tilde{\gamma}^2/2T}\, d\tilde{\gamma} \tag{55}$$

and the remaining integral over $\gamma$ can be expressed as

$$\int_0^\infty \cos(k\tilde{\gamma})\, \tilde{\gamma}^{1/2}\, e^{i(4\mu-1)\tilde{\gamma}/2}\, e^{iI\tilde{\gamma}^2/2\hbar T}\, d\tilde{\gamma}$$

$$= \left(\frac{T}{I}\right)^{3/2} \int_0^\infty \cos(kL_\text{c}T/I)\, (L_\text{c})^{1/2}\, e^{i(4\mu-1)L_\text{c}T/2I}\, e^{i(L_\text{c})^2 T/2\hbar I}\, dL_\text{c} \tag{56}$$

using the integration variable $L_\text{c} = I\tilde{\gamma}/T$. This integral can be evaluated exactly in terms of hypergeometric functions, but we are more interested in its evaluation using stationary-phase analysis in the limit $T \to \infty$. Expressing $\cos(kL_c T/I)$ as $[\exp(ikL_c T/I) + \exp(-ikL_c T/I)]/2$, we then evaluate the integral

$$\int_0^\infty (L_c)^{1/2}\, e^{\pm ikL_c T/I}\, e^{i(4\mu-1)L_c T/2I}\, e^{i(L_c)^2 T/2\hbar I}\, dL_c$$

$$= e^{-i\hbar^2(2\mu\pm k-1/2)^2 T/2I} \int_0^\infty (L_c)^{1/2}\, e^{i[L_c+\hbar(2\mu\pm k-1/2)]^2 T/2\hbar I}\, dL_c \tag{57a}$$

$$\sim e^{-i\hbar^2(2\mu\pm k-1/2)^2 T/2I} \sqrt{\frac{2\pi i\hbar I}{T}} \left(L_c^{(\pm)}\right)^{1/2} \tag{57b}$$

with a stationary phase of $L_c^{(\pm)} = -\hbar(2\mu \pm k - 1/2)$. It is only $L_c^{(\pm)} > 0$ that we need be concerned with according to the limits of the integral over $L$ (a stationary phase of 0 would require special consideration, but this value is not encountered since $\mu$ and $k$ are both integers), in which case the sum over $\mu$ in Eq. (55) can be restricted to a maximum value of $\mu = \ell$.

Thus, combining Eqs. (52) – (57), we find for the time dependence of eigenstates as given by the semiclassical propagator

$$2\pi\, e^{iE_\ell T/\hbar} \int_0^\pi \sin\gamma\, d\gamma\, P_\ell(\cos\gamma)\, K_{sc}^{(S^2)}(\gamma, T)$$

$$\sim e^{iE_\ell T/\hbar} \sum_{\substack{k=0 \\ \{\ell+k,\text{even}\}}}^{\ell} (-1)^{\ell+k+1} \frac{2-\delta_{k0}}{4^{\ell+1}} \binom{\ell-k}{(\ell-k)/2} \binom{\ell+k}{(\ell+k)/2} \sum_{\mu=0}^{\ell} \frac{\Gamma(\mu-1/2)}{\Gamma(\mu+1)}$$
$$\times \Big[(-2\mu-k+1/2)^{1/2}\Theta(-2\mu-k+1/2)\, e^{-i\hbar(2\mu+k-1/2)^2 T/2I}$$
$$+(-2\mu+k+1/2)^{1/2}\Theta(-2\mu+k+1/2)\, e^{-i\hbar(2\mu-k-1/2)^2 T/2I}\Big] \tag{58a}$$

$$= \begin{cases} \sqrt{\dfrac{\pi}{2}}\, e^{-i\hbar T/8I}, & \text{for } \ell = 0 \quad (58b) \\[2ex] \dfrac{\sqrt{6\pi}}{4}\, e^{-i\hbar T/8I}, & \text{for } \ell = 1 \quad (58c) \\[2ex] \dfrac{3\sqrt{10\pi}}{16}\, e^{-i\hbar T/8I} + \dfrac{\sqrt{2\pi}}{32}\, e^{23i\hbar T/8I}, & \text{for } \ell = 2 \quad (58d) \\[2ex] \dfrac{5\sqrt{14\pi}}{32}\, e^{-i\hbar T/8I} + \dfrac{\sqrt{6\pi}}{64}\, e^{39i\hbar T/8I}, & \text{for } \ell = 3 \quad (58e) \\[2ex] \dfrac{105\sqrt{2\pi}}{256}\, e^{-i\hbar T/8I} + \dfrac{5\sqrt{10\pi}}{512}\, e^{55i\hbar T/8I} + \dfrac{29\sqrt{2\pi}}{2048}\, e^{79i\hbar T/8I}, & \text{for } \ell = 4 \quad (58f) \end{cases}$$

where $\Theta$ denotes a Heaviside step function.

Regarding the *leading terms* in the expressions of Eqs. (58b) – (58f), these all arise from the stationary phase $L_c^{(-)}$ (contained in the *second* term within the square brackets of Eq. (58a)) having value given by $k = \ell$ together with $\mu = 0$, so that $L_c^{(-)} = \hbar(\ell + 1/2)$. The associated dominant paths that the particle travels on in a eigenstate with given value of $\ell$ are those with classical energy of $\hbar^2(\ell + 1/2)^2/2I$. It is only by inclusion of an additional $\exp(i\hbar T/8I)$ prefactor (as in the exact propagator of Eq. (4)) that the leading terms in Eqs. (58b) – (58f) will lead to the correct, non-time-dependent result as on the left-hand side of Eq. (51).

The non-leading terms in Eqs. (58b) – (58f), with or with an additional prefactor, have incorrect time dependence for the eigenstates in question. However, the magnitude of those terms is small, typically about $1/20$ of the leading terms for the $\ell$ values displayed in Eq. (58) and becoming progressively smaller for larger $\ell$. It also should be noted that the leading terms of Eq. (58) do not have magnitude equal to 1 (which would be the correct result according to the left-hand side of Eq. (51)). We find that those magnitudes can be written as $\Gamma(\ell + 3/2)/\left[\sqrt{\ell + 1/2}\ \Gamma(\ell + 1)\right]$. This expression approaches 1 for large $\ell$, differs from 1 by less than 10% for $\ell \geq 1$, and is about 25% different than 1 for $\ell = 0$.

In summary, if an $\exp(i\hbar T/8I)$ prefactor were to be included in the semiclassical propagator, it would then produce fairly accurate eigenstates when utilized in Eq. (51) for the case of $\ell \geq 1$, and even for $\ell = 0$ it would yield a correct time dependence although with incorrect magnitude by about 25%. Most importantly, the type of paths that dominate in producing an eigenstate with quantum number $\ell$ are those having their starting and ending locations separated by a path with classical angular momentum of $L_c \approx \hbar(\ell + 1/2)$, with the range of $L_c$ values that contribute significantly to the eigenstate being on the order of $\sqrt{\hbar I/T}$ (in accordance with the method of stationary phase [31,67]).

**Appendix V**

Numerical integration for the results of Section III was performed by dividing the full range of each integration into a series of small, equal-width integration intervals. We use that width, times the value of the integrand at the midpoint of a small interval, as an estimate of the integral over that interval. Hence, for an integration over a variable $u$ ranging between $u_a$ and $u_b$, and with $N$ intervals, the midpoints where the integrand is evaluated are given by $u_i = u_a + (i - 0.5)\Delta u$, $i = 1,2, \dots N$ where $\Delta u = (u_b - u_a)/N$ is the interval width. Results in Figs. 6 and 7 were obtained using 64 intervals for the $\alpha$ integral, 32 for the $\theta_f$ integral, and 1 for the $\phi_f$ integral. For the $\Phi_0$ integral in Eq. (32a) as plotted in Fig. 6 we used 64 intervals, with the same value used for the number of $\theta_{L_c}$ plotting points in each quadrant of the plots of Eq. (37) in Fig. 7. An interval width of 0.001 (employing units of $\hbar = I = 1$) was used for the $L_c$ integrals of Eq. (36), with the same value used in the plotting of Eq. (32a) in Fig. 6. For the $L'$ integral of Eq. (27), the number of integration intervals was taken to be the maximum of $\left\lfloor 5\sqrt{\hbar T/I} \right\rfloor$ or $\left\lfloor 3\sqrt{T/\hbar I}\ |L_c| \right\rfloor$. This value was chosen so that the results of Fig. 6 (even when viewed on an expanded scale) do not significantly change with further increase in this number. Similarly, increasing any of the other numbers of intervals, or decreasing any of the interval widths, does not significantly change the results.

However, this relative insensitivity of the results is dependent upon the extent of the integral over $L_c$ that we actually employ in evaluating Eq. (36), since the associated integrand becomes more and more oscillatory as $|L_c|$ (and hence $|\tilde{\gamma}|$ as obtained from Eq. (25)) increases. This situation is the usual one encountered in path-integral phasor diagrams of the type considered here [17]. Hence, as $|L_c|$ increases, a smaller interval width is then required for $L_c$ as well as for $L'$ (and $\alpha$). The results in Fig. 7 were obtained using an $L_c$ integral extending over $\pm 3\hbar(\ell + 1/2)$ except when $\ell = 0$ and $m = 0$ in which case we used $\pm 4\hbar(\ell + 1/2)$. Again, no significant change in the results was found when increasing the number of integration intervals nor decreasing their width, so long as we restrict ourselves to this range of the $L_c$ integral. This choice is found to produce results that are close to the fully converged value (for an infinite range) as further discussed at the end of this Section.

Our choice of grid for the numerical integrations does not require integrand evaluations at the boundaries $u = u_a$ or $u_b$ of the integration range, which is why we prefer it compared to a grid consisting of $u_i = u_a + (i-1)\Delta u,\ i = 1,2, \dots N+1$. In path-integral analysis of the type performed here, there oftentimes can be discontinuities of an integrand that occur at boundaries of the integration range (due to a divergence or a branch cut), *i.e.* resulting in a so-called improper integral. Our use of a grid that is offset from the boundaries permits direct, numerical evaluation of the integral in a manner that converges to the true value as the interval size goes to zero, even without any further adjustment of the method.

Nevertheless, in cases where a divergence of the integrand occurs at the boundary (but nonetheless the integral, or at least its Cauchy principal value, is finite), then $\Delta u$ times the value of the integrand at the midpoint provides a very poor estimate of the integral over the small interval. Hence, a modified procedure is required, preferably one that does not involve much more computation time than that required for simply evaluating the integrand at the midpoint of the interval. We have developed such a method, which we illustrate here using two examples. First, we discuss the integral over $\alpha$ in Eq. (27), for which an inverse square-root divergence of the integrand occurs when $\alpha \to \gamma$ so long as $\gamma \neq 0$. If $\gamma = 0$ then the nature of the singularity changes: if $n \neq 0$ then it becomes like $1/\alpha$, whereas if $n = 0$, there is no singularity. For the integration interval that borders the $\alpha = \gamma$ value, we wish to construct an estimate for the exact value of the integral

$$\int_{\gamma}^{\gamma+\Delta\alpha} \frac{d\alpha\ (\alpha + 2\pi n)\ e^{iI(\alpha+2\pi n)^2/2\hbar T}}{(\cos\gamma - \cos\alpha)^{1/2}} \tag{59}$$

where $\gamma$ is in the range $0 < \gamma < \pi$, and $\Delta\alpha$ is the interval width, with $\Delta\alpha \ll (\pi - \gamma)$.

As an initial attempt to form an estimate for Eq. (59), we expand the integrand to first order about $\alpha = \gamma$ and then evaluate the integral using that expansion. We thus obtain an estimate of

$$(\gamma + 2\pi n)\ e^{iI(\gamma+2\pi n)^2/2\hbar T}\ \frac{2\sqrt{\Delta\alpha}}{\sqrt{\sin\gamma}}\quad . \tag{60}$$

By numerical comparison of Eqs. (59) and (60) as a function $\gamma$, using $T = 32\pi$ (in units of $\hbar = I = 1$) and considering various values of $\Delta\alpha$, we find that this estimate can be rather inaccurate when $\gamma < \Delta\alpha$. One way to improve the estimate is to form an expansion of the integrand of Eq. (59) to second order; we find that this improves the estimate for $\gamma \gtrsim \Delta\alpha$, but when $\gamma \ll \Delta\alpha$ the estimate actually becomes worse (since the integral of the expansion contains powers of $\Delta\alpha/\tan\gamma$).

We find that a useful, alternate method for improving the estimator of Eq. (60) is to make small adjustments in the values of the terms therein, *i.e.,* by adding fractions of $\Delta\alpha$ to certain occurrences of $\gamma$ in the expression. To motivate this method, let us consider forming an expression like Eq. (60) using different procedure, starting with a first-order expansion of *only* the divergent part of the integrand of Eq. (59), which is given by $(\cos\gamma - \cos\alpha)^{-1/2}$ when both $n \neq 0$ and $\gamma \neq 0$. We integrate that over $\alpha$ varying from $\gamma$ to $\gamma + \Delta\alpha$ to yield $2\sqrt{\Delta\alpha}/\sqrt{\sin\gamma}$, and then we combine that result with an evaluation of the remaining $(\alpha + 2\pi n)\exp[iI(\alpha + 2\pi n)^2/2\hbar T]$ part of the integrand at some specific value of $\alpha$. If this $\alpha$-value is taken to be simply $\gamma$, then we end up with the same expression as in Eq. (60). However, we consider here using a different $\alpha$-value. For example, if we were dealing with a very strong divergence at $\alpha = \gamma$, then an evaluation at $\alpha = \gamma$ would indeed be appropriate, but if we had only a very weak divergence (or no divergence), then a mid-point evaluation at $\alpha = \gamma + \Delta\alpha/2$ would be best. For the present situation of a square-root divergence, it seems reasonable to try a value of $\alpha = \gamma + \Delta\alpha/4$ for evaluating the remaining part of the integrand.

In this manner, we form a revised estimate of Eq. (59), as

$$\left(\gamma + \frac{\Delta\alpha}{4} + 2\pi n\right)\, e^{iI\left(\gamma+\frac{\Delta\alpha}{4}+2\pi n\right)^2/2\hbar T}\, \frac{2\sqrt{\Delta\alpha}}{\sqrt{\sin\gamma}}\ . \tag{61}$$

Let us consider the usefulness of this estimator. Plotting Eqs. (60) and (61) as a function of $\gamma$, for various values of $\Delta\alpha$ and $n$ (with $T$ maintained at $32\pi$, using units of $\hbar = I = 1$) and comparing those results to a high-precision evaluation of Eq. (59), we find that for certain values of $n$ and $\gamma$ (in particular, $n = 0$ and $\gamma \gtrsim \Delta\alpha$) that Eq. (61) is a better estimator than Eq. (60). However, for $n = 0$ and $\gamma \ll \Delta\alpha$, Eq. (61) is a much worse estimator than Eq. (60) since the former diverges as $\gamma \to 0$ whereas the latter does not (nor does Eq. (59)). For all other values of $n$ and/or $\gamma$, we find that the estimates provided by Eqs. (60) and (61) are comparable in their accuracy.

To improve Eq. (61) such that a better estimate is provided for the case of $n = 0$ and $\gamma \ll \Delta\alpha$, we also modify the $\gamma$ value in the $\sqrt{\sin\gamma}$ term of the denominator. Specifically, we take the estimator to be

$$\left(\gamma + \frac{\Delta\alpha}{4} + 2\pi n\right)\, e^{iI\left(\gamma+\frac{\Delta\alpha}{4}+2\pi n\right)^2/2\hbar T}\, \frac{2\sqrt{\Delta\alpha}}{\sqrt{\sin(\gamma + \Delta\alpha/8)}}\ . \tag{62}$$

This form solves the problem of the unwanted divergence in Eq. (61) when $n = 0$ and $\gamma \to 0$. Indeed, Eq. (62) provides an estimate that is very much better than that given by Eq. (60) for any values of $n$, $\gamma$, and $\Delta\alpha$, *except* when $n \neq 0$ and $\gamma \ll \Delta\alpha$. In that case, Eq. (60) diverges as $\gamma \to 0$, as does the exact expression of Eq. (59), but Eq. (62) does not diverge. However, importantly, these divergences in the exact expression all cancel out, due to a Cauchy principal value that ends up being taken in the integrals over $L_{\mathrm{c}}$ and $L'$ whenever $\tilde{\gamma}$ (as given by Eq. (25) and which determines both $n$ and $\gamma$ according to Eq. (11)) crosses an even multiple of $\pi$. (In contrast, when $\tilde{\gamma}$ is an odd multiple of $\pi$, then $\gamma = \pi$ so that the range of the integral over $\alpha$ is zero and hence there is no divergence). Therefore, having the estimator of Eq. (62) produce a finite value rather than a diverging value when $\gamma \to 0$ is no problem at all, and actually it is a convenient feature of Eq. (62) since obviates the need to test for potential numerical overflows (of the estimator) in the computer code.

Our second example of numerical evaluation of an improper integral deals with the integral over $\theta_\mathrm{f}$ in Eq. (35a), for which the $\left(\cos^2\theta_\mathrm{g}/\cos^2\theta_\mathrm{f}-1\right)^{-1/2}$ term of the integrand diverges when $\theta_\mathrm{f}\to\theta_\mathrm{g}$ or $\theta_\mathrm{f}\to\pi-\theta_\mathrm{g}$, corresponding to the boundaries of the integration. Just as in the prior example, one way to evaluate the integral would be to perform a first-order expansion of the integrand about these boundary points. However, in this case the integrand is sufficiently complicated (containing the $\Delta\mathcal{I}_{\ell,m}$ terms) so that this sort of expansion cannot be analytically evaluated in closed form. Hence, we move directly to the alternate procedure introduced in the prior example, considering just a divergent portion of the integrand and attempting to find an estimate for that which is then combined with a single-point evaluation of the remainder of the integral. Thus, we start with the integrals

$$\int_{\theta_\mathrm{g}}^{\theta_\mathrm{g}+\Delta\theta_\mathrm{f}} d\theta_\mathrm{f}\,\frac{\sin\theta_\mathrm{f}}{2\cos^2\theta_\mathrm{f}\sqrt{\hbar I\left(\cos^2\theta_\mathrm{g}/\cos^2\theta_\mathrm{f}-1\right)\left(\tan^2\theta_\mathrm{f}-\cos^2\theta_\mathrm{g}/\cos^2\theta_\mathrm{f}+1\right)/T}} \tag{63a}$$

and

$$\int_{\pi-\theta_\mathrm{g}-\Delta\theta_\mathrm{f}}^{\pi-\theta_\mathrm{g}} d\theta_\mathrm{f}\,\frac{\sin\theta_\mathrm{f}}{2\cos^2\theta_\mathrm{f}\sqrt{\hbar I\left(\cos^2\theta_\mathrm{g}/\cos^2\theta_\mathrm{f}-1\right)\left(\tan^2\theta_\mathrm{f}-\cos^2\theta_\mathrm{g}/\cos^2\theta_\mathrm{f}+1\right)/T}}\ , \tag{63b}$$

with $0\le\theta_\mathrm{g}\le\pi$, and we expand the integrands to first order about $\theta_\mathrm{f}=\theta_\mathrm{g}$ or $\theta_\mathrm{f}=\pi-\theta_\mathrm{g}$, respectively. We then perform the resulting integrals, yielding in both cases an estimate of

$$\sqrt{2\,\Delta\theta_\mathrm{f}\tan\theta_\mathrm{g}}\ . \tag{64}$$

We find that this expression provides a reasonable estimate of Eqs. (63a) and (63b), except when $\theta_\mathrm{g}\ll\Delta\theta_\mathrm{f}$ or $\pi-\theta_\mathrm{g}\ll\Delta\theta_\mathrm{f}$, respectively. In these cases, the $\left(\tan^2\theta_\mathrm{f}-\cos^2\theta_\mathrm{g}/\cos^2\theta_\mathrm{f}+1\right)^{-1/2}$ term in the integrand makes a large contribution to the integral beyond that given by Eq. (64), since this term, when $\theta_\mathrm{g}=0$, itself diverges as $\theta_\mathrm{f}\to 0$.

As for our prior example, we can attempt to improve the estimator of Eq. (64) by including second-order terms in our expansion, but once again we find that, while those do indeed improve the estimate when $\theta_\mathrm{g}\gtrsim\Delta\theta_\mathrm{f}$, they fail when $\theta_\mathrm{g}\ll\Delta\theta_\mathrm{f}$ since the integral of the expansion contains powers of $\Delta\theta_\mathrm{f}/\tan\theta_\mathrm{g}$. Hence, we utilize our alternate method, slightly modifying the value of $\theta_\mathrm{g}$ in Eq. (64). We find that a modification of

$$\sqrt{2\,\Delta\theta_\mathrm{f}\tan\left(\theta_\mathrm{g}+\Delta\theta_\mathrm{f}/2\right)} \tag{65}$$

works well for this purpose, with the same value of $\theta_\mathrm{f}=\theta_\mathrm{g}+\Delta\theta_\mathrm{f}/2$ used in the remaining terms of the integrand for the lower integration boundary, Eq. (63a), or $\theta_\mathrm{f}=\pi-\theta_\mathrm{g}-\Delta\theta_\mathrm{f}/2$ for the upper integration boundary, Eq. (63b).

Computational codes were developed using Mathematica (with default precision) and then translated into double-precision FORTRAN; see https://www.cmu.edu/physics/stm/publ/137/ for both sets of code, with the Mathematica code also listed in an Attachment to this paper. Individual distributions in Fig. 6 or 7 typically require a few hours of processor time using the FORTRAN code, with the Mathematica runs requiring nearly 100× more time (results obtained from the two codes agree to within an absolute error of $10^{-6}$). We note that much shorter runs using many fewer integration intervals yield substantially the same results, with the main influence of the coarser grids being some variability in the overall magnitude of the distributions. Figure 11 shows some results for $\mathcal{P}^{(2)}_{\ell,m}(\theta_{\mathbf{L}_c},T)$ distributions not provided in Fig. 7.

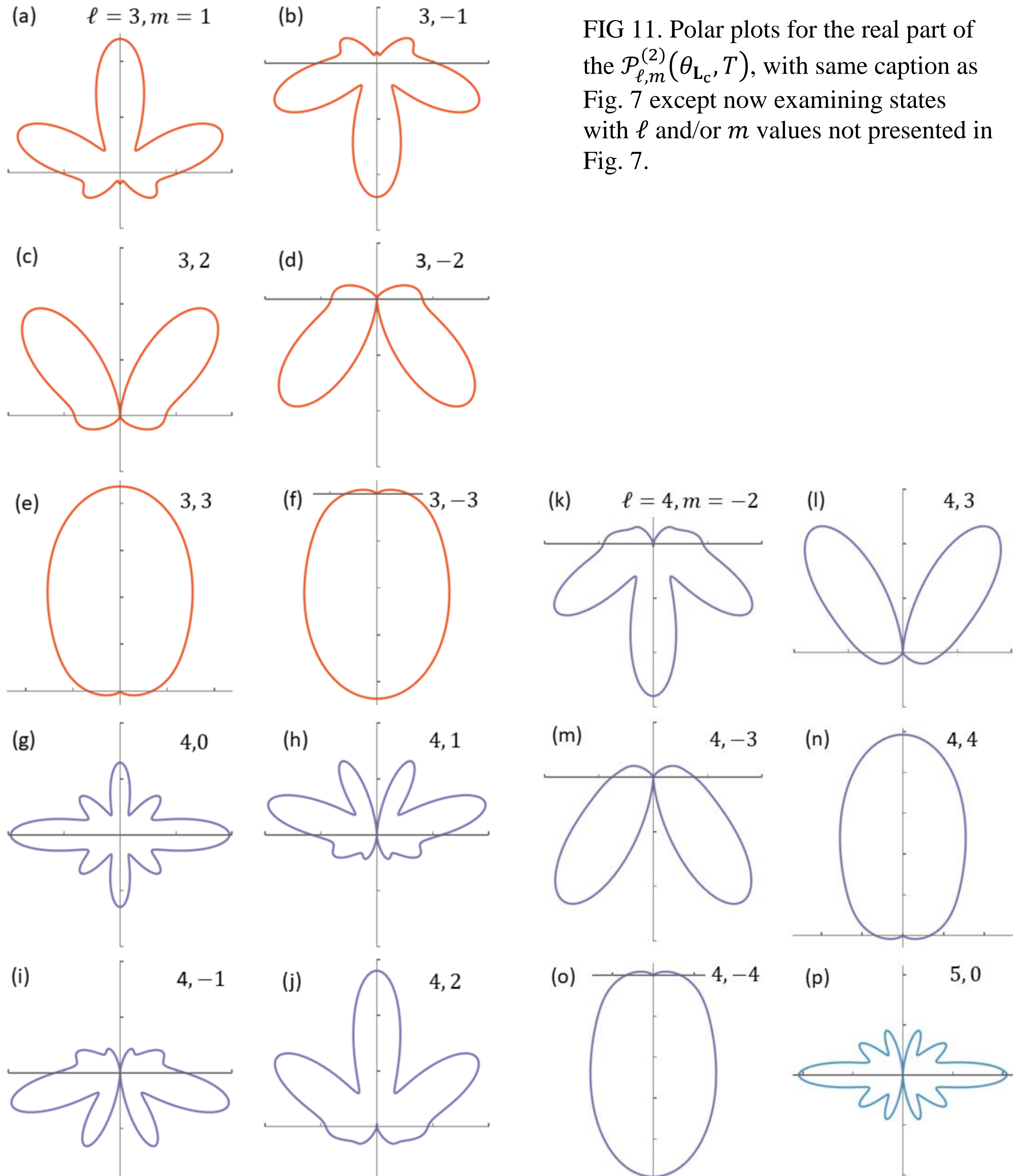

FIG 11. Polar plots for the real part of the $\mathcal{P}^{(2)}_{\ell,m}(\theta_{\mathbf{L_c}}, T)$, with same caption as Fig. 7 except now examining states with $\ell$ and/or $m$ values not presented in Fig. 7.

Finally, we return to further discuss the issue of the range of integration for the $L_c$ integral of Eq. (35a). As commented above, the results of Fig. 7 were obtained using a range of $\pm 3\hbar(\ell+1/2)$ except when $\ell=0$ and $m=0$ in which case $\pm 4\hbar(\ell+1/2)$ was used; we now justify these choices. In Fig. 12 we display computational results for $\mathcal{P}^{(2)}_{\ell,m}(\theta_{\mathbf{L}_c},T)$ in which the range is taken to be a variable, writing an expression for the range as $\pm\kappa\hbar(\ell+1/2)$ and plotting results as a function of $\kappa$. For the $\ell=0$, $m=0$ state, Figs. 12(a) – 12(d), we see that a small oscillation occurs in its overall magnitude (visible upon close examination, as indicated in the figure caption). A much larger oscillation is found to occur for the $\ell=1$, $m=1$ state, Figs. 12(e)

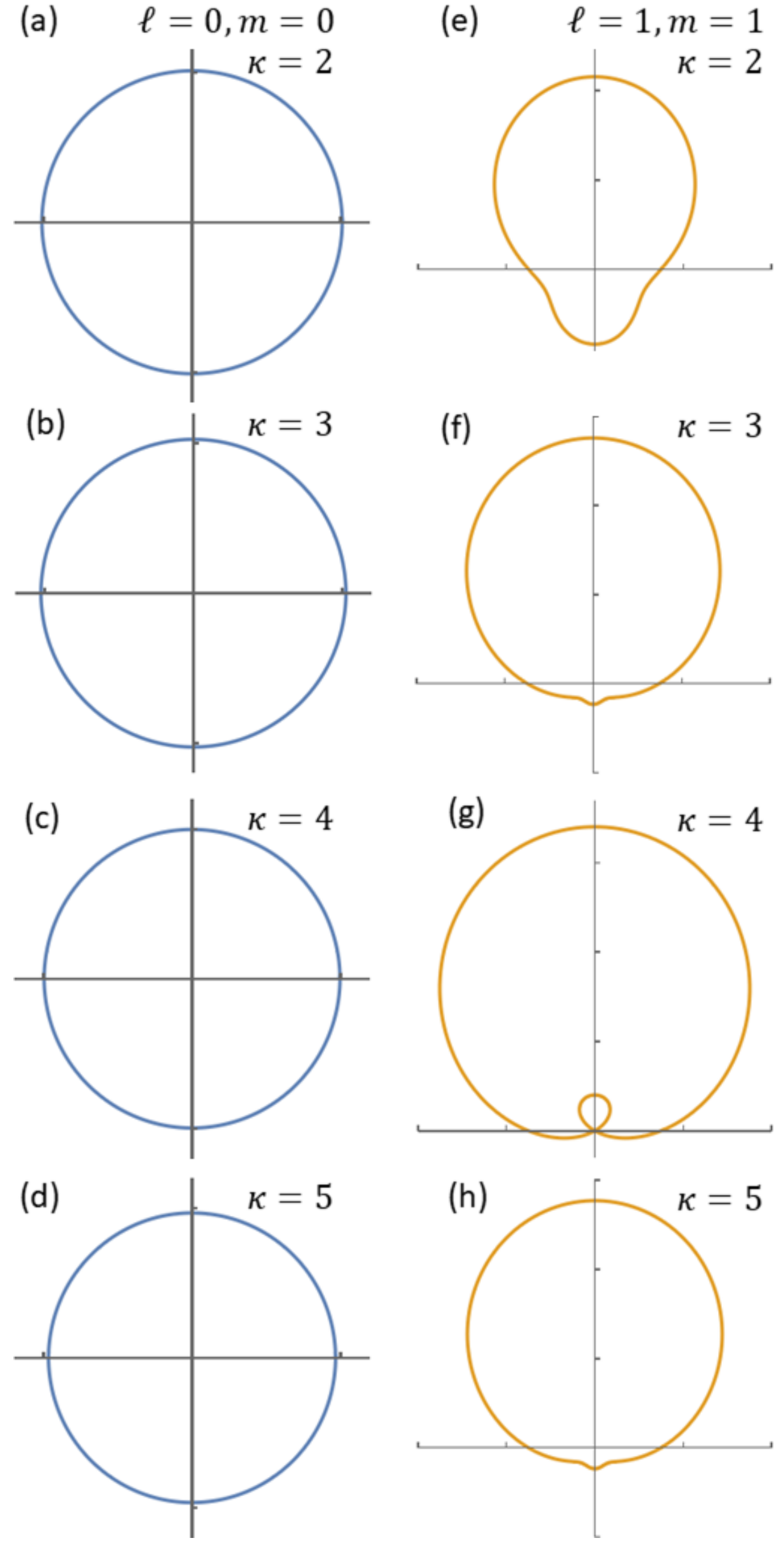

FIG 12. Polar plots for the real part of the $\mathcal{P}^{(2)}_{\ell,m}(\theta_{\mathbf{L}_c},T)$, with same caption as Fig. 7 except now examining the dependence upon the range of the $L_c$ integration. This range is denoted $\pm\kappa\hbar(\ell+1/2)$ with variable $\kappa$ taking the values indicated. For the $\ell=0$, $m=0$ state a very small oscillation is visible in the magnitude of its distributions as seen in panels (a) – (d): the tick marks at 0.5 are slightly *inside* the circular distributions for panel (b) and *outside* for panel (d), whereas the distributions overlap the tick marks for panels (b) and (c). A much larger variation of the distributions with $\kappa$ is apparent for the $\ell=1$, $m=1$ state shown in panels (e) – (h). In panel (g), the small loop of the distribution that is contained *within* the much larger loop corresponds to negative values of $\mathcal{P}^{(2)}_{\ell,m}(\theta_{\mathbf{L}_c},T)$ over a range of $\theta_{\mathbf{L}_c}$ in the lower half-plane of the plot (since the $\mathcal{P}^{(2)}_{\ell,m}(\theta_{\mathbf{L}_c},T)$ values are negative, then the points in the polar plot are plotted in the *upper* half-plane).

– 12(h). Similarly large oscillations occur for all states with $m \neq 0$. In contrast, states with $\ell > 0$ and $m = 0$ display nearly no oscillation whatsoever (more than 10× smaller than the oscillation that is weakly visible for the $\ell = 0$, $m = 0$ state in Fig. 12).

These large oscillations that occur for states with $m \neq 0$ are associated with the "sidebands" discussed in Fig. 6. Inspection of results of the type shown in Fig. 12, but for larger $\kappa$ values, reveal no indication whatsoever of a decrease in the oscillation amplitude as $\kappa$ increases. Nevertheless, we can assign a "converged" value for the oscillating $\mathcal{P}_{\ell,m}^{(2)}(\theta_{\mathbf{L}_c}, T)$ distributions simply by using a mean value of these results, as given by an average over a $\kappa$ range of $\pm 2$ relative to some central value. That is to say, averaging over the $\mathcal{P}_{0,0}^{(2)}(\theta_{\mathbf{L}_c}, T)$ distributions in Figs. 12(a) – 12(d), we achieve a result that is nearly indistinguishable from the even-$\kappa$ results of Fig. 12(a) or 12(c). Similarly, averaging over the $\mathcal{P}_{1,1}^{(2)}(\theta_{\mathbf{L}_c}, T)$ distributions in Figs. 12(e) – 12(h), we achieve a result that is nearly indistinguishable from the odd-$\kappa$ results of Fig. 12(f) or 12(h). We thus identify even-$\kappa$ results as being essentially the same as a fully converged (*i.e.* averaged) distribution for the $\ell = 0$, $m = 0$ state. For states with $m \neq 0$, we identify odd-$\kappa$ results as being essentially the same as a fully converged distribution, and the same situation holds true for states with $\ell > 0$ and $m = 0$ (since oscillations are extremely small in that case).

It is interesting to note that the distributions that lie far from the mean, such as Figs. 12(e) and 12(g), still satisfy Eqs. (38) and (39) with good accuracy. For example, considering $\kappa = 4$, then if we sum up the distribution of Fig. 12(g) together with one for $\ell = 1$, $m = -1$ (which can be obtained from Fig. 12(g) simply by reflecting it about the horizontal axis) along with one for $\ell = 1$, $m = 0$ (which is nearly identical to that shown in Fig. 7(b)), a circular distribution is obtained that satisfies Eq. (39) to the same level of accuracy as occurs for the results of Fig. 7. In any case, the use of a mean distribution, as obtained by averaging over a $\kappa$ range of $\pm 2$ relative to some central value, seems like a reasonable method of obtaining a result that can be viewed as being "converged".

However, closer examination of results as a function $\kappa$ reveals that, in addition to their oscillatory behavior, they can also display a very slight, monotonic variation as $\kappa$ increases, depending on the precise value chosen for the interval sizes of both the $L_c$ and $L'$ integrals (and perhaps for the $\alpha$ integral as well, although we have not investigated that). As already discussed above, smaller and smaller interval widths are required in order to avoid errors as $L_c$ and $L'$ increase. These required changes in the interval width have been made to the $L'$ integrals for the results shown in Figs. 7, 11, and 12, since we used the number of $L'$ intervals given by the maximum of $\left\lfloor 5\sqrt{\hbar T/I} \right\rfloor$ or $\left\lfloor 3\sqrt{T/\hbar I}\,|L_c| \right\rfloor$. However, for the $L_c$ integral, a fixed interval width of 0.001 was employed. Maintaining these values, and making computations for, say, $\kappa$ varying from 2 to 10, reveals a slight downwards trend in the values of $\mathcal{P}_{\ell,m}^{(2)}(\theta_{\mathbf{L}_c}, T)$ averaged over $\kappa$. Alternatively, if we also modify the interval width used from the $L_c$ integral, allowing it to decrease with $\kappa$, then we can achieve a mean $\mathcal{P}_{\ell,m}^{(2)}(\theta_{\mathbf{L}_c}, T)$ value that is practically constant as a function of $\kappa$. In any case, conveniently, we find that results for small, fixed values of $\kappa$ are very close to those converged values. We choose the minimum possible $\kappa$ values such that $|L_c|$-values well above $\hbar(\ell + 1/2)$ are properly included: an even value of $\kappa = 4$ for the $\ell = 0$ state, and an odd value of $\kappa = 3$ for all other states.

Concerning the origin of the oscillations in the curves of Fig. 12 (and the associated sidebands in the distributions of Fig. 6), we note that a large amount of cancellation in the probability amplitudes of the various paths is required when $|L_c|$ is well above $\hbar(\ell + 1/2)$ in order to form an eigenstate. On the one hand, since we are using the propagator of Eq. (4), we are expressing the distributions of Figs. 6 and 7 essentially in terms of geodesic, great-circle motion (together with the *elastica* as labeled by $\alpha$). But on the other hand, since we are describing eigenstates of fixed *z*-component of angular momentum, we are inherently dealing with nonclassical (non-geodesic) orbits. This importance of nonclassical free-particle paths is very apparent when examining the dominant paths associated with the Pöschl-Teller potential in discussed in Section IV and Appendix VI. As the polar angle of the plane of the paths decreases from $\pi/2$ to near 0, then the speed of travel along the paths deviates more and more from a constant (in order to achieve a fixed value for the *z*-component of the angular momentum). In particular, for large $\ell$ and small $|m| > 0$, kinks in the paths develop. Since our use of Eq. (4) fundamentally expresses the distributions of Figs. 6 and 7 in terms of geodesic, great-circle motion (along with the *elastica*), the description necessarily must include all sorts of these geodesics, thereby giving rise to the sidebands with $|L_c|$ values differing from $\hbar(\ell + 1/2)$. It is not surprising in this regard that we must use different descriptions for $m = 0$ and $m \neq 0$ cases (odd-$\kappa$ values for the latter and even-$\kappa$ for the former at least when $\ell = 0$, with either even or odd values of $\kappa$ working fine when $m = 0$ and $\ell > 0$), considering the fact that the latter is described by a Pöschl-Teller potential but the former is not.

**Appendix VI**

In addition to the $S^2$ propagator of Eq. (4), as well as the time-sliced form of the propagator [4,54], there is a separate means of analysis that leads to another, exact, path-based expression. Specifically, an *azimuthal decomposition* is performed in accordance with [3]

$$K^{(S^2)}(\theta_0, \phi_0, \theta_f, \phi_f, T) = \sum_{m=-\infty}^{\infty} K_{\text{ad}}^{(S^2,m)}(\theta_0, \theta_f, T)\, \frac{1}{2\pi} e^{im(\phi_f - \phi_0)} \,, \tag{66}$$

where $K_{\text{ad}}^{(S^2,m)}$ are individual, azimuthally decomposed propagators for each value of $m$. These propagators deal with paths that are described in terms of a single variable, $\theta(t)$, since the $\phi$-dependence of Eq. (59) is formulated in terms of $\exp(im\phi)$ Fourier-expansion terms (which are identical to the azimuthal part of the wave functions), rather than paths. Thus, Eq. (66) as it stands is not suitable for describing in detail the $\big(\theta(t), \phi(t)\big)$ paths that make up the spherical harmonics, since there are no explicit $\phi(t)$ paths here. Nevertheless, here we will perform an approximate, semiclassical analysis in order to deduce $\phi(t)$ paths, so that comparison can be made to the paths of the propagator of Eq. (4).

As a heuristic introduction to the azimuthally decomposed propagators, we do not initially utilize Eq. (66), but rather, we consider how the problem of free-particle motion in the curved space $S^2$ can be transformed into an equivalent problem that occurs in a flat space [3,62]. We start from the angular part of the Schrödinger equation in spherical coordinates, with solutions given by spherical harmonics $Y_{\ell m}(\theta,\phi) = N_{\ell,m} P_\ell^m(\cos\theta)\exp(im\phi)/2\pi$ where the normalization factors are given by $N_{\ell,m} = (-1)^m\sqrt{\pi(2\ell+1)(\ell-m)!/(\ell+m)!}$ and $P_\ell^m$ are associated Legendre polynomials. Evaluation of the $\phi$-derivatives then leads to the well-known result

$$-\left[\frac{1}{\sin\theta}\frac{d}{d\theta}\left(\sin\theta\frac{d}{d\theta}\right)-\frac{m^2}{\sin^2\theta}\right]P_\ell^m(\cos\theta) = \ell(\ell+1)P_\ell^m(\cos\theta) \tag{67}$$

where the values of the integer $\ell$ for given $|m|$ extend over $\ell \geq |m|$. A well-known transformation [3,62] is then made to a different set of eigenfunctions,

$$y_\ell^m(\cos\theta) \equiv \sqrt{\sin\theta}\, P_\ell^m(\cos\theta)\,, \tag{68}$$

which yields (including a multiplier of $\hbar^2/2I$ on both sides)

$$\frac{\hbar^2}{2I}\left[-\frac{d^2}{d\theta^2}-\frac{1}{4}+\frac{m^2-1/4}{\sin^2\theta}\right]y_\ell^m(\cos\theta) = E_\ell\, y_\ell^m(\cos\theta)\ . \tag{69}$$

In this equation there is a second derivative term in $\theta$ but no first derivative term. Therefore, this transformed problem is identical to what occurs in a flat space, and thus a time-sliced form of the propagator can be written down for each integer value of $m$ in just the same way as done for a flat 1-dimensional space (albeit one with a restricted range of the coordinate, $0 \leq \theta \leq \pi$).

Additionally, in terms of $y_\ell^m$ the normalization of the wave function appears as

$$\int_0^\pi d\theta\sin\theta\ |P_\ell^m(\cos\theta)|^2 = \int_0^\pi d\theta\ |y_\ell^m(\cos\theta)|^2 \tag{70}$$

which equals $\left(2\pi/N_{\ell,m}\right)^2$. This result, in which the metric appearing in the integrand on the right-hand side of Eq. (70) is a constant, is again consistent with writing a path integral for the problem just as if it were a problem in a flat space. From Eq. (69), we see that in this transformed problem there is a potential of [3]

$$V(\theta) = -\frac{\hbar^2}{8I}+\frac{\hbar^2(m^2-1/4)}{2I\sin^2\theta}\ . \tag{71}$$

Importantly, an energy of $-\hbar^2/8I$ occurs in the first term on the right-hand side. Since this term appears as a constant (negative) potential, it will contribute to the path integral simply by producing a prefactor of $\exp(i\hbar T/8I)$, *i.e.* precisely the same as occurs for Camporesi's form of the propagator, Eq. (4). In this sense, the two propagators are similar.

Regarding the second term on the right-hand side of Eq. (71), it corresponds to a Pöschl-Teller (PT) potential that the particle in this transformed problem experiences, so long as the integer value of $m$ satisfies $|m| \geq 1$, *i.e.* $m \neq 0$ [3]. When $m = 0$, the potential of Eq. (71) does *not* correspond to a PT potential since there is an attractive rather than a repulsive divergent potential at $\theta = 0$ and $\theta = \pi$, so this case requires special consideration as further discussed below. Setting aside this particular case for a moment, a propagator $K^{(\mathrm{PT},m)}(\theta_0,\theta_\mathrm{f},T)$ for the PT potential of Eq. (71) can be expressed in terms of its spectral representation as

$$K^{(\mathrm{PT},m)}(\theta_0,\theta_\mathrm{f},T) = \sum_{\ell=|m|}^{\infty} \frac{\left(N_{\ell,m}\right)^2}{2\pi}\, y_{\ell,m}^{*}(\theta_0,\phi_0)\, y_{\ell,m}(\theta_\mathrm{f},\phi_\mathrm{f})\, e^{-iE_\ell T/\hbar}\ , \tag{72}$$

which, again, is valid only for $|m| \geq 1$. Multiplying by $e^{im(\phi_\mathrm{f}-\phi_0)}$ and utilizing Eq. (68) then leads to

$$\frac{K^{(\mathrm{PT},m)}(\theta_0,\theta_\mathrm{f},T)}{\sqrt{\sin\theta_0 \sin\theta_\mathrm{f}}} \frac{e^{im(\phi_\mathrm{f}-\phi_0)}}{2\pi} = \sum_{\ell=|m|}^{\infty} Y_{\ell m}(\theta_0,\phi_0)\, Y_{\ell m}(\theta_\mathrm{f},\phi_\mathrm{f})\, e^{-iE_\ell T/\hbar}\ . \tag{73}$$

Again, ignoring the restriction of $m \neq 0$ on $K^{(\mathrm{PT},m)}$, we can then sum both sides of Eq. (73) over $m$ values ranging from $-\infty$ to $+\infty$. The summations over $m$ and $\ell$ can be interchanged on the right-hand side, thus producing the spectral representation for $K^{(S^2)}(\theta_0,\phi_0,\theta_\mathrm{f},\phi_\mathrm{f},T)$. Comparing with Eq. (66), we arrive at an expression for the azimuthally decomposed propagators [3]

$$K_\mathrm{ad}^{(S^2,m)}(\theta_0,\theta_\mathrm{f},T) = \frac{1}{\sqrt{\sin\theta_0 \sin\theta_\mathrm{f}}}\, K^{(\mathrm{PT},m)}(\theta_0,\theta_\mathrm{f},T)\ . \tag{74}$$

To interpret Eq. (74) in terms of explicit paths, ideally we would like to utilize a closed-form, path-based expression for $K^{(\mathrm{PT})}(\theta_0,\theta_\mathrm{f},T)$. Unfortunately, no such expression is presently known for position space (although a fixed-energy form *is* known [68]). Nevertheless, to obtain some *qualitative* insight into the types of paths that we are dealing with, we will examine semiclassical paths associated with the PT potential, taking these to have classical energy equal to the quantized energy $E_\ell$ (in a more detailed analysis, these paths would also include whatever additional paths are required to form an *exact* propagator for the PT potential). Additionally, we will assume that the $\phi(t)$ part of the motion in Eq. (66) can be at least qualitatively described in terms of paths which directly produce the $\exp[im(\phi_\mathrm{f}-\phi_0)]$ term there. This type of analysis leads to very different paths than the great-circle (and *elastica*) paths associated with Camporesi's form of the propagator, as was described in Section II(F) and Appendix III. The essential difference between the two propagators is that the latter focuses on paths that are classical trajectories for free-particle motion on a sphere (or nearly so, for the *elastica*), with fixed value of total classical angular momentum, whereas the former focuses on those specific paths that provide the $\exp(im\phi)$ form for the azimuthal part of the wave function, *i.e.* paths that yield a specific azimuthal portion of the wave function.

Returning to the case of $m = 0$ in the potential of Eq. (71), as already noted above, the potential cannot in that case be understood as a PT one. Hence, the $m = 0$ term of Eq. (74) is not valid in that expression, as written. However, Eq. (66) is still valid as a definition of $K_\mathrm{ad}^{(S^2)}(\theta_0,\theta_\mathrm{f},T)$, and Kleinert has rigorously obtained time-sliced expressions for these propagators for all values of $m$ [3]. Those expressions reduces to the same thing as Eq. (74) for $m \neq 0$, but for $m = 0$ this reduction does not occur. That is to say, a fully correct form of the propagator is obtained in both cases, but when $m = 0$ this propagator cannot be understood in terms of motion in a PT potential.

Let us now attempt to explicitly describe the paths associated with the azimuthally decomposed $S^2$ propagators, restricting ourselves to the case of $|m| \neq 0$ so that the $\theta(t)$ part of the motion can be described by 1-dimensional motion in a PT potential. To describe paths for that motion, we

equate the quantized energy eigenvalue to the sum of classical kinetic plus potential energy with the latter given by Eq. (71), yielding

$$\frac{I}{2}\dot{\theta}^2 + \frac{\hbar^2(m^2 - 1/4)}{2I\sin^2\theta} = \frac{\hbar^2\ell(\ell+1)}{2I} \tag{75}$$

where we have *not* included the $-\hbar^2/8I$ term of Eq. (71) in our treatment here since we know that that term arises from fluctuations of the paths, as discussed in Section II(E). Solving Eq. (75) for $\dot{\theta}$ yields

$$\dot{\theta} = \frac{\hbar}{I}\left[\ell(\ell+1) - \frac{(m^2 - 1/4)}{\sin^2\theta}\right]^{1/2} \tag{76}$$

where $\theta$ varies between $\theta_0 \equiv \sin^{-1}\left[\sqrt{m^2 - 1/4}/\sqrt{\ell(\ell+1)}\right]$ and $\pi - \theta_0$, for given $|m| \geq 1$ (and $\ell \geq |m|$). Integrating $\left(\dot{\theta}\right)^{-1}$ over $\theta$ then provides an expression for $t(\theta)$,

$$t(\theta) = -\frac{1}{\sqrt{\ell(\ell+1)}}\tan^{-1}\left[\frac{\cos\theta}{\sqrt{\sin^2\theta - (m^2 - 1/4)/[\ell(\ell+1)]}}\right] + \frac{\pi}{2\sqrt{\ell(\ell+1)}} \tag{77}$$

with $t(\theta_0) = 0$. This expression describes the first half-period of the motion, ending at $\theta = \pi - \theta_0$. We can extend it to a full period by using $\pi - \theta$ rather than $\theta$ for the second half-period, as well as adding $\pi/\sqrt{\ell(\ell+1)}$ to the time. Multiple periods can then be described simply by adding $2\pi/\sqrt{\ell(\ell+1)}$ to the time for each successive period.

Paths of the PT potential thus have the particle "bouncing back and forth between the poles of the sphere", in just the manner described by Kleinert [3]. It is worth noting that, with the PT potential, we are actually solving a problem with a wave function of $\sqrt{\sin\theta}\,P_\ell^m(\cos\theta)$ of Eq. (68) rather than $P_\ell^m(\cos\theta)$ itself. The stronger zeroes that occur at $\theta = 0$ and $\pi/2$ in the former function compared to the latter will certainly affect the associated paths that describe the wave functions. Nonetheless, this effect is not expected to be too large at least for the case of $|m| = \ell \gg 1$, when the wave functions have their maxima far from $\theta = 0$ and $\pi/2$.

In order to add a $\phi(t)$ portion of the motion, we utilize paths which lead directly to azimuthal component of the wave function of $\exp(im\phi)$, as in Eq. (66). With the total angular momentum squared given by $L^2 = I^2\left(\dot{\theta}^2 + \sin^2\theta\,\dot{\phi}^2\right)$, the azimuthal component squared is given by the second term, so we take

$$I\sin\theta\ \dot{\phi} = m\hbar \tag{78}$$

such that a plot of the probability amplitude associated with the $\phi(t)$ part of a motion will precisely match the $\exp(im\phi)$ wave function (in the same manner as for motion in $\mathbb{R}^1$ [17], where the momentum $M\dot{x}$ of the dominant type of path for a state labeled by momentum eigenvalue $k$ equals $\hbar k$ [17]). Using Eq. (76) for $\dot{\theta}$, we then find for $\phi$

$$\phi = \int_{\theta_0}^{\theta}\frac{\dot{\phi}}{\dot{\theta}}\,d\theta = \int_{\theta_0}^{\theta}\frac{m\hbar}{I\,\dot{\theta}\sin\theta}\,d\theta \tag{79a}$$

$$= -\frac{m}{\sqrt{m^2 - 1/4}} \left\{ F\left[\theta, \frac{\ell(\ell+1)}{(m^2 - 1/4)}\right] - F\left[\theta_0, \frac{\ell(\ell+1)}{(m^2 - 1/4)}\right] \right\} \quad (79b)$$

$$= -\frac{m}{\sqrt{m^2 - 1/4}} \mathrm{Im}\left\{ F\left[\theta, \frac{\ell(\ell+1)}{(m^2 - 1/4)}\right] \right\} \quad (79c)$$

where $F(\theta, k^2)$ is an elliptic integral of the first kind. This expression describes the first half-period, ending at $\phi_h \equiv -m\,\mathrm{Im}\{F[\pi - \theta_0, \ell(\ell+1)/(m^2 - 1/4)]\}/\sqrt{m^2 - 1/4}$. The second half-period is obtained by using an argument of $\pi - \theta$ rather than $\theta$ in the elliptic integral, and adding $\phi_h$ to $\phi$. Successive periods are then obtained by further additions of multiples of $\phi_h$.

Figure 13 displays the these dominant paths for $|m| = \ell = 1, 2, 3,$ and 4. We see that the paths appear essentially as precessing great-circle orbits. The angle of the plane of the apparent great-circles relative to the vertical *z*-axis (*i.e.* the polar angle, prior to any precession) can be seen to be increasing as $|m| = \ell$ increases. Qualitatively, this type of behavior follows what is expected from the vector model for orbital angular momentum [13], and indeed, the curves in Fig. 13, in particular for $\ell = 3$ or 4, are quite similar to those of Fig. 1(b) as well as to Thaller's semiclassical, graphical presentation of this problem [15]. However, if we consider values of $|m|$ closer to zero, say $|m| = 1$ for large $\ell$, then the paths become quite different than great circles. Kinks form in the paths near $\theta \approx 0$ and $\theta \approx \pi$ (as faintly seen in Fig. 13(a), and much more evident for $|m| = 1$ with large $\ell$). Moreover, the speed along the paths deviates from a constant, being largest at $\theta = \pi/2$ (the azimuthal component of the speed, $\sin\theta\ \dot{\phi}$, is a constant whereas the polar component, $|\dot{\theta}|$, varies from 0 up to some maximum value at $\theta = \pi/2$).

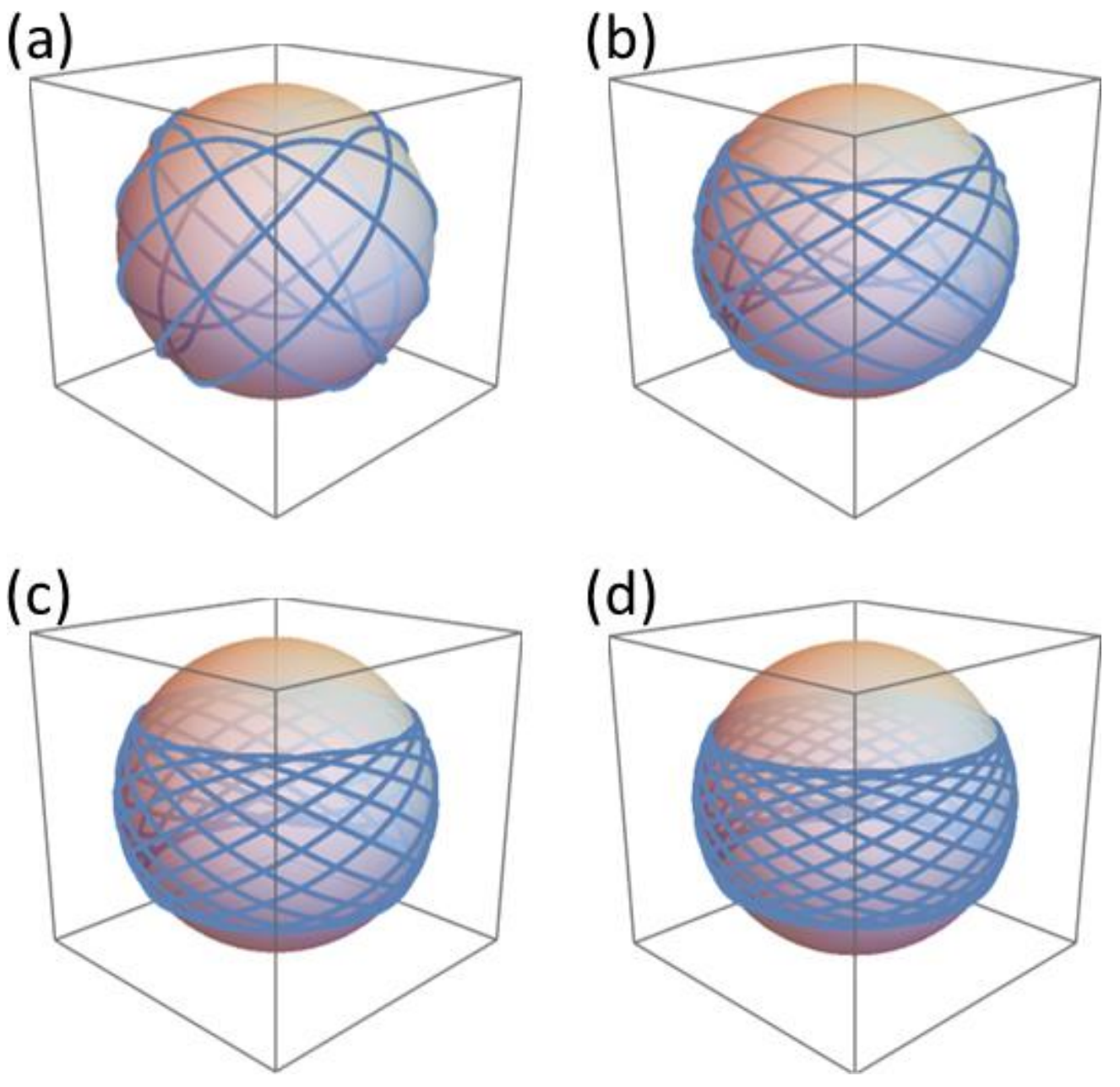

FIG 13. Qualitative, semiclassical paths associated with the azimuthally decomposed propagator. The $\theta(t)$ part of the paths is in accordance with Eq. (77), associated with the Pösch-Teller potential, whereas the $\phi(t)$ part is assumed to be given by Eq. (79). Results are for $|m| = \ell = 1, 2, 3,$ and 4 in (a) – (d) and are plotted using 7, 11, 15, and 19 periods, respectively, in the $\theta(t)$ part of the motion.

[26] C. Morette, “On the Definition and Approximation of Feyman's Path Integrals”, Phys. Rev. **81**, 848 (1951).
[27] L. S. Schulman (2005) *ibid.*, Chapter 23.
[28] P. Cartier and C. DeWitt-Morette (2006) *ibid.*, Chapter III-8.
[29] P. A. Horváthy, “The Maslov correction in the semiclassical Feynman integral”, Cent. Eur. J. Phys. **9**, 1, (2011).
[30] V. P. Maslov, M. V. Fedoriuk: *Semi-classical approximation in quantum mechanics, Mathematical Physics and Applied Mathematics Vol. 7*, ed. by M. Flato et al. (D. Reidel, Dordrecht, 1981).
[31] G.-L. Ingold, *Path Integrals and Their Application to Dissipative Quantum Systems*, Chap. 1 in *Coherent Evolution in Noisy Environments*, A. Buchleitner and K. Hornberger, Eds. (Springer-Verlag, Berlin, 2002). Note that there is a misprint in Eq. (75); there clearly should be a factor of $\alpha$ multiplying all occurrences of $f''(x_0)$.
[32] W. K. Burton and A. H. DeBorde, “The Evaluation of Transformation Functions by means of the Feynman Path Integral”, Nuovo Cim. **2**, 197-202 (1955).
[33] M. C. Gutzwiller, “Phase-Integral Approximation in Momentum Space and the Bound States of an Atom. II”, J. Math. Phys. **10**, 1004 (1969).
[34] R. Landauer, “Associated Legendre Polynomial Approximations”, J. Appl. Phys. **22**, 87 (1951).
[35] A. Norcliffe “Relating the classical and quantal theories of angular momentum and spin”, J. Phys. A: Gen. Phys. **5,** 231 (1972), and references therein.
[36] R. More, “Semiclassical matrix-mechanics. II. Angular momentum operators”, J. Phys. II France **1**, 97 (1991), and references therein.
[37] M. C. Gutzwiller, “The Geometry of Quantum Chaos”, Phys. Scr. **T9**, 184 (1985).
[38] E. Abdalla and R. Banerjee, “Quantisation of the Multidimensional Rotor”, Brazilian J. Phys. **31**, 80 (2001).
[39] L. S. Schulman (2005) *ibid.*, p. 214.
[40] H. Kleinert (2009) *ibid.*, note at end of Chapter 8, drawing a distinction between the solution of a short-time propagator and the solution of a path integral.
[41] H. Kleinert (2009) *ibid.*, note at the end of Chapter 10 pertaining to DeWitt’s work (Ref. [43]) and associated works of other authors.
[42] P. Cartier and C. DeWitt-Morette (2006) *ibid.*, p. 138.
[43] B. S. DeWitt, “Dynamical Theory in Curved Spaces. I. A Review of the Classical and Quantum Action Principles”, Rev. Mod. Phys. **29**, 377 (1957).
[44] C. Grosche and F. Steiner, “Path integrals on curved manifolds”, Z. Phys. C **36**, 699 (1987).
[45] C. Grosche and F. Steiner, “The Path Integral on the Pseudosphere”, Ann. Phys. (N.Y.) **182**, 120 (1988).
[46] F. Bastianelli and P. van Nieuwenhuizen, *Path Integrals and Anomalies in Curved Space* (Cambridge University Press, Cambridge, 2006).
[47] M. Ding and X. Xing, “Time-Slicing Path-integral in Curved Space”, Quantum **6**, 694 (2022), and references therein.
[48] R. P. Feynman and A. R. Hibbs (2005), *ibid.*, p. 177.
[49] L. F. Abbott and M. B. Wise, “Dimension of a quantum path”, Am. J. Phys. **49**, 37-39 (1981).
[50] C. DeWitt-Morette, K. D. Elworthy, B. L. Nelson, and G. S. Sammelman, “A stochastic scheme for constructing solutions of the Schrödinger equations”, Ann. Inst. Henri Poincaré A **32**, 327 (1980).

[51] D. J. Toms, “The Schwinger Action Principle and the Feynman Path Integral for Quantum Mechanics in Curved Space”, arXiv:hep-th/0411233v1 (2004).

[52] L. Parker, “Path integrals for a particle in curved space”, Phys. Rev. D **19**, 438 (1979).

[53] F. Langouche, D. Roekaerts, and E. Tirapegui, *Functional integration and semiclassical expansions,* Mathematics and its Applications, Vol. 10 (Springer Science+Business Media, Dordrecht, 1982), Section 6.2.

[54] C. Grosche and F. Steiner, “Path integrals on curved manifolds”, Z. Phys. C **36**, 699 (1987).

[55] A. Zee, *Einstein Gravity in a Nutshell* (Princeton University Press, Princeton, 2013), p. 590.

[56] C. Truesdell, “The Influence of Elasticity on Analysis: The Classical Heritage”, Bull. Amer. Math. Soc. **9**, 293 (1983).

[57] G. D. Birkhoff, “Dynamical systems with two degrees of freedom”, Trans. Amer. Math. Soc. **18,** 199 (1917).

[58] G. Brunnett and P. E. Crouch, “Elastic Curves on the Sphere”, Adv. Comput. Math. **2**, 23 (1994).

[59] S. Matsutani, “Euler’s Elastica and Beyond”, J. Geom. Symmetry Phys. **17**, 45 (2010).

[60] P. Fonda, V. Jejjala, and A. Veliz-Osorio, “On the shape of things: From holography to elastica”, Ann. Phys. **385**, 358 (2017).

[61] L. C. Biedenharn and J. D. Louck, *Angular Momentum in Quantum Physics* (Cambridge University Press, Cambridge, 1985).

[62] T. D. Lee, *Particle Physics and Introduction to Field Theory,* updated first ed. (Harwood Academic Publishers, London, 1988), Section 19.3, p. 480.

[63] D. H. McIntyre, *Quantum Mechanics: A Paradigms Approach* (Pearson, Boston, 2012).

[64] W. Pauli, “Über ein Kriterium für Ein- oder Zweiwertigkeit der Eigenfunktionen in der Wellenmechanik” Helv. Phys. Acta **12**, 147 (1939).

[65] E. Merzbacher, “Single Valuedness of Wave Functions”, Am. J. Phys. **30**, 237 (1962).

[66] J. S. Dowker and R. Critchley, “Scalar effective Lagrangian in de Sitter space”, Phys. Rev. D **13**, 224 (1976).

[67] C. M. Bender and S. A. Orzag, *Advanced Mathematical Methods for Scientists and Engineers I – Asymptotic Methods and Perturbation Theory* (Springer, New York, 1999).

[68] H. Kleinert and I. Mastapic, “Summing the spectral representations of Pöschl-Teller and Rosen-Morse fixed-energy amplitudes”, J. Math. Phys. **33**, 643 (1992).

Attachment

- fig 6, compute P(1) distribution as a function of characteristic angular momentum Lc
- in variable names below, an appended ‘p’ means prime, so e.g. Lpmin and Lpmax are min and max values of Lp = L’ (L-prime); an appended ‘t’ means tilde, so e.g. $\gamma$t = $\gamma$-tilde ($\gamma$ with tilde mark over it)
- coarse grid is shown below, with $n\Phi 0=8$; $n\alpha=8$; $n\theta f=8$; $n\phi f=1$; delL=0.025; requires about 15 minutes to run using Mathematica
- fine grid is $n\Phi 0=64$; $n\alpha=64$; $n\theta f=32$; $n\phi f=1$; delL=0.001; requires about 1 day to run using FORTRAN code, with that code running about 100x faster than Mathematica code having the same parameters (although part of that difference arises from the faster processor employed for the FORTRAN run)

```
In[•]:= Clear["Global`*"];
el = 1; m = 0; T = N[32 π]; nΦ0 = 8; nα = 8; nθf = 8; nϕf = 1; delL = 0.025;
delΦ0 = N[π / nΦ0]; delθf = N[π / nθf]; delϕf = N[2 π / nϕf];
Lpmin = -N[1 / Sqrt[T]]; Lpmax = N[1 / Sqrt[T]];
Lmin = -4.5; Lmax = 4.5; nL = Round[(Lmax - Lmin) / delL];
θ[γt_, θf_, ϕf_, Φ0_] :=
  N[(stf = Sin[θf]; ctf = Cos[θf]; ArcCos[-stf Cos[Φ0] Sin[γt] + ctf Cos[γt]])];
ϕ[γt_, θf_, ϕf_, Φ0_] := N[(stf = Sin[θf]; ctf = Cos[θf]; spf = Sin[ϕf]; cpf = Cos[ϕf];
    ArcTan[-spf Sin[Φ0] Sin[γt] + cpf (ctf Cos[Φ0] Sin[γt] + stf Cos[γt]),
     cpf Sin[Φ0] Sin[γt] + spf (ctf Cos[Φ0] Sin[γt] + stf Cos[γt])])];
n[γt_] := N[Floor[γt / (2 π)] - (1 + 2 Floor[γt / (2 π)]) Mod[Floor[γt / π], 2]];
γ[γt_] := N[ Abs[Mod[γt - π, 2 π] - π]];
prefact[T_] := (Sqrt[T] / 2) N[(1/ (√2 π ⅈ)) (1 / (2 π ⅈ T))^(1/2) Exp[ⅈ (el + 1 / 2)^2 T / 2]];
tab1 = Table[L = Lmin + (iL - 0.5) delL; L, {iL, nL}];
tab2 = prefact[T] × Table[L = Lmin + (iL - 0.5) delL;
    nLp = Max[Floor[5 Sqrt[T]], Floor[3 Sqrt[T] Abs[L]]]; delLp = N[(Lpmax - Lpmin) / nLp];
    delLp Total[Table[Lp = L + Lpmin + (iLp - 0.5) delLp; γt = -Lp T; Abs[Sin[γt]] delθf Total[
        Table[ θf = (iθf - 0.5) delθf; Sin[θf] delϕf Total[Table[ ϕf = (iϕf - 0.5) delϕf;
            delΦ0 Total[Table[Φ0 = (iΦ0 - 0.5) delΦ0;
                Conjugate[SphericalHarmonicY[el, m, θf, ϕf]]
                 SphericalHarmonicY[el, m, θ[γt, θf, ϕf, Φ0], ϕ[γt, θf, ϕf, Φ0]]
                 (-1)^n[γt] (αmin = γ[γt]; αmax = π; delα = (αmax - αmin) / nα;
                  Total[Table[α = αmin + (iα - 0.5) delα; αn = (α + 2 N[π] × n[γt]);
                    If[iα ≠ 1, delα αn Exp[ⅈ αn^2 / (2 T)] / (Cos[γ[γt]] - Cos[α])^(1/2),
                     α1n = γ[γt] + delα / 4 + 2 N[π] × n[γt];
                     α1n Exp[ⅈ α1n^2 / (2 T)] 2 ((√delα) / (√Sin[γ[γt] + delα / 8]))],
                    {iα, nα}]]), {iΦ0, nΦ0}]],
            {iϕf, nϕf}]], {iθf, nθf}]], {iLp, nLp}]], {iL, nL}];
fname = "tab1.dat"; Print[fname]; stmp = OpenWrite[fname]; Write[stmp, tab1];
Close[stmp];
fname = "tab2.dat"; Print[fname];
stmp = OpenWrite[fname]; Write[stmp, tab2]; Close[stmp];
Show[{ListPlot[Transpose[{tab1, Re[tab2]}], PlotRange → {{-4.5, 4.5}, {-3, 5}},
   PlotStyle → {ColorData[97, "ColorList"]〚1〛}], ListPlot[Transpose[{tab1, Im[tab2]}],
   PlotStyle → {ColorData[97, "ColorList"]〚2〛}]}, ImageSize → Medium]
```

- figs 7, 11, 12 - compute P(2) angular distribution
- same comments here as for preceding fig 6 code, except replace nΦ0 in the comments there with nθg as applicable here

```
Clear["Global`*"];
el = 1; m = 0; T = N[32 π]; nθg = 8; nα = 8; nθf = 8; nϕf = 1; delL = 0.025;
delθg = N[(π / 2) / nθg];  delϕf = N[2 π / nϕf];
Lpmin = -N[1 / Sqrt[T]]; Lpmax = N[1 / Sqrt[T]];
θ[γt_, θf_, ϕf_, Φ0_] :=
  N[(stf = Sin[θf]; ctf = Cos[θf]; ArcCos[-stf Cos[Φ0] Sin[γt] + ctf Cos[γt]])];
ϕ[γt_, θf_, ϕf_, Φ0_] := N[(stf = Sin[θf]; ctf = Cos[θf]; spf = Sin[ϕf]; cpf = Cos[ϕf];
    ArcTan[-spf Sin[Φ0] Sin[γt] + cpf (ctf Cos[Φ0] Sin[γt] + stf Cos[γt]),
     cpf Sin[Φ0] Sin[γt] + spf (ctf Cos[Φ0] Sin[γt] + stf Cos[γt])])];
n[γt_] := N[Floor[γt / (2 π)] - (1 + 2 Floor[γt / (2 π)]) Mod[Floor[γt / π], 2]];
γ[γt_] := N[ Abs[Mod[γt - π, 2 π] - π]];
prefact[T_] := N[(1 / (√2 π ⅈ)) (1 / (2 π ⅈ T))^(1/2) Exp[ⅈ (el + 1 / 2)^2 T / 2]];
tab[Lmin_, Lmax_, nL_] := Table[θg = (iθg - 0.5) delθg;
   θfmin = θg;  θfmax = N[π] - θg; delθf = N[(θfmax - θfmin) / nθf];
   (Sqrt[T] prefact[T] delL / 2) Total[Table[L = Lmin + (iL - 0.5) delL;  nLp =
       Max[Floor[5 Sqrt[T]], Floor[3 Sqrt[T] Abs[L]]];  delLp = N[(Lpmax - Lpmin) / nLp];
      delLp Total[Table[Lp = L + Lpmin + (iLp - 0.5) delLp;  γt = -Lp T; Abs[Sin[γt]] ×
          Total[Table[ θf = θfmin + (iθf - 0.5) delθf;
            delϕf Total[Table[ ϕf = (iϕf - 0.5) delϕf;
                Φ0 = ArcTan[((Tan[θf])^2 / ((Cos[θg] / Cos[θf])^2 - 1)  - 1)^(1/2)];  Φ0p = π - Φ0;
                If[iθf ≠ 1 && iθf ≠ nθf, delθf Sin[θf]
                   (Sin[θg] / ((Cos[θf])^2 ((Tan[θf])^2 - ((Cos[θg] / Cos[θf])^2 - 1) )^(1/2)
                       ((Cos[θg] / Cos[θf])^2 - 1)^(1/2))), (2 delθf Abs[Tan[θf]])^(1/2)] ×
                 Conjugate[SphericalHarmonicY[el, m, θf, ϕf]]
                 (SphericalHarmonicY[el, m, θ[γt, θf, ϕf, Φ0], ϕ[γt, θf, ϕf, Φ0]] +
                   SphericalHarmonicY[el, m, θ[γt, θf, ϕf, Φ0p], ϕ[γt, θf, ϕf, Φ0p]])
                 (-1)^n[γt] (αmin = γ[γt]; αmax = π; delα = (αmax - αmin) / nα;
                  Total[Table[α = αmin + (iα - 0.5) delα; αn = (α + 2 N[π] × n[γt]);
                    If[iα ≠ 1, delα αn Exp[ⅈ αn^2 / (2 T)] / (Cos[γ[γt]] - Cos[α])^(1/2),
                     α1n = γ[γt] + delα / 4 + 2 N[π] × n[γt];
                     α1n Exp[ⅈ α1n^2 / (2 T)] 2 ((√delα) / (√Sin[γ[γt] + delα / 8]))],
                    {iα, nα}]]), {iϕf, nϕf}]],
           {iθf, nθf}]], {iLp, nLp}]], {iL, nL}]], {iθg, nθg}];
tab1 = Table[θg = (iθg - 0.5) delθg; θg, {iθg, nθg}];
Lmin = 0; Lmax = 3 (el + 1 / 2);  nL = Round[(Lmax - Lmin) / delL];
tab2 = tab[Lmin, Lmax, nL];
tabx1 = Table[θg = (iθg - 0.5) delθg;
```

```
    ang = θg;
    rad = tab2⟦iθg⟧;
    rad Cos[ang], {iθg, nθg}];
taby1 = Table[θg = (iθg - 0.5) delθg;
    ang = θg;
    rad = tab2⟦iθg⟧;
    rad Sin[ang], {iθg, nθg}];
tabx2 = Table[θg = (iθg - 0.5) delθg;
    ang = θg;
    rad = tab2⟦iθg⟧;
    -rad Cos[ang], {iθg, nθg, 1, -1}];
taby2 = Table[θg = (iθg - 0.5) delθg;
    ang = θg;
    rad = tab2⟦iθg⟧;
    rad Sin[ang], {iθg, nθg, 1, -1}];
Lmin = -3 (el + 1 / 2); Lmax = 0;  nL = Round[(Lmax - Lmin) / delL];
tab3 = tab[Lmin, Lmax, nL];
tabx3 = Table[θg = (iθg - 0.5) delθg;
    ang = θg;
    rad = tab3⟦iθg⟧;
    -rad Cos[ang], {iθg, nθg}];
taby3 = Table[θg = (iθg - 0.5) delθg;
    ang = θg;
    rad = tab3⟦iθg⟧;
    -rad Sin[ang], {iθg, nθg}];
tabx4 = Table[θg = (iθg - 0.5) delθg;
    ang = θg;
    rad = tab3⟦iθg⟧;
    rad Cos[ang], {iθg, nθg, 1, -1}];
taby4 = Table[θg = (iθg - 0.5) delθg;
    ang = θg;
    rad = tab3⟦iθg⟧;
    -rad Sin[ang], {iθg, nθg, 1, -1}];
tabx = Join[tabx1, tabx2, tabx3, tabx4, {tabx1⟦1⟧}];
taby = Join[taby1, taby2, taby3, taby4, {taby1⟦1⟧}];
fname = "tab1.dat"; Print[fname];
stmp = OpenWrite[fname]; Write[stmp, tab1];  Close[stmp];
fname = "tab2.dat"; Print[fname];
stmp = OpenWrite[fname]; Write[stmp, tab2];  Close[stmp];
fname = "tab3.dat"; Print[fname];
stmp = OpenWrite[fname]; Write[stmp, tab3];  Close[stmp];
ListLinePlot[Transpose[{tabx, taby}], PlotRange → {{-1, 1}, {-1, 1}}, AspectRatio → 1]
```